\documentclass[twoside,letterpaper,phd]{macthesis}
\usepackage{fancyhdr}
\usepackage{deluxetable}
\usepackage{natbib} 
\usepackage{aastex_hack} 
\usepackage{rotating}

\usepackage{amsmath}
\usepackage{amssymb}
\usepackage{amsthm}
\usepackage{exscale}
\usepackage[mathscr]{eucal}
\usepackage{bm}
\usepackage{eqlist} 

\usepackage{epigraph}
\usepackage{lettrine}
\usepackage{keyval}

\usepackage[dvipsnames]{color}

\usepackage{multirow}

\usepackage[Lenny]{fncychap}
\ChTitleVar{\Huge\sffamily\bfseries}

\title{The Physics of Mergers:  Theoretical and Statistical Techniques \\
  Applied to Stellar Mergers in Dense Star Clusters}
\author{Nathan W. C. Leigh}
\dept{Physics and Astronomy}
\degreedate{June 2011}
\copyrightyear{2011}

\documenttype{Thesis}

\submittedto{The Graduate School}

\numberofreaders{2}

\advisor[Thesis Advisor]
        {Dr. Alison Sills}
        {Professor, McMaster University}

\readerone[Committee Member]
        {Dr. William Harris}
        {Professor, McMaster University}

\readertwo[Committee member]
          {Dr. Alan Chen}
          {Professor, McMaster University}

\begin{document}
\parindent=0.5in


\frontmatter

%



\mactitlepage

%
\pagestyle{plain}
    \begin{onehalfspace}




\pagestyle{empty}
\vspace*{0.5in}

\noindent 
DOCTOR OF PHILOSOPHY (2010) \hspace{2.0cm} McMaster University \\
(Physics and Astronomy)     \hspace{4.2cm} Hamilton, Ontario 
\vspace{0.5in}

\noindent
TITLE: The Physics of Mergers:  Theoretical and Statistical Techniques Applied to Stellar Mergers in Dense Star Clusters \\ \\
AUTHOR: Nathan W. C. Leigh, B.Sc. (University of Toronto), M.Sc. (McMaster University)\\ \\
SUPERVISOR: Professor Alison Sills\\ \\
NUMBER OF PAGES: xvi, 184

    \end{onehalfspace}
\newpage


\pagestyle{plain}
    \begin{onehalfspace}
\addcontentsline{toc}{chapter}{Abstract}
        \thispagestyle{fancy}

\textrm{}\\\\
\noindent\textbf{\huge\textsf{Abstract}}\\\\

\noindent In this thesis, we present theoretical and statistical
techniques broadly related to systems of dynamically-interacting
particles.  We apply these techniques to observations of dense 
star clusters in order to study gravitational interactions between
stars.  These include both long- and short-range interactions, as well
as encounters leading to direct collisions and mergers.  The latter
have long been suspected to be an important formation channel for
several curious types of stars whose origins are unknown.  The former
drive the structural evolution of star clusters and, by leading to
their eventual dissolution and the subsequent dispersal of their
stars throughout the Milky Way Galaxy, have played an important role
in shaping its history.  Within the last few decades, theoretical work
has painted a comprehensive picture for the evolution of star
clusters.  And yet, we are still lacking direct observational
confirmation that many of the processes thought to be driving this
evolution are actually occuring.  The results presented in
this thesis have connected several of these processes to real
observations of star clusters, in many cases for the first time.  This
has allowed us to directly link the observed
properties of several stellar populations to the physical
processes responsible for their origins.

We present a new method of quantifying the frequency of encounters
involving single, binary and triple stars using an adaptation of the
classical mean free path approximation.  With this technique, we have
shown that dynamical encounters involving triple stars occur commonly
in star clusters, and that they are likely to be an important
dynamical channel 
for stellar mergers to occur.  This is a new result that has important
implications for the origins of several peculiar types of stars (and
binary stars), in
particular blue stragglers.  We further present several new
statistical techniques that are broadly applicable to systems of
dynamically-interacting particles composed of several different types
of populations.  These are applied to observations of star clusters in
order to obtain quantitative constraints for the degree to 
which dynamical interactions affect the relative sizes and spatial
distributions of their different stellar populations.  To this end, 
we perform an extensive analysis of a large sample of colour-magnitude
diagrams taken from the ACS Survey for Globular Clusters.  The results
of this analysis can be summarized as follows:  (1) We have compiled a
homogeneous catalogue of stellar populations, including main-sequence,
main-sequence turn-off, red giant branch, horizontal branch and blue
straggler stars.  (2) With this catalogue, we have quantified the
effects of the cluster dynamics in determining the relative sizes
and spatial distributions of these stellar populations.  (3) These
results are particularly interesting for blue stragglers since they
provide compelling evidence that they are descended from binary stars.
(4) Our analysis of the main-sequence populations is consistent with 
a remarkably universal initial stellar mass function in old massive
star clusters in the Milky Way.  This is a new result with 
important implications for our understanding of star formation in the
early Universe and, 
more generally, the history of our Galaxy.  Finally, we describe how
the techniques presented in this thesis are ideally suited for
application to a number of other outstanding puzzles of modern
astrophysics, including chemical reactions in the interstellar medium
and mergers between galaxies in galaxy clusters and groups.  

\newpage
\thispagestyle{empty}
\mbox{}
    \end{onehalfspace}
\newpage

\thesisdedication{SupplementaryMaterial/Dedication}{}

%
\addcontentsline{toc}{chapter}{Co-Authorship}
\begin{onehalfspace}
    \thispagestyle{fancy}

\textrm{}\\\\
\noindent\textbf{\huge\textsf{Co-Authorship}}\\\\

\noindent 
Chapters 2, 3, 4, and 5 of this thesis all represent original papers written by
myself, Nathan W. C. Leigh and published in the \textit{Monthly
Notices of the Royal Astronomical Society} (\textit{MNRAS}).  The
reference for Chapter 
2 is Leigh N., Sills A. 2010, \textit{MNRAS}, 410, 2370, and Dr. Alison Sills is
the only co-author.  The reference for Chapter 3 is Leigh N., Sills
A., Knigge C. 2009, \textit{MNRAS}, 399L, 179, and Dr. Alison Sills and
Dr. Christian Knigge are co-authors.  Chapters 4 and 5 have both been
accepted for publication in the \textit{MNRAS}, however they had not
yet been assigned journal codes at the time of publication of this
thesis.  Dr. Alison Sills and Dr. Christian Knigge are co-authors on 
both of these two publications.  Dr. Knigge provided a great deal of guidance in
choosing the appropriate statistical techniques throughout all of our
analyses, and offered
considerable advice in analyzing and interpreting our results.  He
also had a number of noteworthy 
creative contributions to the analytic model presented in Chapter 5.
All previously published material has been 
reformatted to conform to the required thesis 
style.  I grant an irrevocable, non-exclusive license to McMaster
University and the National Library of Canada to reproduce this
material as part of this thesis.
\newpage
\thispagestyle{empty}
\mbox{}
\end{onehalfspace}

%
\addcontentsline{toc}{chapter}{Acknowledgments}
    \begin{onehalfspace}
        \thispagestyle{fancy}

\textrm{}\\\\
\noindent\textbf{\huge\textsf{Acknowledgements}}\\\\

\noindent The work presented in this thesis represents the culmination
of over 6 years of research.  I did not embark upon this endeavor
alone, and here I wish to thank the people who supported me the most
along the way.

First and foremost, I owe a profound thank you to my thesis
supervisor Dr. Alison Sills. Her patient guidance, support, and
friendship made my journey through graduate school not only
successful, but also a hell of a lot of fun.  
I am grateful for all of her creative input and meticulous advice, and
this thesis has been made immeasurably stronger because of it.  Her
dedicated support of my education and her relentless encouragement of
my pursuits were unwavering.  The woman is a hero.  A heart-felt thank
you must also be extended to Dr. Christian Knigge and Dr. Robert
Mathieu whose guidance, advice and tutelage have played a significant
role in shaping my abilities and identity as a researcher and scientist.

I also wish to thank my parents.  I am incredibly fortunate to be able
to look upon my parents as my greatest and most trusted friends.
There is nobody in this world who 
knows me better, and your acceptance and support have meant everything
to me.  This thesis is for you.  A sincere thank you to my brother
Peter as well, for setting 
the bar high when it came to a great many things.  Despite being the
big brother, I have looked up to you often.

I also wish to thank my thesis advisors and committtee members
Dr. Bill Harris, Dr. Alan Chen, Dr. Michael Shara, Dr. Christine
Wilson and Dr. Ethan Vishniac.  Your insight and guidance
have helped me greatly, and this thesis would not be what it is 
without you.

Finally, I wish to thank my friends and collaborators.  You have
influenced me in countless ways, and have helped shape my perception
of the world.  Not only have your words given me inspiration, they
have made me laugh, smile and cry time and time again.  For this
I will always be grateful.  There are too many of you to list here by
name, however Dr. Aaron Geller, 
Dr. Evert Glebbeek, Dr. Charles-Philippe Lajoie and Dr. David
Chernoff must be mentioned 
explicitly for their contributions to the work presented in this
thesis, and for their exhaustive efforts in dumbing-down concepts to a
sufficiently basic level that I can understand them.   

    \end{onehalfspace}

%
\begin{singlespace}
    \thispagestyle{empty}
\setlength{\epigraphwidth}{5.in}

\linespread{1.5}
\epigraph{\vspace{1.5in}\normalsize ``Imagination is more important
than knowlege.''}
  {\normalsize \textsc{\\Albert Einstein} (1879-1955)}

\linespread{1.5}
\epigraph{\vspace{1.5in}\normalsize ``The surest way to corrupt a
youth is to instruct him to hold in higher
esteem those who think alike than those who think differently.''}
  {\normalsize \textsc{\\Friedrich Nietzsche} (1844-1900)}

\linespread{1.5}
\epigraph{\vspace{1.5in}\normalsize ``When the going gets weird, the
weird turn pro.''}
  {\normalsize \textsc{\\Hunter S. Thompson} (1937-2005)}


\linespread{1.5}
\epigraph{\vspace{1.5in}\normalsize ``Instead of building newer and
larger weapons of mass destruction, I think mankind should try to get
more use out of the ones we have.''}
  {\normalsize \textsc{\\Jack Handey} (1949-present)}


\newpage
\thispagestyle{empty}
\mbox{}


\end{singlespace}
\newpage
VC
\thesistableofcontents

\thesislistoffigures

\thesislistoftables





\thesismainmatter

\allowdisplaybreaks{
%
\begin{doublespace}
\pagestyle{fancy}
\headheight 20pt
\lhead{Ph.D. Thesis --- N. Leigh }
\rhead{McMaster - Physics \& Astronomy}
\chead{}
\lfoot{}
\cfoot{\thepage}
\rfoot{}
\renewcommand{\headrulewidth}{0.1pt}
\renewcommand{\footrulewidth}{0.1pt}

\chapter{Introduction} \label{chapter1} 
\thispagestyle{fancy} 

%
The vast majority of the stars in our
Universe are found in galaxies, each of which is typically populated
by billions of members.  The hierarchy of stellar communities within
galaxies can be understood if an analogy is made between stars and
people.  In this context, parallels can be drawn between galaxies and
countries, star clusters and cities, as well as multiple star systems
and families.  It is now known that at least half of the stars in
galaxies were born in star clusters \citep[e.g.][]{kroupa02a,
  kroupa02b}, spherical conglomerates 
composed of anywhere from a few hundred to a few million members all
orbiting their common center of mass (see
Figure~\ref{fig:NGC7078}).  Within these stellar 
nurseries, around 50\% of all stars are thought to spend the
majority of their lives in 
close proximity to one or more of their siblings by forming multiple 
star systems \citep[e.g.][]{durisen94, bate97a, bate97b, kroupa01,
  sollima10}.  Most of these are 
binary star systems, which consist of two stars orbiting their common 
centre of mass under the influence of their mutual gravitation
attraction.  Although less common, stable configurations of triple,
quadruple and quintuple star systems also exist, along with even
higher order multiples \citep[e.g.][]{latham05, eggleton08,
  tokovinin08}.  Despite the 
fact that stars typically spend at most a small fraction of their
considerably long lives in even relative isolation, our understanding
of the mechanisms via which the presence of their siblings can affect
their evolution is far from complete.  

\begin{figure} [!h]
  \begin{center}
 \includegraphics[scale=0.5]{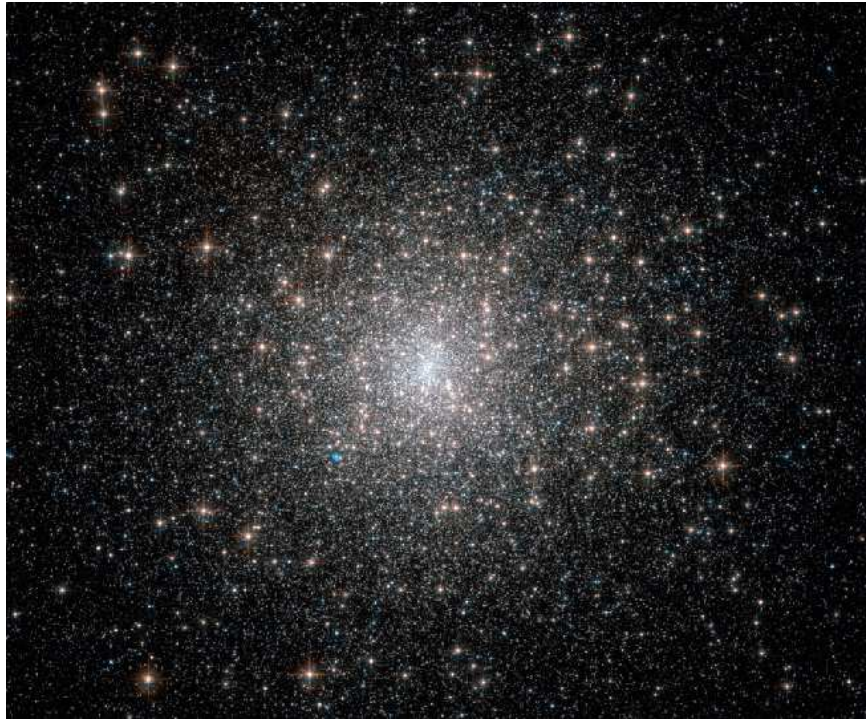}
  \caption[An image taken with the \textit{Hubble Space Telescope} of
  the globular cluster NGC 7078]{An image taken with the
    \textit{Hubble Space Telescope} of NGC 7078 (M15), a globular
    cluster in the Milky Way galaxy.  Image credit:  NASA, ESA, and
    The Hubble Heritage Team.
      \label{fig:NGC7078}}
    \end{center}
\end{figure}  

Gravity mediates a wide variety of interactions between stars.  Within
the dense cores of star clusters, for instance, strong gravitational 
encounters often occur in which the distance of closest approach is
comparable to the radii of the stars \citep[e.g.][]{spitzer72,
  spitzer75, heggie75, hut83b}.  These encounters often lead to
complex dances that can result in stars being ejected from their host
clusters \citep[e.g.][]{henon69, hut92, kroupa02b, demarchi10}, or even
stellar collisions \citep[e.g.][]{leonard89, sills01}.
Binary stars are also at the
mercy of gravity and, if their orbital separation becomes sufficiently
small, this can lead to the transfer of mass from the surface of one
component to the other \citep{mccrea64}, or even their complete
coalescence \citep[e.g.][]{andronov06}.
Not only are gravitational interactions important for stellar mergers,
they also drive the evolution of star clusters.  Gravity slowly acts
to dissolve clusters, continuously ejecting stars so that they may
join the rest of the Galactic population
\citep[e.g.][]{portegieszwart01}.  It follows that a complete 
understanding of the
dominant physical processes operating in clusters, including how their
stars are formed and the dynamical interactions that cause their stars
to escape, is a key ingredient in piecing together the history of our
Galaxy.  

Mergers are a particularly interesting and important outcome of
stellar interactions.  Observations have revealed 
numerous examples of curious stars and mysterious astrophysical
processes that are thought to be related to stellar mergers
\citep[e.g.][]{sandage53, webbink84, paczynski86, troja10,
  miroshnichenko07, farrell09}.  
Many
of these objects are commonly found in dense star clusters, hinting at
the complexity of their dynamical evolution and the potential
importance of dynamics for stellar mergers.  
Although their origins remain unknown, some of these 
objects have served as important tools for furthering our 
understanding of the Universe.  For instance, Type Ia supernovae
have provided the most robust constraints to date for the rate of
Universal expansion \citep[e.g.][]{perlmutter99}.  Despite their 
considerable astrophysical significance, we are far from completely
understanding the physical mechanisms that cause mergers to occur.  

Stellar mergers are but one example of this ubiquitous
physical process, which occurs between objects on many scales.
From colliding galaxies to fusing atoms,
mergers occur on spatial and temporal scales ranging by
many orders of magnitude.  Familiar principles such as conservation of
energy and momentum often apply, however, regardless of scale.  
This is certainly the case for the various forms of dynamical
interactions that lead to mergers.  
Examples include chemical reactions in the interstellar medium, strong
gravitational interactions involving massive black holes and even 
encounters between galaxies in galaxy clusters.  
It follows that the development of tools designed to further
our understanding of stellar dynamics and mergers also have
applications for a number of other physical sub-disciplines.

In this thesis, we present statistical and theoretical techniques
related to the physics of mergers.  These are applied to observations
of dense star clusters in order to study the mergers of stars.  In
this chapter, we will outline our motivation for conducting this 
research.  In
Section~\ref{SPs}, we briefly review our current
understanding of stellar evolution theory, and discuss the observed
properties of star clusters and their stellar populations, including
the various types of stellar exotica.  In
Section~\ref{dynamics_intro}, we provide a brief description of the dominant
physical processes that drive the dynamical evolution of star
clusters.  All of these issues are connected in Section~\ref{wheretogo},
where we describe how they have motivated the development of the
techniques that will be presented in this thesis, along with their
application to observations of star clusters.

\section{Stellar Populations in Star Clusters} \label{SPs}

\subsection{Single Star Evolution} \label{standard}

By serving as sites for star formation, star clusters have played a
crucial role in shaping the present-day features of our Galaxy.  And
yet, this is but one example of their 
astrophysical significance.  Star clusters are also ideal
laboratories for learning about stellar evolution.  Historically,
their importance in this regard has stemmed from the often-adopted
assumption that all of the stars in a given cluster were born 
from the same gas cloud at more or less the same time.  If true, star
clusters offer a large sample of stars with a wide range of masses,
but a very narrow range in both age and chemical composition.
Therefore, observations of star clusters offer robust constraints for
stellar evolution theories by providing direct tests for their
predictions for a diverse spectrum of stellar masses of known age and
composition. 

Colour-magnitude diagrams (CMDs) are one of the most
important tools available to astronomers for studying stellar
evolution \citep{hertzsprung09}.  In these diagrams, the
observed surface temperature is 
plotted against luminosity (or brightness) for every star in the 
cluster.  The latter quantity is plotted backwards, and provides a
proxy for colour -- hot stars are blue and cool stars are red.
This creates a characteristic appearance that is observed in all
cluster CMDs.  An example is shown
in Figure~\ref{fig:CMD} for the GC NGC 
6205.  This characteristic shape can be understood by
considering two principles of stellar 
evolution.  The first is called the Vogt-Russell Theorem and states
that the mass, composition and age of a star uniquely determine its radius,
luminosity and internal structure, in addition to its subsequent
evolution \citep{vogt25, russell25}.  The second is simply that the
rate at which a star evolves, and in so doing changes its brightness
and colour, is inversely proportional to some power of its
mass \citep[e.g.][]{iben91}.  It follows that, for a cluster composed
of a large number of stars with a distribution of masses
but similar compositions (at birth) and ages, the most massive stars
are also the most evolved.  Therefore, every evolutionary phase in
the life of a star is typically represented in at least moderately old
clusters, and this causes the distribution of stars in the 
CMD to adhere to the characteristic shape shown in Figure~\ref{fig:CMD}.  
Our understanding of single star evolution is sufficiently
complete that it is now known how this shape changes as a
function of composition and age.  In general, the majority of the features
characteristic of CMDs are well understood, although exceptions
do exist.  Before getting to these in Section~\ref{extra}, we
will review the life cycle of a typical star in our Galaxy and connect
each evolutionary phase to its corresponding location in the CMD.

\begin{figure} [!h]
  \begin{center}
 \includegraphics[scale=0.5]{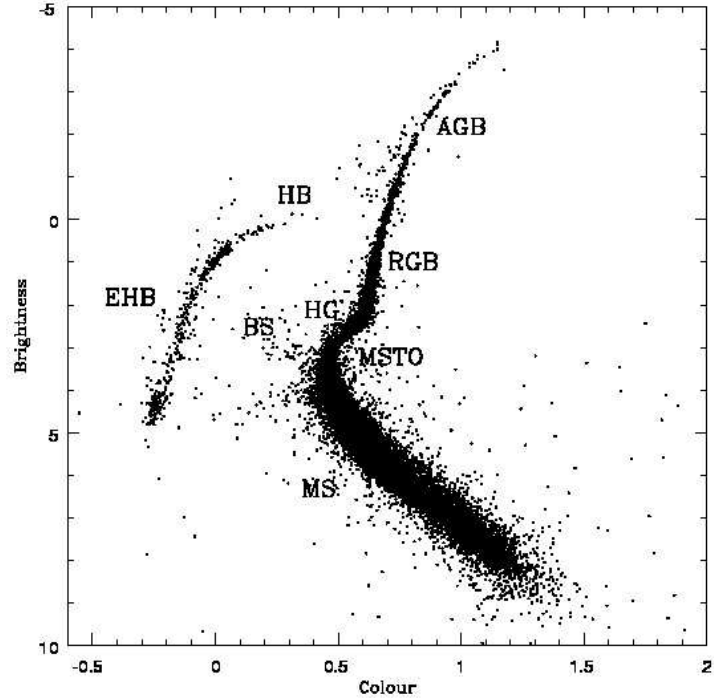}
    \caption[CMD for the Milky Way globular cluster NGC
    6205]{Colour-magnitude diagram for the Milky Way globular cluster 
      NGC 6205.  The data used to create this plot were taken from
      \citet{sarajedini07}.  
      As described in the text, the x-axis can be read as temperature
      (or, equivalently, colour), 
      which increases from right to left (red to blue).  On the
      y-axis, brightness increases from bottom to top.  
      Labels for blue straggler (BS), red giant branch (RGB), horizontal
      branch (HB), extended horizontal branch (EHB), asymptotic giant
      branch (AGB), Hertzsprung gap (HG), main-sequence turn-off (MSTO) and
      main-sequence (MS) stars are shown.  Stars with large photometric errors
      have been omitted from this plot. 
      \label{fig:CMD}}
    \end{center}
\end{figure}

Most of the star clusters in our Galaxy were born from gas clouds
composed primarily of hydrogen and, to a lesser extent, helium (as
well as heavier elements in trace amounts) \citep[e.g.][]{lada85, pringle89}.  
Gravity contracted the gas into clumps, which grew increasingly
hot and dense as they accreted material from the surrounding medium.
When the central temperature of a clump becomes sufficiently high (on
the order of $10^7$ K), hydrogen is ignited at its centre.  A star is
born.  This marks 
the beginning of the main-sequence (MS) phase, during which
time hydrogen is fused into helium in the stellar core.

Let us consider the life of a typical low-mass star in the Milky Way, from
beginning to end.  Most of the stars in our Galaxy have low
masses roughly spanning the range $0.08 \lesssim$ m $\lesssim 2.0$
M$_{\odot}$ \citep[e.g.][]{kroupa02a}.  There are two reasons for this.
First, the
least massive stars are also the longest lived.  Second, 
\citet{salpeter55} showed that the initial distribution of
stellar masses for stars in the solar-neighbourhood with masses in the
range $0.4 - 10$ M$_{\odot}$ can be described
by a power-law form with an index $\alpha \sim 2.35$.  Today, we know
that the distribution flattens below $\sim 0.5$ M$_{\odot}$, however
the primary conclusion is the same:  most of the stars in our Galaxy
have masses lower than that of our Sun.  What's more, most of these
are MS stars since, for stars of very low-mass, the time-scale for this
evolutionary phase exceeds the age of the Universe.  

Nuclear reactions in the stellar core release heat, and are the
primary source of energy generation within stars, driving their high
luminosities \citep[e.g.][]{clayton68}.  The transport of energy
outward within stars in turn 
causes their compositional and structural profiles to evolve.  This
occurs on a typically slow time-scale on the order of millions of
years or longer \citep[e.g.][]{kippenhahn90, maeder09}.  Both the rate
and outcome of the complex interplay of physical processes occurring
within stars to drive their evolution depend somewhat sensitively on
their mass.  As a 
result, stellar evolution varies considerably between low- and
high-mass stars.  Below, we focus our description of this evolution to
stars with masses $\lesssim 2.0$ M$_{\odot}$.

The MS crosses the CMD from bottom right to top left, as shown in
Figure~\ref{fig:CMD}.  During the MS phase, the radius (and hence 
surface temperature) and luminosity
of a star are primarily determined by its mass, however its age and
initial composition also contribute \citep[e.g.][]{iben91, tout96}.  As a
result, the lowest mass stars in a cluster occupy 
the bottom right end of the MS, with stellar mass increasing 
upward and blue-ward.  Most stars spend the majority of
their lives on the MS slowly converting hydrogen into helium in their
cores.  After slowly increasing their
radius and luminosity by a factor of about 2.5-3, stars reach the end
of the MS phase of their evolution \citep{eggleton06}.  This can
roughly be defined as the point at which the hydrogen fuel within a
central core containing about 10\% of the star's mass has run out.  
Found at the left-most tip of the MS in the CMD, called the
main-sequence turn-off (MSTO), the most massive MS 
stars are in the process of depleting their hydrogen fuel in and near
their centres.  At this point, loosely
referred to as the terminal-age main-sequence (TAMS), stars leave the
MS and start moving to the right across
the CMD, becoming increasingly red.  

The stellar evolution time-scales shorten considerably at this point, and
the helium core begins to contract while the envelope expands.  This
causes a drop in the surface temperature of the star, causing it to
move horizontally across the CMD, crossing what
is called the Hertzsprung gap (Figure~\ref{fig:CMD})
\citep[e.g.][]{popper80}.  The reason for 
this rapid reddening can be understood as follows.  Once the
surface temperature drops well below about $10^4$ K, the primary
mechanism of energy transport in the envelope changes
\citep{eggleton06}.  This is
because the drop in temperature
allows free electrons to recombine with free protons to form
hydrogen atoms, which 
can then act to absorb out-going radiation.  This presents a challenge
for the star since energy still needs to be
transported outward from its nuclear-burning centre, yet radiation can
no longer leak out freely.  To compensate,
convection takes over as the dominant form of energy transport
\citep[e.g.][]{bohm-vitense58}.  As
the convective 
base of the envelope deepens, stars of a given luminosity and mass
converge to an approximately unique radius.  The change in stellar
radius that occurs during this short-lived phase of evolution can
range from a factor of less than two for very low-mass stars to a
factor of $\gtrsim 100$ for very massive stars \citep{iben91}.  This
marks the beginning 
of the red giant branch (RGB) phase of evolution, and the base of its
corresponding sequence in the CMD (Figure~\ref{fig:CMD}).  

Due to the unique envelope structure created by the deepening of the
convective envelope, stars of a given mass but different luminosities
must lie on a roughly vertical locus in the CMD
\citep{hayashi62}.  This structure also 
causes the stellar luminosity to increase steeply as a function of the
mass of the helium core \citep[e.g.][]{iben68}.  As hydrogen burning
progresses in 
a shell immediately outside the core, the helium that is produced
rains down onto it, slowly increasing its mass.  This in turn causes
the star to brighten by up to several orders of 
magnitude and, in so doing, ascend the RGB in the CMD.  

Eventually, the mass and temperature of the helium core reach $\sim
0.47$ M$_{\odot}$ and $\sim 10^8$ K, respectively, although the
precise values depend on both the total stellar mass and chemical
composition \citep{eggleton06}.  It is at this 
point that helium ignites, producing mainly carbon at first, although
later the production of oxygen takes over.  The horizontal branch (HB) in
CMDs (Figure~\ref{fig:CMD}) corresponds to
the core-helium burning phase.  It is not clear why HBs show such a
large spread in their colours since, in principle, both the luminosity
and surface temperature should be roughly constant for core
helium-burning stars.  This mysterious and ubiquitous feature of CMDs
marks a significant gap in our understanding of stellar evolution
theory.  We will return to the curious HB morphologies observed in
the CMDs of old star clusters in Section~\ref{HBs_intro}.

Within about $10^8$ years of the onset of core helium-burning,
low-mass stars will typically possess a core composed of carbon and
oxygen with a mass of $\sim 0.55$ M$_{\odot}$ \citep{iben74}.  The
core is surrounded by a much less massive shell composed primarily of
helium and next to no hydrogen.  This, in turn, is surrounded by an
envelope of unprocessed material extending out to 30-50 R$_{\odot}$
from the core \citep{maeder09}.  Two burning shells now power the
star as it continues 
its rise in luminosity above the RGB.  The sequence in the CMD
corresponding to this phase of evolution is called the asymptotic
giant branch (AGB).  As this phase progresses, the radius and
luminosity of the star
continue to grow and severe mass-loss typically occurs due to
powerful stellar winds.

The life of a typical low-mass star ends when the last of its
envelope is burnt, leaving its luminosity to plummet by several
orders of magnitude and its surface temperature to increase
dramatically.  The product is an incredibly dense remnant composed
of carbon and oxygen called a C/O white dwarf (WD) \citep{iben74}.  
This the case for stars with masses up to $\sim 2$ M$_{\odot}$ when
mass-loss on the AGB is factored in \citep{eggleton06}.  For stars
more massive than 
this, stellar death can be considerably more dramatic, ending in the
ignition of carbon and a subsequent thermonuclear explosion.  As
previously explained, these stars are short-lived and do not occur
commonly in old star clusters, which will be the focus of this
thesis.  Therefore, we refer the interested reader to
\citet{clayton68} and \citet{maeder09} for descriptions of the
evolution of massive stars.

\subsection{Additional Features of Colour-Magnitude Diagrams} \label{extra}

There remain several features characteristic of CMDs that cannot be
explained by standard single star evolution.  In order to
account for their existence, it is 
necessary to invoke the aid of other physical processes known to be
operating in star clusters, such as 
binary star evolution and stellar dynamics.  As described in the
subsequent sections, these processes are thought to be responsible for
producing various types of exotic stellar populations and multiple
star systems.  These curious objects typically appear as outliers in
CMDs, and do not fall on any single star evolution tracks.  
We will discuss the various mysterious features of CMDs and types of
stellar exotica known to populate star
clusters, and review the currently favoured hypotheses for their
origins.

\subsubsection{Horizontal Branch Morphology} \label{HBs_intro}

The horizontal branches of star clusters in the Milky Way have been
observed to display a range of morphologies.  That is, most CMDs have
a horizontal sequence that extends blue-ward from the top of the red
giant branch, and this is referred to as the horizontal branch.
However, the length and appearance of this sequence varies
considerably from cluster-to-cluster and, in some cases, it makes a
sharp vertical transition to dimmer luminosities at its blue-most 
tip.  
CMDs are shown in Figure~\ref{fig:HB_CMDs} for four
different clusters, each of which has a distinct HB morphology.  The
source of this 
variance is unknown, however our current best guess is mass-loss due
to stellar winds, most likely on the RGB \citep[e.g.][]{iben74, dotter10}.
The more mass-loss that occurs, the 
smaller the mass of the remaining envelope.  The surface temperature
increases with decreasing envelope mass since this leaves more of the
hot core exposed.  Consequently, the amount of mass-loss is thought to
determine the colours of HB stars, with stars that experience the
greatest loss in mass ending up the bluest.  In order to reproduce the
observed surface temperatures of the hottest HB stars, the envelope
must lose a mass of 0.2-0.3 M$_{\odot}$ \citep{eggleton06}.  This
seems reasonable and agrees rather well with empirical estimates
\citep[e.g.][]{judge91}.  It is
not yet clear, however, why stars of comparable mass in the same
cluster would experience such different degrees of mass-loss.  It is
also unknown how or why this mass-loss distribution should change from
cluster-to-cluster, although the rate of mass-loss is thought to 
depend on metallicity \citep[e.g.][]{reimers75}.

\begin{figure} [!h]
  \begin{center}
 \includegraphics[scale=0.5]{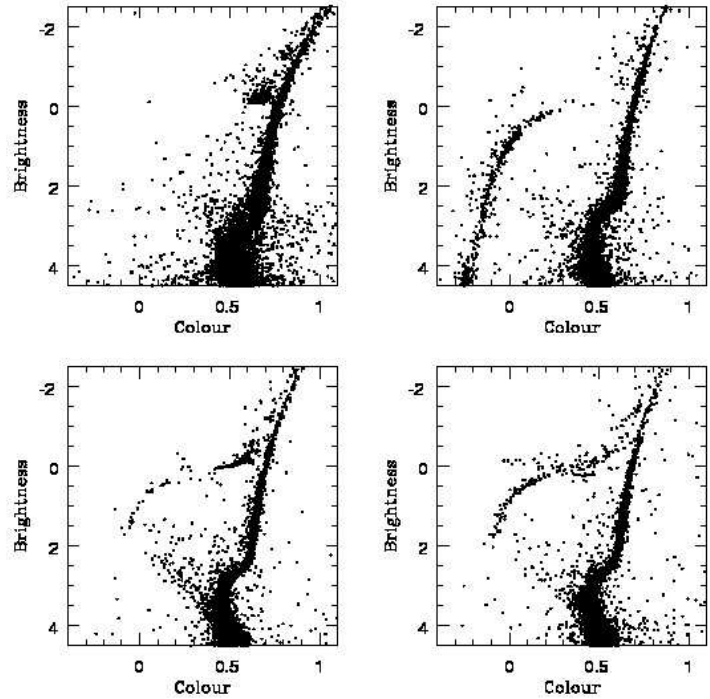}
    \caption[CMD sections for the Milky Way globular clusters NGC
    6205, NGC 104, NGC 1261 and NGC 6934]{Colour-magnitude diagram sections
      for the Milky Way globular clusters NGC 104 (top left inset),
      NGC 6205 (top right inset), NGC 1261 (lower left inset) and NGC 6934
      (lower right inset).  Note the distinct HB morphology 
      characteristic of each cluster.  Blue stragglers are also
      clearly visible just brighter and bluer than the MSTO.  The data
      used to create this plot were taken from 
      \citet{sarajedini07}.  
      \label{fig:HB_CMDs}}
    \end{center}
\end{figure}

Previous studies have confirmed that the 
observed differences in the HBs of Milky Way clusters
are related to metallicity \citep{sandage60}.  This does not tell the
whole story, however, since at least one
additional parameter is required to explain the spread in their
colours (see Figure~\ref{fig:HB_CMDs}).  Many cluster properties have been
suggested as possible 
Second and Third Parameters.  These include age, the 
cluster luminosity and central density.  Unfortunately, no 
definitive candidates have as of yet been 
identified \citep[e.g.][]{rood73, fusi93}.  Notwithstanding,
observations have revealed several peculiar trends for this curious
stellar population in individual clusters.  For instance,
\citet{saviane98}
presented evidence that blue HB stars could be more centrally
concentrated than red HB stars in the cluster NGC
1851.  Conversely, \citet{cohen97} showed that blue HB stars could be
centrally depleted relative to other stellar types in the cluster NGC
6205.  To date, no clear evidence has been found linking the spatial 
distributions of HB stars to any global cluster properties.

The bluest HB stars are often called extreme horizontal branch (EHB)
stars.  It has been
suggested that RGB stars can shed their entire envelopes and still
ignite helium as EHB stars \citep{dcruz96}.  Moreover, it is thought
that EHB stars could skip the AGB phase and evolve into WDs directly
\citep[e.g.][]{maeder09}.  
EHB stars also go by the name of sub-dwarf B (sdB) stars.  Both 
are core-helium burning stars with very small outer envelopes.  
Interestingly, EHB stars in the field tend to have binary companions
\citep{maxted01}.  This could be interpreted as evidence that the 
presence of a binary companion is responsible for causing the dramatic
loss of envelope mass.  On the other hand, most
EHB stars in clusters have been shown to lack a binary companion
\citep[e.g.][]{monibidin06, monibidin09, monibidin11}.  Consequently,
the mechanism responsible for (at least) \textit{their} extreme mass-loss is
still less clear. 


\subsubsection{Blue Stragglers} \label{BSs}

First discovered by \citet{sandage53}, blue straggler stars occupy the
region of the CMD that is just brighter and bluer than the MSTO (see
Figure~\ref{fig:HB_CMDs}).  That
is, they appear as an extension of the main-sequence.  And yet, if all
of the stars in a cluster were born at more or less the same time,
then standard single star evolution predicts that this region
of the CMD should be bare.  Normal MS stars that are of sufficiently
high mass to occupy this region should have long ago evolved onto the
RGB and beyond.  The presence of BSs in star clusters therefore
defies the predictions of single star evolution theory, and signifies 
another one of its mysteries.

This puzzle can perhaps be solved if blue stragglers are produced 
via the addition of fresh hydrogen to the cores of normal 
main-sequence stars \citep[e.g.][]{sills01}.  This is the
currently favoured origin for BSs.  It can occur
via multiple channels, most of which involve the mergers of low-mass
MS stars.  Stars in binaries can be driven to merge if enough orbital
angular momentum is lost.  This can be
mediated by dynamical interactions with other stars
\citep[e.g.][]{leonard89, leonard92}, magnetized
stellar winds \citep[e.g.][]{ivanova03}, tidal dissipation
\citep[e.g.][]{cleary90, chen08b} or even an outer triple companion
\citep[e.g.][]{fabrycky07, perets09}.  Alternatively, stars
can collide directly, although this is also thought to usually be
mediated by multiple star systems \citep{leonard89}.  Stars in close
binaries
can transfer mass if their orbital separations become sufficiently
small for one of the components to over-fill its Roche lobe.  This 
can also deliver fresh hydrogen to normal MS stars and in so doing
produce BSs.  

Whatever the
dominant BS formation mechanism(s) operating in dense star clusters,
dynamical interactions are sufficiently common that they should play
at least some role.  For example,
even if blue stragglers are formed as a result of binary evolution
processes such as mass-transfer, the progenitor binaries themselves
are expected to have experienced at least one dynamical interaction
over the course of their lifetime \citep[e.g.][]{hut83b, leonard89,
  davies04}.  
It follows that the study of BSs in star clusters offers an indirect 
means of probing the interplay between stellar evolution and stellar
dynamics. 

In order to explain recent observations of BS populations in star
clusters, it has been suggested that several BS formation mechanisms
operate simultaneously.  For example,
several studies have reported a bi-modal radial distribution for blue
stragglers \citep[e.g.][]{ferraro97, ferraro99, ferraro04, lanzoni07,
  geller08}.  That is, the number of BSs is the highest in the central
cluster regions, then falls off as the distance from the cluster
centre increases but eventually rises
again in the cluster outskirts.  This has primarily been observed in
globular clusters (GCs), which are particularly massive, dense and old star
clusters 
usually found in the outer reaches of our Galaxy.  One theory proposed
to explain this
result is that the blue stragglers found in the cluster core were
formed from collisions, whereas
those in the cluster outskirts were formed from binary mass-transfer
\citep{ferraro04}.  This hypothesis is motivated by the fact that the
time-scale for collisions to occur between stars is very short in the
core, but drops off considerably in the cluster outskirts where the
stellar densities are much lower \citep{leonard89}.  Alternatively, it
has been suggested that the blue stragglers in the cluster
outskirts could have been formed in the core but were later kicked out
as a result of dynamical encounters involving binary stars
\citep[e.g.][]{sigurdsson93, mapelli06}.


A spectroscopic survey of the cluster NGC 188 performed by
\citet{mathieu09} revealed that most, if not all, of its BSs have
binary companions.  \citet{mathieu09} proposed that their results are
consistent with the general picture that both mass-transfer 
and mergers are simultaneously producing BSs in NGC 188, which is also
consistent with the results of \citet{chen08a}.  The latter authors fit
theoretical stellar evolution tracks to the observed colour-magnitude
diagram for NGC 188.  Based on their results, they argue that most BSs
in this cluster are too massive to have formed from mass-transfer
alone.  If true, this suggests that many BSs in NGC 188 must be the
products of stellar mergers.  In support of this, \citet{perets09}
argued that the results of \citet{mathieu09} can be
explained if triple stars play an important 
role in BS formation by acting as catalysts for mergers
\citep{perets09}.  Triple stars
can evolve internally by transferring angular momentum between their
inner and outer orbits.  This can cause the eccentricity of the inner
binary to increase dramatically \citep{kozai62}.  Tidal 
friction can then act to reduce the orbital separation by removing
orbital energy from the inner binary at each periastron passage
\citep{fabrycky07} (the term periastron refers to the point of closest
approach for an eccentric binary orbit).

Blue stragglers have also been found in the field of our Galaxy.
Observations of these BSs suggest that both mass-transfer and binary
coalescence are occurring.  For example,
\citet{distefano10} recently reported the discovery of two BSs with
white dwarf companions in the field of our Galaxy.  The authors
suggest that these BSs were
formed from mass-transfer when the white dwarf progenitors
evolved to ascend the RGB and in so doing over-filled their
Roche lobes.  Moreover, \citet{brown10} discovered a $\sim 9.1$
M$_{\odot}$ hypervelocity MS star in the field.  The
authors argue that it must be a blue straggler since its flight time
from the Milky Way (MW) exceeds its MS lifetime.  They conclude that
this BS must 
be the product of a close binary that coalesced sometime after being
ejected from the Galactic centre.  This could explain its
hypervelocity provided its ejection was caused by a dynamical
interaction involving the central massive black hole.

Several spectroscopic studies have also been performed to study individual
BSs in clusters.  For example, at
least two BSs in the globular cluster 47 Tuc have been found with masses
exceeding twice that of the main-sequence turn-off (MSTO)
\citep{shara97, knigge08}.  The large masses of these BSs suggest that
they are the products of the mergers of two or more low-mass MS stars.
Another similarly massive BS was reported by \citet{vandenberg01} in
the old open cluster M67.  This BS is thought to be a member of a
triple star system that contains not one, but two BSs
\citep{sandquist03}.  BSs have been identified in both binaries and
triples in several other clusters as well, including the open cluster
NGC 6819 \citep{talamantes10}.

\subsubsection{Binary Stars}

Unresolved binary stars pollute the CMDs of star clusters, which 
become peppered with peculiar outliers that do not fall on any single 
star evolution tracks.  This is because clusters are located at
sufficiently great distances from us that the components of a given 
binary pair are indistinguishable and they appear as a single object.  
The combined light of the binary components blends together, producing a 
combination of colour and brightness that ordinary single stars are
never expected to display over the course of their
evolution.  This typically creates a secondary sequence in cluster CMDs
located above the MS \citep[e.g.][]{milone08}.  The reason for this is that
most cluster binaries are composed of normal MS stars 
\citep[e.g.][]{geller08}, and the combined 
light of their components makes these binaries appear slightly
brighter than ordinary single MS stars.

There are many examples of objects that are almost certainly bonafide
cluster members but appear in curious locations in the CMD.  Examples
include 
red stragglers \citep[e.g.][]{kaluzny03}, yellow stragglers
\citep[e.g.][]{latham05}, and 
the sub-subgiant branch stars found in M67 \citep{mathieu03}.  Red 
stragglers appear to the right (red-ward) of the RGB, yellow stragglers
appear to the left (blue-ward) of the RGB and above the Hertzsprung
gap, and sub-subgiant branch stars appear below the Hertzsprung gap.
Most of these are 
thought to be unresolved binaries or triples and, in some cases,
their curious CMD locations are speculated to be due to a recent
episode of mass-transfer.  

\section{Stellar Dynamics in Star Clusters} \label{dynamics_intro}

Star clusters are
in many ways analogous to stars themselves.  They are both, in
effect, self-gravitating spheres of interacting particles.  In both
cases, gravity 
plays a key role in driving the structural and compositional
changes that characterize their evolution.  Virial equilibrium is a central
physical principle that deepens this analogy.  This condition must be
satisfied in order for any self-gravitating system to achieve a 
state of dynamic equilibrium, at which point the inward pull of
gravity is balanced by the outward push of the pressure endowed to the
system by the relative motions of its particles.  The condition for
virial equilibrium in a star cluster can be expressed in the following form:
\begin{equation}
\label{eqn:virial}
2T + W = 0,
\end{equation}
where $T$ is the total kinetic energy of the system and $W$ is its total
gravitational energy.  Although additional physics must be factored
in, such as the energy generated from nuclear-burning, a similar
criterion for virial equilibrium exists for stars
\citep[e.g.][]{chandrasekhar39}.  It follows that
star clusters are virialized 
systems composed of objects which are themselves virialized.    

Both short- and long-range gravitational interactions between stars
act to evolve the structures and compositions of star clusters on
time-scales ranging from millions to billions of years.  In the
following sections, we describe the primary dynamical processes
responsible for this evolution.

\subsection{Two-Body Relaxation} \label{two-body}

Although many of the details are still unclear, star clusters are
thought to be born 
from massive gas clouds that fragment on several length
scales to form a large number of stars \citep[e.g.][]{lada95,
  mckee07}.  It ends up
that the presently observed structures of star clusters are largely
independent of the details of this collapse.  This is because
gravity will quickly act to mix the stars into a gravitationally-bound
spherical distribution for almost any initial configuration
\citep{heggie03}.  Once settled, 
most of the stars in a cluster will traverse stable orbits, often with
rosette-shapes, throughout it \citep{heggie03}.  
Their paths will typically remain more or less undisturbed during a 
single orbit.  
Over longer time-scales, however, the motions of the 
stars are affected by the cumulative effects of distant gravitational
encounters as well as the odd close encounter.

Like stars, star clusters evolve via the slow diffusion of heat from
their centres to their outer edges.  Gravitational encounters between
pairs of stars act as the mechanism for heat transport in much the
same way collisions between pairs of molecules govern the flow of heat
throughout an ideal gas.  In the case of star clusters, however, it 
is the cumulative effect of many weak, and therefore
distant, encounters that dominates, as opposed to the odd
strong or close encounter \citep{heggie03}.  This process is called
two-body relaxation.  The time-scale for it to occur, called the
relaxation time, is 
considerably longer than the time required for a given star to cross the
length of the cluster, called the crossing time.  A useful measure of the
relaxation time for the entire cluster comes from the half-mass
relaxation time t$_{rh}$, which is calculated from average quantities inside
the half-mass radius r$_h$ (defined as the distance from the cluster centre
containing half its total mass).  This is given by:
\begin{equation}
\label{eqn:t-rh1}
t_{rh} \sim \frac{0.138N^{1/2}r_h^{3/2}}{(Gm)^{1/2}ln\Lambda},
\end{equation}   
where $m$ and $N$ are the average stellar mass and the total number of
stars inside r$_h$, respectively.  The Coulomb logarithm ln$\Lambda$
is the factor by which small-angle encounters (i.e. encounters for
which the angle between
the initial and final velocity vectors of the deflected star is small)
are more effective than large-angle encounters in a star cluster
with given density and velocity distributions.  

For a typical Milky Way GC, t$_{rh}$ is on the order of a
billion years.  GCs are among the oldest objects in the Universe, with
ages usually ranging from $\sim$ 9-12 Gyrs
\citep[e.g.][]{deangeli05}.  Consequently, 
we expect that most will have had sufficient time for two-body
relaxation to have played a significant role in shaping their
present-day features.  Indeed, two-body
relaxation has been shown to dominate cluster evolution for a
significant fraction of the lives of old MW GCs
\citep[e.g.][]{gieles11}.  Other effects also play their part.  For
example, mass-loss due to stellar evolution has 
also been shown to affect the dynamical evolution of star
clusters, although its primary role is played during their early
evolutionary phases when massive stars with powerful stellar winds 
are still present \citep[e.g.][]{applegate86, chernoff90,
  fukushige95}.  The gas expelled by these winds can be significant.
It escapes from the cluster, which expands in response to the loss in
mass.  This serves to delay the evolutionary progression induced by
two-body relaxation.

Equation~\ref{eqn:t-rh1} provides only a rough guide since the
cumulative effects of two-body encounters depend on the stellar mass.
This brings us to another central concept of stellar dynamics:
equipartition of kinetic energies \citep[e.g.][]{henon69, giersz96}.  
This is the tendency for all stars in a cluster to end up with
comparable kinetic energies.  This results from the fact that, during
an individual gravitational encounter, the more massive star will
impart a net positive acceleration to the less massive star,
increasing its speed.  This comes at the expense of the kinetic energy
of the more massive star, which receives a corresponding net
deceleration and slows down.  In systems where stars of widely
differing mass occur, 
which is the case for real star clusters, the cumulative effects of
these gravitational encounters cause stars with masses greater than
the average stellar mass to lose kinetic energy.  As a result, massive
stars tend to slow down and sink to lower orbits within the 
cluster potential.  At the same time, 
stars with masses less than the average stellar mass tend to gain
kinetic energy and speed up, causing them to rise within the potential
well of the cluster.  In conjunction with close encounters occurring
primarily within the dense cluster core, this process contributes to a
steady stream of stars escaping from the system.  The important point
is that the stars that comprise clusters are all born with comparable
\textit{velocities}, but over time they tend toward a state in which they all
have comparable \textit{kinetic energies}.

The tendency for the most massive stars in a cluster to
accumulate in the central regions and low-mass stars to be
dispersed to wider orbits is called mass segregation.  The time-scale
for it to occur is typically short compared to the half-mass
relaxation time.  This can be understood as follows.  The time-scale
on which two-body encounters operate on an average star of mass $m$
has been shown to be well-approximated by Equation~\ref{eqn:t-rh1}
\citep{spitzer87}.  In general, however, the time-scale on which this
process operates on a star of mass $M >> m$ is given by
\citep{vishniac78, spitzer87}: 
\begin{equation}
\label{eqn:mass-segregation}
t_r \sim \frac{m}{M}t_{rh}.
\end{equation}
This implies that the most massive stars in a cluster will segregate
into the core on a time-scale much shorter than t$_{rh}$.  It follows
that most GCs will become fully mass segregated very early on in their
lifetimes \citep[e.g.][]{gaburov08, mcmillan07}.  This means that the
spatial distributions of their stars become stratified according to
their mass.  In other words, the probability of finding a star in the
cluster outskirts is inversely proportional to its mass.  It is
important to note, however, that we still see a range of masses at all
cluster radii.

Two-body relaxation is one of the primary mechanisms responsible for
the transfer of heat outward within a cluster.  This causes the inner
cluster regions to contract, and the outer regions to expand.
To see why this is the case, consider what occurs if energy is
transferred from the inner part of 
an isolated system of gravitationally-interacting particles to the
outer part.  The inner part should cool.  However, gravity then causes
these stars to drop to lower orbits within the cluster potential and,
in so doing, speed up.  
The net effect of this is that the average velocity of stars in the
inner part increases, causing its temperature to increase.  The energy
transferred to the outer part causes it to heat up as well, but it
expands in response to the addition of heat.  In general, the increase
in the temperature of the outer part is larger when it is smaller in
size.  This creates a temperature gradient between the inner and 
outer parts.  As a result, more heat flows outward and the temperature
gradient is enhanced.  This leads to a 
runaway effect that has been dubbed the gravothermal catastrophe.
For typical MW GCs, a phenomenon known as core collapse occurs within
about 10-20 half-mass relaxation times \citep{henon69, spitzer75, hut92}.
This marks an enhancement in the central density by several orders of
magnitude. 

Theoretical models suggest that the time-scale for core collapse to occur is
often longer than the age of the Universe.  Therefore, most MW GCs should 
currently be in a phase of core contraction.  This evolutionary phase
will only come to an end once the central 
density becomes sufficiently high for encounters involving
binaries to halt the process \citep[e.g.][]{hut83c, goodman93, fregeau09}.  
The net effect of these interactions 
is for single stars to steal energy from the orbits of the
binaries.  This imparts additional kinetic energy to the single stars
at the expense of reducing the orbital separations of the binaries.
In turn, this provides a heat source for the cluster, and ultimately
halts the collapse of the core.

%

\subsection{Small-N Dynamics} \label{small-N}

Our discussion of cluster evolution has naturally brought us to the
issue of small-N gravitational dynamics.  These are short-range
interactions involving only a few stars for which the distance of
closest approach can be 
comparable to the stellar radii.  During single-binary
and binary-binary encounters, resonant interactions often occur in
which the stars remain bound for many crossing times.  Two
objects (where an object can refer to a single, binary or even triple 
star) approach one another and a series of close, or strong,
gravitational interactions ensue.  
A number of outcomes are possible.  These include exchanges between
the components of binaries or even their complete dissociation, the
formation of triples, as well as stellar collisions and mergers.

Numerous scattering experiments have been performed to
explore the outcomes of binary-binary and, in particular,
single-binary encounters
\citep[e.g.][]{mcmillan86, sigurdsson93, fregeau04}.  Most of the
earliest of these studies were
performed in the point-particle limit.  Consequently, they ignored the
the stars' finite sizes, and so neglected the importance of
taking into account 
the dissipative effects of tidal interactions and direct contact
between stars \citep[e.g.][]{hut83a, mikkola83}.  This was later
remedied by, for instance, \citet{mcmillan87} and \citet{cleary90}.
Encounters involving four or more stars require longer integration times
to run the simulations to completion, and involve a large number of
free parameters.  As a result, few studies have been
conducted to explore the outcomes of binary-binary encounters or
interactions involving triple systems.  To date, none of these have
considered the finite sizes of the stars in a completely realistic way.

It has been known for some time that encounters between stars, and
even direct collisions, can occur frequently in dense stellar
systems \citep[e.g.]{hills76, hut83a, leonard89}.  In the cores of
globular clusters (GCs), the time between collisions
involving two single stars can 
be much shorter than the cluster lifetime \citep{leonard89}.  
The time between encounters involving binary stars can be
considerably shorter still given their much larger cross-sections for
collision.  In globular and, especially, open clusters with high
binary fractions, mergers are thought to occur frequently during
resonant interactions involving binaries \citep[e.g.][]{leonard92}.
What's more, collision products have a significant probability of
undergoing a second or even third collision 
collision during a given single-binary or binary-binary interaction.
This is because the impact causes the collision product to expand, 
increasing the cross-section for a
subsequent collision to occur \citep[e.g.][]{fregeau04}.

Small-N dynamical interactions play a number of important roles in
star cluster evolution.  By acting as an important heat
source for clusters, binaries become modified by dynamical
interactions.  Although the details are not yet clear, this can change
the distribution of orbital parameters of binary populations in
clusters \citep[e.g.][]{hut83b, sigurdsson93}.  In turn, this could have important implications for
binary star evolution by, for example, stimulating mass-transfer.
Observational evidence has been found in support of this.  Previous
studies have reported evidence that the sizes of some binary 
populations thought to be undergoing mass-transfer are correlated with
the rate of stellar collisions \citep[e.g.][]{pooley06}.

\section{Where To Go From Here} \label{wheretogo}

We have a good working knowledge of both stellar evolution and stellar
dynamics in star clusters.  But our understanding is far from
complete.  This brings us to the question:  Where
do we go from here?  The interaction between stellar evolution
and stellar dynamics in star clusters remains a largely unexplored
frontier.  What observational effects might 
we expect to find?  How will these affect star cluster evolution, and
what implications could they have for the history of our Galaxy?
The search for answers to these questions will be the primary focus of
this thesis.  

The techniques presented in this thesis are motivated by our current
understanding of both stellar evolution and stellar dynamics, in
addition to evidence that suggests that their interaction could
account for the origins of several mysterious stellar populations.
Gravitational dynamics is the common theme unifying all of these
methods.  They will be applied to 
observations of dense star clusters in order to study, among other
things, stellar mergers.  An emphasis will be placed on the development of
statistical and theoretical tools that can be applied to a number of
other astrophysical sub-disciplines with analogous dynamical
processes and population statistics.  These tools will be used in this
thesis to further our understanding of 
the various channels via which stellar evolution and stellar dynamics
interact in star clusters, as well as to quantify
the implications of these effects for observations of stellar
populations.  

In Chapter~\ref{chapter2}, we introduce a new adaptation of the
classical mean free path approximation.  With it, we compare the
rates of gravitational encounters occurring between single, binary and
triple stars using observations of real star clusters.  This has
allowed us to outline a systematic methodology that can be used to
constrain the dynamical origins of observed multiple star systems
containing merger products.  In Chapter~\ref{chapter3}, we
introduce a statistical technique that can be used to compare the
relative sizes of the different stellar populations in a large sample
of star clusters spanning a diverse range of properties.  In order to
study the effects had by the cluster dynamics on each of the different
stellar populations, we apply our 
method to a large sample of 56 Milky Way GCs.  In
Chapter~\ref{chapter4}, we refine this technique and apply it to a 
new sample of 35 GCs compiled using much more sophisticated
observations taken from the ACS Survey for Globular Clusters.  Using
this new data, we present a homogeneous catalogue
for the different stellar populations, along with a simple prescription to
select stars out of the CMD belonging to each population.  In
Chapter~\ref{chapter5}, we
present an analytic model for blue straggler formation in globular
clusters.  We compare its predictions to observed BS numbers taken
from our stellar population catalogue using a new
statistical technique.  With this method, we constrain the dominant
blue straggler formation mechanism operating in GCs.  In
Chapter~\ref{chapter6}, we present a new method of quantifying
cluster-to-cluster differences in the stellar mass functions of a
large sample of clusters.  We apply our technique to the ACS data,
and use it to constrain the degree of universality of the initial
mass function for our sample.  Given the very old ages of Milky Way globular
clusters, this has important implications for our understanding
of star formation in the early Universe.  
Finally, in Chapter~\ref{chapter7}, we summarize our results and
discuss directions for future research.




\pagestyle{fancy}
\headheight 20pt
\lhead{Ph.D. Thesis --- N. Leigh }
\rhead{McMaster - Physics \& Astronomy}
\chead{}
\lfoot{}
\cfoot{\thepage}
\rfoot{}
\renewcommand{\headrulewidth}{0.1pt}
\renewcommand{\footrulewidth}{0.1pt}




\chapter{An Analytic Technique for Constraining the 
  Dynamical Origins of Multiple Star Systems Containing Merger
  Products} \label{chapter2}
%
\thispagestyle{fancy}

\section{Introduction} \label{intro2}

It has been known for some time that encounters, and even direct
collisions, can occur frequently between stars in dense stellar systems
\citep[e.g.]{hills76, hut83a, leonard89}.  In the cores of globular
clusters (GCs), the time between collisions
involving two single stars can 
be much shorter than the cluster lifetime \citep{leonard89}.
The time between encounters involving binary stars can be
considerably shorter still given their much larger cross sections for
collision.  In globular and, especially, open clusters (OCs) with high 
binary fractions, mergers are thought to occur frequently during
resonant interactions involving binaries \citep[e.g.][]{leonard92}.
What's more, collision products have a significant probability of
undergoing more than one 
collision during a given single-binary or binary-binary interaction
since the initial impact is expected to result in shock heating
followed by adiabatic expansion, increasing the cross section for a
second collision to occur \citep[e.g.][]{fregeau04}.    

Several types of stars whose origins remain a mystery are speculated
to be the products of stellar mergers.  Blue stragglers (BSs) in
particular are thought to be produced via the addition of fresh
hydrogen to the cores of low-mass main-sequence (MS) stars.  Recent
evidence has shown that, whatever the dominant BS formation 
mechanism(s) operating in both globular and open clusters, it
is likely to in some way depend on binary stars \citep{knigge09,
  mathieu09}.  The currently favored mechanisms include collisions
during single-binary and binary-binary encounters
\citep[e.g.][]{leonard89}, mass transfer 
between the components of a binary system \citep[e.g.][]{chen08a,
  chen08b} and the coalescence of two 
stars in a close binary due to perturbations from an orbiting
triple companion \citep[e.g.][]{eggleton06, perets09}.

A handful of spectroscopic
studies have revealed that in some GCs there exist BSs with
masses exceeding twice that of the MS turn-off \citep[e.g.][]{shara97,
  knigge08}.  Such massive BSs must have been formed from the
mergers of two or more low-mass MS stars since they
are too massive to have been formed from mass transfer.  In a few 
cases, this can also
be argued for entire BS populations using photometry.  For instance,
\citet{chen08b} performed detailed binary evolution calculations to
study dynamical stability during mass transfer from an evolving
giant star onto a MS 
companion.  Based on their results, it can arguably be inferred that
most BSs in NGC 188 are sufficiently bright that they probably could
not have formed from 
mass transfer alone.  If true, this suggests that most of these BSs
must be the products of stellar mergers.  Regardless of
the dominant BS formation mechanism(s) operating in dense star clusters,
dynamical interactions should play at least some role.  For example,
even if blue stragglers are formed as a result of binary evolution
processes such as mass transfer, the progenitor binaries themselves
should have been affected by at least one dynamical
interaction over the course of their lifetime.

Numerous scattering experiments have been performed to
explore the outcomes of binary-binary and, in particular,
single-binary encounters 
\citep[e.g.][]{mcmillan86, sigurdsson93, fregeau04}.  Most of the
earliest of these studies were 
performed in the point-particle limit, ignoring altogether the often
non-negligible implications of the stars' finite sizes
\citep[e.g.][]{hut83b, mikkola83}.  Later, more realistic
simulations clearly demonstrated the importance of taking into account
the dissipative effects of tidal interactions and direct contact
between stars \citep[e.g.][]{mcmillan87, cleary90}.  As a result of
the increased number 
of free parameters for the encounters and the longer integration times
required to run the simulations to completion, few studies have been
conducted to explore the outcomes of binary-binary encounters or
interactions involving triple systems.

In this chapter, we introduce an analytic technique to 
constrain the most probable dynamical origin of an observed binary or
triple system containing one or more merger products.  Provided the
observed system is found within a moderately dense cluster
environment with binary and/or triple fractions of at least a few
percent, the probability is often high that it formed from a merger
during 
an encounter involving one or more binary or triple stars.  In
Section~\ref{method2}, we present an equation for energy conservation 
during individual stellar encounters and outline the process for
applying our technique.  Specifically, we present a step-by-step
methodology to evaluate whether or not an assumed dynamical history
could have realistically produced an observed system and describe how
to determine the most probable dynamical formation scenario.  
In Section~\ref{results2}, we apply 
our technique to a few observed binary and triple systems thought to
contain merger 
products, in particular a triple system that is thought to contain two BSs and
the peculiar period-eccentricity distribution of the BS binary
population in NGC 188.  We discuss the implications of our results in
Section~\ref{discussion2}. 

\section{Method} \label{method2}

In this section, we present a general prescription for conservation of
energy during stellar encounters.  We will limit the
discussion to typical interactions thought to occur in globular and
old open clusters, although our technique can be generalized to any
choice of parameter space.  The types of encounters of interest
in this chapter will predominantly involve low-mass MS stars with 
relative velocities at infinity ranging from $\lesssim 1$ km/s
to $\sim 10$ km/s \citep[e.g.][]{leonard89, sigurdsson93}.  Our technique
describes how to isolate the most probable dynamical formation history
for an observed binary or triple containing one or more merger products
by providing an estimate for the time required for a given interaction
to occur in a realistic cluster environment.  

We begin by assuming that an observed system was formed
directly from a dynamical interaction (or sequence of
interactions).  In this case, the observed parameters of the system
provide the final distribution of energies for the system resulting 
from the interaction(s).  After choosing an appropriate dynamical
scenario (i.e. whether the objects involved in the interaction(s) are
single, binary or triple stars), we can work backwards using 
energy conservation to constrain the initial
energies going into the encounter.  
This provides an estimate for the initial orbital
energies and therefore semi-major axes of any binaries or 
triples going into the interaction.  This in turn gives the
cross section for collision and hence the time required for the
hypothesized interaction(s) to occur.  

Since the formation event must have happened in the last 
$\tau_{BS}$ years, where $\tau_{BS}$ is the lifetime of the merger
product, a formation scenario is likely only if the derived encounter
time-scale is shorter than the lifetime of the merger product(s).
Conversely, if the derived encounter time-scale is longer than the
lifetime of the merger product(s), then that dynamical formation scenario 
is unlikely to have occurred in the last $\tau_{BS}$ years.  In
general, the shorter the derived encounter time-scale, the
more likely it is that one or more such encounters actually took place
within the lifetime of the merger product(s).  Finally, if the
derived encounter time-scale is longer than $\tau_{BS}$ for every
possible dynamical formation scenario, then a dynamical origin is
altogether unlikely for an observed multiple star system containing
one or more BSs.  Either that, or the encounter time-scales must have
been shorter in the recent past (or, equivalently, the central cluster
density must have been higher). 

\subsection{Conservation of Energy} \label{energy}

Consider an encounter in which at least one of the two bodies involved
is a binary or triple star.  Though a complex exchange of energies occurs,
energy must ultimately be
conserved in any dynamical interaction.  The total energy that goes
into the encounter must therefore be equal to the total energy
contained in the remaining configuration:
\begin{equation}
\begin{gathered}
\label{eqn:energy-conserv}
\sum_i^{N_i} \Omega_i(I_i,\omega_i) + \sum_i^{N_i} W_i(m_i,X_i,Z_i,\tau_i) \\ 
\sum_i^{N_i} U_i(m_i,X_i,Z_i,\tau_i) + \sum_j^{S_i} T_j(M_{cl},M_j) + \\ 
\sum_k^{M_i} \epsilon_k(\mu_k,M_k,a_k) =
    \sum_{ii}^{N_f} \Omega_{ii}(I_{ii},\omega_{ii}) + \\
 \sum_{ii}^{N_f} W_{ii}(m_{ii},X_{ii},Z_{ii},\tau_{ii}) + \sum_{ii}^{N_f} U_{ii}(m_{ii},X_{ii},Z_{ii},\tau_{ii})
 \\
+ \sum_{jj}^{S_f} T_{jj}(M_{jj}) + \sum_{kk}^{M_f} \epsilon_{kk}(\mu_{kk},M_{kk},a_{kk}) +
  \Delta, \\
\end{gathered}
\end{equation}
where $N_i$, $S_i$ and $M_i$ are the total number of stars, objects
(single, binary, triple or even quadruple stars) and orbits,
respectively, that went into the encounter.  Similarly, $N_f$, $S_f$ and
$M_f$ are the total number of stars, objects and orbits remaining after the
encounter. 

We let $\Omega_i = \Omega_i(I_i,\omega_i)$ represent the
bulk rotational kinetic energy in star $i$, which is a function of the star's
moment of inertia $I_i$ and angular rotation rate $\omega_i$:
\begin{equation}
\label{eqn:Omega}
\Omega_i(I_i,\omega_i) = \frac{1}{2}I_i\omega_i^2
\end{equation} 
The moment of inertia is in turn a function of the density profile within a
star, which changes along with its internal structure and composition
as the star evolves.  The moment of inertia is given by \citep{claret89}: 
\begin{equation}
\begin{split}
\label{eqn:mom-inertia}
I &= \frac{8{\pi}}{3}\int_{0}^{R} \rho(r)r^4dr \\
  &= \beta^2mR^2, \\
\end{split}
\end{equation}
where $\beta$ is the radius of gyration.  For example, a typical 1 M$_{\odot}$
star in an old open cluster having an age of $\sim$ 6 Gyrs has $\beta
= 0.241$.  For comparison, a 1 M$_{\odot}$ star in a
typical Milky Way GC having an age of $\sim$ 11 Gyrs has $\beta =
0.357$ \citep{claret89}.  A large spread of rotation speeds have been
observed for MS stars in 
open and globular clusters, with measured values ranging from  $\sim$ 1 -
20 km s$^{-1}$ \citep{mathieu09}.  A 1 M$_{\odot}$
star having a radius of 1 R$_{\odot}$ and a rotation speed of 2 km
s$^{-1}$ with $\beta=0.241$ has $\Omega \sim 10^{36}$ Joules.

We let $W_i = W_i(m_i,X_i,Z_i,\tau_i)$ represent the gravitational
binding energy 
of star $i$, where $m_i$ is the star's mass, $X_i$ is its initial
hydrogen mass fraction, $Z_i$ is its initial metallicity and $\tau_i$
is its age.  For a spherical mass with a density distribution
$\rho_i(r)$, the gravitational binding energy is given by:
\begin{equation}
\begin{split}
\label{eqn:binding}
W_i(m_i,X_i,Z_i,\tau_i) &= \frac{16\pi^2G}{3}\int_0^{R_i}
\rho_i(r)r^4dr \\
                        &= -{\delta}(m_i,X_i,Z_i,\tau_i)G\frac{m_i^2}{R_i}, \\
\end{split}
\end{equation}
where the parameter $\delta$ is chosen to reflect the structure of the
star and is therefore a function of its mass, age and chemical
composition.  For instance, a typical 1 M$_{\odot}$
star with $(X,Z) = (0.70,0.02)$ in an old open cluster with an age
of $\sim$ 6 Gyrs has $\delta = 1.892$.  For
comparison, an older but otherwise identical star in a typical Milky
Way GC with an age of $\sim$ 11 Gyrs has $\delta = 6.337$
\citep{claret89}.  Roughly regardless of age, this gives $|W| \sim 10^{42}$
Joules for a 1 M$_{\odot}$ star with a radius of 1 R$_{\odot}$.

We let $U_i = U_i(m_i,X_i,Z_i,\tau_i)$ represent the total internal
energy contained in star $i$ (i.e. the star's thermal energy
arising from the random motions of its particles).  By solving the
equations of stellar 
structure, the total internal energy of a purely isolated single star
is uniquely determined by its mass, initial composition and age.  
Stars are made up of a more or less virialized fluid so
that, ignoring magnetic fields, the gravitational binding energy of a
star in hydrostatic
equilibrium is about twice its internal thermal energy
\citep{chandrasekhar39}.  Using this version of the virial theorem, a
1 M$_{\odot}$ star having a radius of 1 R$_{\odot}$ has $U \sim 5
\times 10^{41}$ Joules. 

The total translational kinetic energy of object $j$ (single, binary,
triple, etc. star) is represented by $T_j$:
\begin{equation}
\label{eqn:trans}
T_j(M_{cl},M_j) = \frac{1}{2}M_jv_j^2,
\end{equation} 
where $M_j$ is the total mass of the object, 
and $v_j$ is its bulk translational speed.  The
translational velocities of stars in clusters for which energy
equipartition has been achieved as a result of two-body relaxation
obey a Maxwell-Boltzmann distribution, with the heaviest stars
typically having the lowest velocities and vice versa
\citep{spitzer87}.  According to
the virial theorem, the root-mean-square velocity $v_{rms}$ of the
distribution depends on the total mass of the cluster $M_{cl}$.  It 
follows that the velocities of stars in a fully relaxed cluster
are approximately determined by their mass and the total mass of the
cluster.  Assuming that the typical velocity of 
a $<m>$ star is roughly equal to $v_{rms}$,
where $<m>$ is the average stellar mass in the cluster, 
energy equipartition can be invoked in some clusters to approximate
the translational 
kinetic energy, and hence velocity, of a star or binary of mass $M_j$:
\begin{equation}
\label{eqn:v-typ}
v_j = \Big( \frac{<m>}{M_j} \Big)^{1/2}v_{rms}.
\end{equation}
We note that Equation~\ref{eqn:v-typ} can only be applied in clusters
for which the half-mass relaxation time is much shorter than the
cluster lifetime.  At the same time, the tidal truncation of the
velocity distribution must not be significant.  We will return to this
in Section~\ref{general2}.  

For a 1 M$_{\odot}$ star with a 
translational speed of 1 km s$^{-1}$ (typical of stars in old open
clusters), we find $T \sim 10^{36}$ Joules.  For comparison, the same
star traveling at a speed of 10 km s$^{-1}$ (typical of stars in
globular clusters) has $T \sim 10^{38}$ Joules.  We can put this into
perspective by equating Equation~\ref{eqn:trans} with
Equation~\ref{eqn:binding}, which shows that
a direct collision between two 1 M$_{\odot}$ stars would require an
impact velocity of $\sim$ 1000 km s$^{-1}$ in order to completely
unbind the merger remnant.

The total orbital energy of orbit $k$ is denoted $\epsilon_k$, 
and is given by:
\begin{equation}
\label{eqn:orbital}
\epsilon_k(\mu_k,M_k,a_k) = -\frac{G{\mu_k}M_k}{2a_k},
\end{equation}
where $\mu_k = m_1m_2/(m_{1}+m_{2})$ is the reduced mass of the orbit, $M_k =
m_1+m_2$ is the total mass and $a_k$ is the orbital semi-major axis.  A
binary composed of two 1 M$_{\odot}$ stars with a period of 1000 days
has $|\epsilon| = 10^{38}$ Joules.  For comparison, an otherwise identical
binary with a period of 1 day has $|\epsilon| = 10^{40}$ Joules.  Since
most stable triples are observed to have outer periods of $\sim$ 1000
days with close inner binaries \citep{tokovinin97, perets09}, it follows
that the total orbital energies of stable triples will be dominated
by the orbital energies of their inner binaries.  

After the encounter occurs, the energies that collectively define the
state of the newly formed system include the rotational kinetic
($\Omega_{ii}(I_{ii},\omega_{ii})$), thermal
($U_{ii}(m_{ii},X_{ii},Z_{ii},\tau_{ii})$) and gravitational binding
($W_{ii}(m_{ii},X_{ii},Z_{ii},\tau_{ii})$) energies of the stars that
are left-over, as well as the total translational kinetic
($T_{jj}(m_{jj})$) and orbital
($\epsilon_{kk}(\mu_{kk},M_{kk},a_{kk})$) energies of any left-over
stars, binaries or triples.  The internal and gravitational binding 
energies of the left-over stars once again depend on their mass,
composition and age.  The mass, composition and evolutionary status of
a star will decide how it responds to tidal interactions (and
therefore how much tidal energy is deposited) since these are the
principal factors that determine its internal structure, in particular
whether or not its envelope is radiative or convective
\citep{podsiadlowski96}.  Finally, we let $\Delta$ represent the
energy lost from the system due to radiation and mass loss.  We do not
expect this term to be significant for the majority of the encounters
we will consider since the rate of mass-loss from low-mass MS stars is
small and the time-scales under consideration are
relatively short compared to the lifetimes of the stars.  Moreover,
the velocity dispersions characteristic of the clusters we will
consider are sufficiently small that we expect most
collisions to have a relatively low impact velocity.  Though
significant mass loss can occur for high impact collisions as often
occur in the Galactic centre, mass loss is only $\lesssim$ 5\% for the low
impact velocity collisions expected to occur in open and globular
clusters \citep{sills01}.  Therefore, we will henceforth assume
$\Delta \sim 0$.

Depending on the initial parameters of the
encounter, one or more terms in Equation~\ref{eqn:energy-conserv} can
often be neglected.  For example, given that the rotational energies
of typical MS stars in 
both open and globular clusters are several orders of magnitude
smaller than the other terms in Equation~\ref{eqn:energy-conserv}, we
can neglect these terms for the types of encounters of interest.  Even
a 1 M$_{\odot}$ star rotating at a rate of 100 km s$^{-1}$ has a
rotational energy of only $\sim 10^{39}$ Joules.  This is considerably 
higher than even the highest rotation rates observed for both BSs and
normal MS stars \citep[e.g.][]{mathieu09}.  From this, we expect
typical rotational energies to be significantly smaller than
the orbital energies of even moderately hard binaries.  
Furthermore, Equation~\ref{eqn:trans} can be combined with
Equation~\ref{eqn:v-typ} and equated to Equation~\ref{eqn:orbital} in
order to define the hard-soft boundary for a given cluster.  If a
binary is hard, its initial orbital energy will outweigh its 
initial translational kinetic energy during typical stellar encounters
\citep{heggie75}.  We do not expect soft binaries to survive for very
long in dynamically-active clusters \citep{heggie75} (i.e. most
binaries are hard), so that 
the translational kinetic energies of the 
stars, binaries and triples typically found in OCs can usually be
neglected when applying Equation~\ref{eqn:energy-conserv}.  This
suggests $T_j = 0$.  However, the final translational kinetic energies
of the stars, 
binaries and triples left-over should not be neglected for encounter
outcomes in which one 
or more stars are ejected with very high velocities.  This can leave
the remaining stars in a much more tightly bound configuration since
stars ejected with high velocities carry off significant amounts of
energy.  Finally, provided there are no very high 
impact collisions and tides dissipate a negligible amount of energy
from the system, we also expect that: 
\begin{equation}
\label{eqn:int-conserv}
\sum_i^{N_i} U_i + \sum_i^{N_i} W_i \sim \sum_{ii}^{N_f} U_{ii} + \sum_{ii}^{N_f} W_{ii}.
\end{equation}

We can simplify our
energy conservation prescription considerably for the majority of
encounters occurring in old OCs.  Neglecting the rotational kinetic
energies of the stars, Equation~\ref{eqn:energy-conserv} becomes: 
\begin{equation}
\begin{gathered}
\label{eqn:energy-conserv2}
\sum_k^{M_i} \epsilon_k(\mu_k,M_k,a_k) = \sum_{kk}^{M_f} \epsilon_{kk}(\mu_{kk},M_{kk},a_{kk}) + \\
\sum_{jj}^{S_f} T_{jj}(M_{jj}) - \Delta_{m}, \\
\end{gathered}
\end{equation}
where we have assumed $T_j = 0$ and $\Delta_{m}$ is the amount of
energy deposited in any merger products formed during the encounter.
In other words, the term $\Delta_{m}$ provides the required 
correction to Equation~\ref{eqn:int-conserv} resulting from 
internal energy being deposited in the merger product(s) as a result
of collisions and tides.  If no mergers occur or all of the orbital
energy of the merging binary is imparted to the other interacting stars,
$\Delta_{m} = 0$. 

Both the total linear and angular momenta must also be conserved
during any dynamical interaction.  This provides two additional
constraints that must also be satisfied, however the total linear and
angular momenta are vector quantities that depend on the angle of
approach as well as the relative orientations of the objects.  As a result,
there are more free parameters to fit when trying to constrain the
initial parameters of an encounter using conservation of linear and
angular momentum.  Therefore, it is considerably more difficult to
extract information pertaining to the initial orbital parameters of
an encounter using conservation of momentum than it is with
conservation of energy. 

We have assumed that exchange interactions do not occur in applying
our technique.  Clearly, this assumption could be invalid for some
systems.  If this is the case, then our method is also invalid.  This
suggests that our technique is ideally suited to clusters for which the
encounter time-scales are comparable to the lifetime of a typical
merger product.  This maximizes the probability that BSs do not experience
any subsequent dynamical interactions after they are formed.  
Of course, most clusters of interest are unlikely to satisfy this
criterion.  
We can assess the probability that an exchange encounter has 
occurred for a particular system by calculating the different
encounter time-scales and comparing to the lifetime of the merger
product.  If the time-scales are sufficiently short, then
the possibility of an exchange interaction having occurred after the
system's formation must be properly addressed.  We will discuss this
further in Section~\ref{discussion2}. 

\subsection{Generalized Approach} \label{general2}

In this section, we outline a step-by-step
methodology to constrain realistic dynamical formation scenarios
that could have resulted in the production of an observed
stellar system containing one or more merger products.  These steps
are: 

\begin{enumerate}

\item We must first find \textit{qualitative} constraints for the
system's dynamical history and, in so doing, converge on the most
probable formation scenario.  The choice of 
formation history should be guided by the observed properties of the
binary or triple system containing the merger product(s).  The following
guidelines can be applied to find the most probable scenario.

First, the analytic rates for single-single (1+1), single-binary (1+2),
single-triple (1+3), binary-binary (2+2), binary-triple (2+3) and
triple-triple (3+3) encounters can be
compared to obtain a rough guide as to which of these encounter
types will dominate in a given cluster.  The total rate of
encounters of a given type in a
cluster core is well approximated by \citep{leonard89} (see
Appendix~\ref{Appendix-A} for a more generalized form for this
equation):
\begin{equation}
\label{eqn:coll-rate}
\Gamma = N_0n_0\sigma_{gf}(v_{rel,rms})v_{rel,rms},
\end{equation}
where $N_0$ is the number of single, binary or triple stars in the
core and $n_0$ is the mean stellar, binary or triple number density in
the core.  The gravitationally-focused cross section for collision
$\sigma_{gf}$ is given by Equation 6 of \citet{leonard89}.
Gravitationally-focused cross sections for the various encounter types
are provided in Appendix~\ref{Appendix-A} along with the values assumed
for their pericenters.

In general, the number of single, binary and triple stars are 
given by, respectively, $(1-f_b-f_t)N_c$, $f_bN_c$ and $f_tN_c$,
where $N_c = 2/3{\pi}n_0r_c^3$ is the total number of objects in the
core \citep{leonard89}, $f_b$ is the
fraction of objects that are binaries and $f_t$ is the fraction of
objects that are triples.  Assuming for simplicity that 
$v_{rel}$ is roughly equal for all types of encounters, the rates for
two types of encounters can be compared to find the parameter space for
which one type of encounter will dominate over another.  These
relations can be plotted in 
the $f_b-f_t$ plane in order to partition the parameter space for
which each of the various encounter types will occur with the greatest
frequency, as illustrated in Figure~\ref{fig:fb-ft}.  Given a
cluster's binary and triple fractions, this provides a 
simple means of finding the type of encounter that will occur with the
greatest frequency.  Our results
are in rough agreement with those of \citet{sigurdsson93} who found
that single-binary interactions dominate over binary-binary
interactions in clusters having core binary fractions $f_b \lesssim
0.1$, and may dominate for $f_b$ up to 0.25-0.5 in some cases.

\begin{figure} [!h]
  \begin{center}
 \includegraphics[scale=0.5]{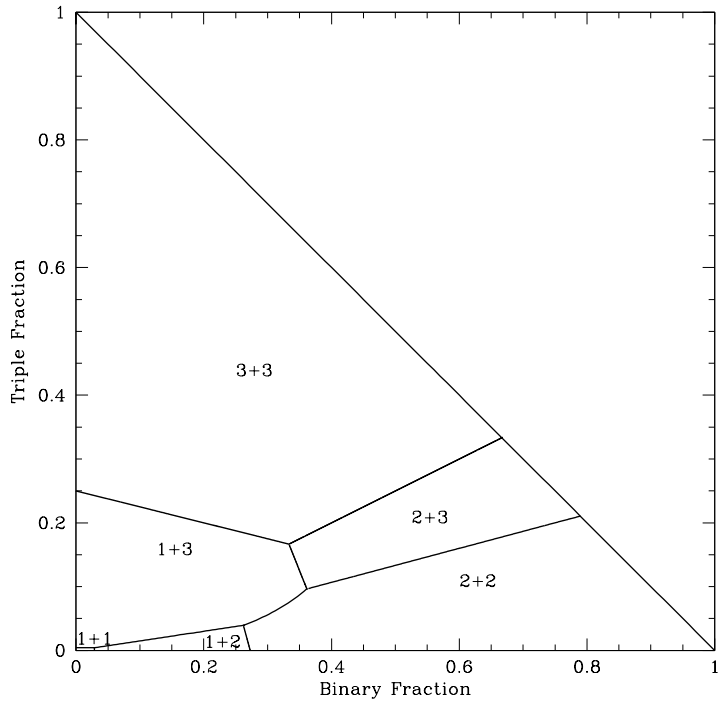}
   \caption[Plot showing the parameter space in the f$_b$-$f_t$ plane for
which each of the various encounter types dominate]{Plot showing
  the parameter space in the $f_b-f_t$ plane for
which each of the various encounter types dominate.  Boundaries
between regions are indicated by solid lines, each segment of which is
obtained by equating two particular encounter rates using
Equation~\ref{eqn:coll-rate} and the relevant cross sections derived
using Equation 6 of \citet{leonard89}.  We assume $a_t =
5a_b$ and $a_b = 90 R$ in obtaining the relations between encounter
rates.  This is a reasonable choice for the ratio between
the average binary and triple geometric cross sections given that
the ratio between the outer and inner orbital semi-major axes of
triples must be relatively large (by a factor of $\gtrsim$ 10) in
order for them to be stable \citep{mardling01}.
   \label{fig:fb-ft}}
  \end{center}
\end{figure}

Second, the masses of the components of an
observed binary or triple system provide a lower limit for the
number of stars that could have gone into its formation.  The minimum
number of 
stars that must have merged to form a given collision product, labelled
$N_{min,i}$, is equal to the integer nearest to and larger than the
quantity $m_{rem}/m_{TO}$, where $m_{rem}$ is the mass of the merger
remnant and $m_{TO}$ corresponds to the mass of the main-sequence
turn-off (MSTO).  This assumes that $m_{TO}$ has not changed
significantly since the dynamical formation of the system, which
should be valid provided the merger products are significantly more
massive than the turn-off so that their lifetimes are relatively
short.  The number of stars that went 
into an encounter $N_i$ must therefore satisfy:
\begin{equation}
\label{eqn:Nmerge}
N_i \ge \sum_i^M N_{min,i} + M_{0},
\end{equation}
where $M_{0}$ is the number of normal stars (i.e. not formed from
mergers) and $M$ denotes the number of merger products. 

Third, an estimate for the average stellar mass, and therefore the
masses of typical stars expected to undergo encounters, should be
guided by a realistic stellar mass function for 
the host cluster.  First of all, observations have shown that a
significant depletion of 
low-mass stars occurs in dynamically evolved clusters (i.e. those for
which $t_{rh} \ll t_{age}$, where $t_{rh}$ is the half-mass
relaxation time and $t_{age}$ is the cluster age) since they are
preferentially ejected from the cluster during close encounters
\citep[e.g.][]{vonhippel98, bonatto05, demarchi10}.  This suggests
that OCs and the least massive Milky Way GCs should 
have stellar mass functions that, at least in their central cluster
regions, appear eroded at the low-mass end.   From this, we expect
$<m> \lesssim m_{TO}$.  Finally, most of the stars
involved in dynamical encounters should have masses close to or even
slightly greater than the average stellar mass $<m>$.  This is because
gravitational focusing is strongest for the most massive objects,
contributing to a shorter encounter time-scale. 

Finally, the most likely formation scenario will, strictly speaking, 
minimize the total or cumulative encounter time-scale.  However, the
total time-scale required for a dynamical 
formation scenario that involves more than one encounter will
typically be dominated by the second encounter.  This is because,
after an initial encounter has occurred between any two suitable
objects to form a new stellar configuration, the time-scale for a
second encounter to occur is given by the time required for
the product of the initial encounter to experience a
subsequent encounter.  \textit{This increases the encounter time-scale
  by a factor $N_0$.}  In other words, 
Equation~\ref{eqn:coll-rate} provides the time required for
\textit{any} two of the specified objects to experience an
encounter (two binaries, a binary and a triple, etc.).  It can be
multiplied by $N_0$ to obtain an estimate for the time
required for a \textit{specific} object to experience an encounter.
For example, to find the time required for a specific binary to
experience an encounter with another single, binary or triple star, we
must multiply by a factor $f_bN_c$.  It follows that the 
time-scale for multiple encounters to occur is considerably longer than
any of the single encounter time-scales.  Therefore, unless the number
of either binary or triple stars is very low, 
scenarios involving the fewest number of encounters are generally
preferred since this tends to minimize the total time required for
the encounter(s) to occur in a realistic cluster environment.  

The times \textit{between} the various types of encounters can be
derived using Equation~\ref{eqn:coll-rate} and have been provided in
Appendix~\ref{Appendix-A}.  These time-scales can be multiplied by
$(1-f_b-f_t)N_c$, $f_bN_c$ or $f_tN_c$ to find the time required for a
\textit{particular} single, binary or triple star, respectively, to
encounter another object.  This yields time-scales that are in rough
agreement with Equation 8-125 of \citet{binney87}.  As an example, we
can multiply Equation~\ref{eqn:coll2+2} by $f_bN_c$ to find the time
for a particular binary to experience an encounter with another
binary.  This gives:
\begin{equation}
\begin{gathered}
\label{eqn:tau-2-22}
\tau^b_{2+2} = 2.7 \times 10^{10}f_b^{-1} \Big(\frac{10^3
  pc^{-3}}{n_0}\Big) \\
\Big(\frac{v_{rms}}{5 km/s}\Big)\Big(\frac{0.5
  M_{\odot}}{<m>}\Big)\Big(\frac{1
  AU}{a_{b}}\Big) \mbox{years},
\end{gathered}
\end{equation}

Provided the derived encounter time-scale is shorter than the
lifetime of
the merger product(s), this suggests that the encounter scenario in
question could be realistic and is therefore a candidate formation
history.  Conversely, if the
derived encounter time-scale is longer than the lifetime of
the merger product(s), then the dynamical formation history is unlikely
to have actually occurred.  In general, the shorter the derived
encounter time-scale, the
more likely it is that one or more such encounters actually took place
within the lifetime of the merger product(s). 

The preceding guidelines specify a narrow range of allowed formation
scenarios.  In particular, they constrain the 
number of stars involved in the encounter(s), the types
of objects (i.e. single, binary or triple stars) involved, and the
number of encounters that took place.  In most cases, these guidelines
will converge on a 
single qualitative formation history that is unique up to the possible
initial distribution of energies that describe suitable interactions.

\item Next, we must assign an approximate value based on the
observations to every parameter possible in
Equation~\ref{eqn:energy-conserv}.  Nearly all of the required
information pertaining to the final distribution of energies in
Equation~\ref{eqn:energy-conserv} can be found from spectroscopy
alone, though repeated measurements spread out over a sufficiently
long time-line will typically be required to obtain orbital 
solutions and to detect outer triple companions whenever they are
present.  This gives the final orbital energies $\epsilon_{kk}$ as well
as the gravitational binding energies $W_{ii}$ of the stars according to
Equation~\ref{eqn:orbital} and Equation~\ref{eqn:binding},
respectively.  Since merger products have been shown to typically be
in hydrodynamic  
equilibrium \citep[e.g.][]{sills01}, the stars' internal energies
$U_{ii}$ can then be approximated using the virial theorem.  Alternatively,
stellar models can be used together with photometry.  

The measured broadening of spectral lines gives an estimate for the
stars' rotation speeds (although there is a strong dependence on the
angle of inclination of the stars' axis of rotation relative to the
line of sight), which in turn provides their rotational 
kinetic energies $\Omega_{ii}$ according to Equation~\ref{eqn:Omega}.  In
conjunction with proper motions, radial 
velocity measurements also provide an estimate for the systemic
velocity of the final stellar configuration relative to the
cluster mean, which in turn gives its translational kinetic energy $T_{jj}$
according to Equation~\ref{eqn:trans}.  
Finally, the cluster velocity dispersion provides an estimate of the
relative velocity at infinity for a typical encounter, which in turn
decides the initial translational kinetic 
energies $T_j$ of stars or binaries involved in the encounter.  Under
the assumption of energy equipartition for both single and binary
stars, the initial velocities of the impactors can be approximated by
Equation~\ref{eqn:v-typ}.  The assumption of energy equipartition
should be valid in clusters for which the half-mass relaxation time is
considerably shorter than the cluster lifetime, and this is the case
for most GCs \citep{harris96, deangeli05}.  On the other hand, this
assumption is likely invalid for most open clusters.  This is because
the tidal truncation of the velocity distributions are significant in
OCs since they are much less centrally concentrated.  In this case,
the initial velocities of the impactors can be approximated from the
velocity dispersion, which is nearly independent of mass.

\item We can now obtain \textit{quantitative} constraints for the initial
encounter(s).  Once we have decided on a qualitative encounter
scenario that could have produced the observed merger product(s), we
can estimate the orbital energies of the initial binaries 
or triples going into the encounter using
Equation~\ref{eqn:energy-conserv2}.  This gives us an equation
that relates the initial orbital semi-major axes of all orbits going
into the encounter (to each other).  If only one orbit goes into the
encounter, then we 
can solve for it explicitly.  From this, we can constrain the initial
collisional cross sections for realistic encounters.  If the derived 
cross section is significantly smaller than the average semi-major
axes of all binaries and/or triples in the cluster, then
we can infer that the time required for the encounter to occur is
significantly longer than the corresponding time-scale given in
Appendix~\ref{Appendix-A}.  If we use our derived
cross section found from 
Equation~\ref{eqn:energy-conserv2} instead of this average semi-major
axis, this should give us an idea of the time required for that 
particular type of encounter to occur.  Strictly speaking, however,
this is only a rough approximation since these are typical time-scales
found using the average period (or, equivalently, semi-major axis and
hence cross section). 

If we find that the derived encounter time-scale is 
longer than the lifetime of a typical merger product, then the chosen
formation history is unlikely to have actually occurred.  In this
case, it becomes necessary to 
re-evaluate the possible dynamical formation histories of the observed
binary or triple system, choosing the
next most likely qualitative scenario for further quantitative
analysis.  These
steps can be repeated until either a suitable formation history is
found or the list of possibilities is exhausted so that the only
remaining conclusion is that the observed binary or triple system is
unlikely to have a dynamical origin.

\end{enumerate}

\section{Results} \label{results2}

In this section, we apply our technique to two particular
cases of observed binaries and triples containing merger products.  The 
first is an observed triple system in the old open cluster M67 that is
thought to contain two BSs \citep{vandenberg01,
  sandquist03}.  The second is the 
period-eccentricity distribution of the BS binary
population of the OC NGC 188, which bears a remarkable
resemblance to M67 \citep{mathieu09}.  After determining the most
probable qualitative formation 
scenarios, we obtain quantitative constraints for suitable initial
conditions that could have produced the observed orbital parameters.

\subsection{The Case of S1082} \label{s1082}

S1082 is believed to be a triple system in the old OC M67
\citep{vandenberg01, sandquist03}.  The observations suggest that a
distant triple companion orbits a close binary containing a BS and
another peculiar star.  The companion to the BS has a photometric
appearance that puts it close to the MSTO in the CMD and yet,
curiously, its derived mass is significantly greater than that of the
turn-off.  The outer companion is a BS in its own right, so that S1082
is thought to be composed of two BSs.  Although both the 
inner and outer components of this suspected triple have systemic
velocities that suggest they are both cluster members, it is important
to note that there is no direct evidence proving a dynamical link
between the two \citep{sandquist03}.  Assuming for the time being that
a dynamical link does exist, we can apply the procedure outlined in
Section~\ref{general2} to the case of S1082:

\begin{enumerate}

\item Before applying our technique, it is important to convince
  ourselves that a dynamical origin is possible for the observed
  system.  This is certainly the case for S1082 since no known BS
  formation mechanism could have produced the observed stellar configuration
  without at least some help from dynamical interactions.  The first
  step of our procedure is to find qualitative constraints and,
  in so doing, isolate the most probable encounter scenario.  

First, we need to know the cluster binary and triple fractions in
order to use Figure~\ref{fig:fb-ft} to find the encounter type
occurring with the greatest frequency.  From this, we find that 2+2
encounters presently dominate in M67.  \citet{fan96} showed
that observations of M67 are consistent with a 
cluster multiple star fraction $\sim 50\%$.  More recent
studies report a lower limit for $f_b + f_t$ that is consistent with their
results \citep[e.g.][]{latham05, davenport10}.  Radial
velocity surveys and simulations
of dynamical interactions suggest that old OCs like
M67 are likely to host a number of triples with outer periods
$\lesssim$ 4000
days \citep[e.g.][]{latham05, ivanova08}.  We assume 
$f_t/f_b \sim 0.1$ and, using the result of \citet{fan96}, this gives
$f_b \sim 0.45$ and $f_t \sim 0.05$.  Our assumed ratio $f_t/f_b$ is
slightly lower than found for the field, or $f_t/f_b \sim 0.2$
\citep{eggleton08}. 

Second, we need to constrain the number of stars that went into the
encounter.  The mass of the MSTO in M67 is estimated to be $\sim$ 1.3
M$_{\odot}$ \citep{mathieu09}.  The total mass of S1082 is $\sim$ 5.8
M$_{\odot}$ \citep{sandquist03}.  From Equation~\ref{eqn:Nmerge}, its
formation must therefore have involved at least 5 stars.  

At this point, we can conclude that a single 2+2 encounter could not
have produced S1082 since this scenario involves only 4 stars and we
know that at least 5 stars are needed.  We must therefore consider
either a single 2+3 or 3+3 encounter, or a multiple encounter
scenario.  In order to isolate the most probable of these
possibilities, we must calculate and compare their encounter
time-scales.  To do this, we require
estimates from the observations for a few additional 
cluster parameters.  For M67, the core radius is $r_c$ $\sim$ 1.23
pc \citep{bonatto05, giersz08}.  From this and the central velocity
dispersion, we can calculate the central mass density using Equation
4-124b of
\citet{binney87}, which gives $\rho_0$ $\sim$ $10^{1.9}$ M$_{\odot}$
pc$^{-3}$.  The central stellar number density can then be
approximated according to:
\begin{equation}
\label{eqn:num-density}
n_0 = \frac{\rho_0}{<m>}\frac{M}{L},
\end{equation}
where M/L is the cluster mass-to-light ratio and should be
around 1.5 for an OC as old as M67 \citep[e.g.][]{degrijs08}.  We take
$f_t \sim 0.05$ and assume most stable triples have 
outer periods of $P \sim 1000$ days so that $a_t \sim 3$ AU (assuming
all three stars have a mass of 1 M$_{\odot}$).  We also take $f_b \sim
0.45$ and assume an average binary period of $P \sim 100$ days so that
$a_b \sim 0.6$ AU (assuming both components have a mass of 1
M$_{\odot}$).  We assume an average stellar mass of $<m> \sim 
1.0$ M$_{\odot}$, which is in reasonable agreement with
the observations \citep{girard89}.  Assuming
that the average mass of merger remnants is equal to $2<m>$
and extrapolating the results of
\citet{sills01} for solar metallicity and more massive parent stars,
we will assume that the typical lifetime of a merger product is
$\tau_{BS} \sim 1.5$ Gyrs.

A comparison of the relevant encounter time-scales suggests that the
most probable dynamical formation scenario for S1082 is a single 2+3 
encounter, although a single 3+3 encounter is almost equally as
probable.  Given our assumptions, we find 
$8.9 \times 10^8$ years and $3.3 \times 10^9$ years for the times between
2+3 and 3+3 encounters, respectively, in the core of M67.  From this,
we expect approximately 
two and zero 
2+3 and 3+3 encounters, respectively, to have occurred within the last
$\tau_{BS}$ years.  

From Equation~\ref{eqn:tau-2-22}, we find that the time for a particular
binary to encounter another binary is $7.3 \times 10^{10}$ years.
Similarly, the time for a particular quadruple to encounter another
binary is $8.9 \times 10^9$ years (using Equation~\ref{eqn:tau-2-22}
and assuming the quadruple has a mass
of $4<m>$ and its geometric cross section is twice as large as the
average outer semi-major axis of triples).  Since these time-scales
are considerably longer than the
cluster lifetime, this suggests that a formation scenario for S1082
involving back-to-back 2+2 encounters is unlikely, even if the second
encounter occurred sufficiently soon after the first that all four
stars comprising the initial pair of interacting binaries are still
gravitationally bound.  If we replace one of the 3 binaries involved
in this scenario with a triple system, the total encounter time
remains longer than the cluster lifetime.  

\item Before more quantitative constraints can be found, we
  must refer to the literature in order to obtain estimates 
for every term in Equation~\ref{eqn:energy-conserv2}.  The
observations suggest that a binary system composed of a 2.52
M$_{\odot}$ blue straggler (component 
Aa) and a 1.58 M$_{\odot}$ MS star (component Ab) has a period of $P
\sim 1.068$ days \citep{vandenberg01}.  There is also evidence for a
1.7 M$_{\odot}$
blue straggler companion (component B) that acts as a stable outer
triple with a period of $P \sim 1188.5$ days \citep{sandquist03}.
From this, we can calculate the final orbital energies ($\epsilon_A$
and $\epsilon_B$) of the triple:
\begin{equation}
\label{eqn:orb-f}
|\epsilon_A| + |\epsilon_B| \sim |\epsilon_A| \sim 10^{41} J,
\end{equation}
where we have used the fact that $|\epsilon_A| \gg |\epsilon_B|$ since
$|\epsilon_B| \sim 10^{39}$ J.  
It is important to note that these peculiar stars are often
under-luminous, so the inferred mass of the tertiary companion should
be taken with caution \citep{vandenberg01}.  

The systemic radial velocity of the system is $33.76 \pm 0.12$ km/s
\citep{sandquist03}.  Although this only provides us with an estimate
for the systemic velocity of S1082 in one dimension, it is consistent
with the cluster mean 
velocity of 33.5 km/s.  Therefore, the available knowledge for the
systemic velocity of S1082 is consistent with its final
translational kinetic energy being negligible.  From this, we take
$T_{A,B} \sim 0$.  The central 
velocity dispersion in M67 is only 0.5 km s$^{-1}$ \citep{mathieu86,
  mathieu90}.  This suggests that the relative velocity at infinity, 
and therefore the typical impact velocities of collisions, should be
small compared to the significant orbital energy of component A in
S1082 (and any very hard binaries that went into the encounter).  From
this, we take $\Delta_m \sim 0$.

\item We are now ready to obtain quantitative constraints for the
formation of S1082.  We can use Equation~\ref{eqn:energy-conserv2} 
since M67 is an old OC with a small central velocity
dispersion so that the assumptions used to derive this equation are valid.  
At this point, we must consider the details of our chosen formation
scenario more carefully in order to choose a set of initial conditions
that will satisfy Equation~\ref{eqn:energy-conserv2} as well as
reproduce the observed parameters of S1082.  In doing so, we find that
it is not possible to form S1082 with a single 2+3 encounter.  This is
because 5 stars with masses $m < m_{TO}$ are insufficient to form both
an inner binary with a total 
mass $\sim 4.1$ M$_{\odot}$ and an outer tertiary companion with a
mass 1.7 M$_{\odot}$.  That is, the total mass of the inner binary is
larger than three times the mass of the MSTO so that its formation
must have involved four or more stars.

A single 3+3 encounter is therefore the most probable formation
scenario for S1082.  Although there are a number of ways we can
reproduce the observed parameters of S1082 with a 3+3 encounter,
including the component masses, we must use 
Equation~\ref{eqn:energy-conserv2} to identify the most probable of
these scenarios.  To do this, we can solve for 
the initial orbital energies of all binaries and/or triples going
into the encounter, which will in turn constrain their
initial semi-major axes and therefore cross sections for collision.  
In applying Equation~\ref{eqn:energy-conserv2}, we are only concerned
with the initial and final orbital energies of the inner binaries of
the triples.  This is because the orbital energies of any outer triple
companions will be significantly outweighed by the orbital energies of
their inner binaries.  It follows that the contribution from the outer
orbital energies of triples to the total energy of the encounter will
be negligible. 

From Equation~\ref{eqn:energy-conserv2}, we find 
that the formation of S1082 should have involved at least one hard
binary with 
$|\epsilon_i| \sim 10^{41} J$ whose components did not merge (with each
other) during the encounter in order to account for the significant
orbital energy of component A.  
This need not be the case, however, if one or more stars were ejected
from the system with an escape velocity of $\gtrsim 100$ km/s.  Of
course, this would require an increase in the total number of
stars involved in the interaction and therefore a single encounter
scenario involving 7 or more stars, and hence one or more quadruple
systems.  Although very few observational constraints for the fraction
of quadruple systems in clusters exist, this seems unlikely.

The orbital energies of the inner binaries of the two triples
initially going into the encounter should both be on the order of
$10^{41}$ J.  Although we have found
that only one hard binary with $|\epsilon_i| \sim 
10^{41}$ J is required, triples are only dynamically stable if the
ratio of their inner to outer orbital periods is large
(roughly a factor of ten or more) \citep[e.g.][]{mardling01}.  We do
not expect very wide binaries to survive for very long in dense
cluster environments, which suggests that the inner binary of every
triple is very hard.  We expect from this that these inner binaries
should, to within an order of magnitude, all have roughly the same
orbital energy.  

The presence of outer triple companions is required in order for most
very hard binaries to actually experience encounters within the
lifetime of a typical merger product.  Assuming masses of 1
M$_{\odot}$ for both components, an orbital energy of $10^{41}$ J
corresponds to a period $\sim 0.4$ days, or a semi-major axis $\sim
0.02$ AU.  Therefore, the cross section for collision for a 2+2 encounter
in which both binaries are very hard is smaller than the average
cross section for a 2+2 encounter (found from the observed binary
period distribution) by a factor of $\sim 30$.  This suggests that the
time required for an 
encounter to occur between two very hard binaries is considerably
longer than the cluster lifetime.  This is not the case if the hard
binaries have triple companions, however, since the outer orbit
significantly increases the collisional cross section and hence
decreases the encounter time-scale.  

Energy conservation also suggests that if S1082 did form from a single 3+3
encounter, it is likely that the close inner binaries of the two triples
collided directly.  
If two separate mergers then subsequently occurred during this
interaction, this could have reproduced the observed orbital
parameters 
of the close inner binary of S1082 (component A).  The formation of
the outer triple companion is more difficult to explain via a
single 3+3 encounter since it also appears to be the product of a
stellar merger.  Nonetheless, if the outer companions of both triples
undergoing the encounter end up orbiting the interacting (or merged) 
pair of close inner binaries at comparable distances, it is possible
that their orbits would overlap as a result of gravitational focussing.
Although this seems unlikely, it could produce a blue straggler of the
right mass and orbital period to account for component B.  We 
will return to this possibility in Section~\ref{discussion2}. 

\end{enumerate}

Even though the analytic estimates presented here are only approximate,
they serve to 
highlight the low probability of a system such as S1082 having formed
dynamically in M67.  Based on our results, the most likely formation
scenario for S1082 is 
a single 3+3 encounter, although we expect very few, if any, to have
occurred in the lifetime of a typical merger product.  This need not
be the case, however, provided M67 had a 
higher central density in the recent past since this would increase
the total encounter frequency.  This will be discussed further in
Section~\ref{discussion2}.  

\subsection{The Period-Eccentricity Distribution of the Blue Straggler
  Binary Population in NGC 188}

\citet{mathieu09} found 21 blue stragglers in the old open cluster NGC
188.  Of these, 16 are known to have a binary companion.  Orbital
solutions have been found for 15 of these known BS binaries.  From
this, \citet{mathieu09} showed that the BS binary population in NGC
188 has a curious 
period-eccentricity distribution, with all but 3 having periods of
$\sim 1000$ days.  Of these three, two have periods of $\lesssim 10$
days (binaries 5078 and 7782).  Interestingly, one of these
short-period BS binaries has a non-zero
eccentricity.  The normal MS binary population, on the other hand, shows
no sign of a period gap for $10 \lesssim P \lesssim 1000$ days
\citep{mathieu09}.  We can apply the procedure outlined in
Section~\ref{general2} to better understand how we expect mergers
formed during dynamical encounters to contribute to the BS binary
population in NGC 188.  Although the method described in
Section~\ref{general2} treats one system at a time, we will apply our
technique to the BS binary population of NGC 188 as a whole.

\begin{enumerate}

\item Before applying our technique, we must satisfy ourselves
  that a dynamical origin is possible for a large fraction of the
  observed BS population.  
  Several examples of evidence in favour of a dynamical origin exist.
  For one thing, most BSs in NGC 188 have been
  found to have binary companions \citep{mathieu09}.  This should not
  be the case if most BSs are the products of the coalescence of
  isolated binary systems.  On the other hand, a binary companion
  should be expected if the BSs were formed from mass transfer.  However,
  the results of \citet{chen08b} could suggest that most of the BSs in
  NGC 188 were not formed from mass transfer alone, providing indirect
  evidence that many BSs were formed from mergers.  A binary companion
  should also be expected if the coalesced binary had a tertiary 
  companion to begin with.  Although the Kozai mechanism can act to decrease the
  orbital separation of the inner binary of a triple, an additional
  perturbation is often required in order to fully induce the binary
  to merge \citep{perets09}.  All of this suggests that
  dynamical interactions likely played at least some role in the formation
  of a significant fraction of the BS binary population in NGC 188.

The first step of our technique is to find the
most commonly occurring type(s) of encounter(s).  \citet{geller08}
found a completeness-corrected multiple star fraction $f_b + f_t
\sim 0.27$ out to a period of $\sim 4000$ days.  This represents a
lower limit since it does not include binaries with $P \gtrsim 4000$
days.  Using the same lower limit for the ratio
$f_t/f_b \sim 0.1$ found by \citet{latham05}, Figure~\ref{fig:fb-ft}
suggests that 1+2 encounters should currently dominate in NGC 188.

Second, we must constrain the minimum number of stars required to form
the observed systems.  
From this, we can show that a merger occurring during all but a 1+1
encounter could produce a BS in a binary since this is the only type
of encounter that involves less than 3 stars.  By dividing the total
mass of an observed 
system by $m_{TO}$, we can find an estimate for the minimum number of
stars required to have been involved in its formation.  Since all but
one BS binary contain only a single merger product with a mass $\ge
m_{TO}$, realistic dynamical formation scenarios for these systems 
require 3 or more stars.  Binary 7782, on the other hand, is thought
to contain two BSs so that if its formation involved two separate
mergers it must have required 4 or more stars.

A more quantitative comparison of the different encounter types is
required since we do not yet know if \textit{enough} encounters
occurred in the last $\tau_{BS}$ years to account for all 16 observed
BS binaries.  To do this, we require estimates from the observations
for a few additional 
cluster parameters in order to calculate and compare the relative
encounter time-scales.  NGC 188 has been found to
have a central density of $\rho_0 = 10^{2.2}$ 
L$_{\odot}$ pc$^{-3}$ \citep{sandquist03} and a core radius of $r_c \sim
1.3$ pc \citep{bonatto05}.  
We found an average stellar mass for the cluster of $<m> \sim 0.9$
M$_{\odot}$.  This was done by determining the cluster luminosity
function using only those stars known to be cluster members from the
proper-motion study of \citet{platais03}.  With this, a theoretical
isochrone taken from \citet{pols98} was used to determine the cluster
mass function and the average stellar mass was calculated.  

A sufficient number of suitable dynamical interactions 
should have occurred in the last $\tau_{BS}$ years for the formation
of every BS binary in NGC 188 to have been directly mediated by the
cluster dynamics.  Assuming once again that $\tau_{BS} \sim 1.5$ Gyrs,
the encounter time-scales derived in Appendix~\ref{Appendix-A} suggest
that a minimum of nine, eight, eight, three and one 1+2, 2+2, 1+3,
2+3 and 3+3 encounters, respectively, 
occurred within the lifetime of a typical merger product.  It follows
that at least $\sim 29$ dynamical encounters
should have occurred in NGC 188 in the last $\tau_{BS}$ years.  Of
these 29 encounters, 12 should have involved triples.  In deriving these time-scales,
we have taken $f_b + f_t \sim 0.27$ from \citet{geller09} and have
adopted the same ratio $f_t/f_b \sim 0.1$ as found for M67.
Consequently, these represent lower 
limits for $f_b$ and $f_t$, so that our derived encounter rates are
also lower limits.  
We have also assumed an 
average outer semi-major axis for triples of $a_t \sim 3$ AU 
(corresponding to a period of $\sim$ 1000 days for a binary composed
of two 1 M$_{\odot}$ stars).

\item The next step is to apply our energy conservation prescription
  to the observed BS population.  This will allow us to constrain the orbital
  energies of typical binaries or triples expected to form BS binaries
  via dynamical interactions.  First, we must refer to the literature
  in order to obtain estimates for every term in
  Equation~\ref{eqn:energy-conserv2}. 
From the observed BS period-eccentricity distribution in NGC 188, we know
that most BS binaries have periods of $\sim 1000$ days (we will call
these long-period binaries), although there
are a couple with very short periods of $\sim 10$ days (which we will
call short-period binaries).  This provides 
an estimate for the final orbital energies $\epsilon_{kk}$ of BS binaries
formed during dynamical interactions.  Specifically, we find
$|\epsilon_{kk}| \sim 10^{39}$ J and $|\epsilon_{kk}| \sim 10^{40}$ J for the
final absolute orbital energies of the long- and short-period BS
binaries, respectively.

Every BS in NGC 188 with a high cluster membership probability
has both radial and proper motion velocities that, to within their
respective uncertainties, are consistent with the observed central
velocity dispersion of $\sigma_0 = 0.41$ km s$^{-1}$ \citep{platais03,
  geller08, geller09}.  From this, we assume that the
final translational kinetic energies of any binaries or triples formed
from dynamical interactions will be negligible, or $T_{jj} \sim 0$.  As
for M67, we assume $\Delta_m 
\sim 0$ since we expect low impact velocities for collisions as a
result of the low velocity dispersion in NGC 188.

\item We are now equipped with estimates from the literature that will
  allow us to obtain quantitative constraints for specific encounter
  scenarios.  In particular, we can use
  Equation~\ref{eqn:energy-conserv2} to constrain the initial 
  orbital energies of all binaries and/or triples going into an 
  encounter.  We can also constrain the specific details of
  interactions for which the encounter outcome reproduces the observed
  parameters of the BS binary.  We will consider 
  energy conservation separately for two different classes of BS
  binaries, namely short- and long-period. 

The short-period BS binaries have large (absolute) orbital energies.
Equation~\ref{eqn:energy-conserv2} suggests that this energy requires
that at least one hard binary was involved in the encounter.
Alternatively, an encounter which involved only softer binaries must
have resulted in at least one star escaping with a significant
velocity ($\gtrsim 100$ km/s).  However, in order to produce a merger
product in a short-period binary, encounters involving triples
are the most favoured.  Binary-binary (and especially 1+2) encounters
which involve at least one hard binary will have smaller
cross sections for collision than those encounters involving wide
binaries.  But energy conservation suggests that 2+2 encounters
involving wide binaries are unlikely to produce a merger product in a close
binary.  Stable triples, on the other hand, contain a hard inner
binary \citep{mardling01} which will naturally account for the large
orbital energy of 
the resulting BS binary.  Stable triples also have a large cross
section for collision because of the wide orbit of the tertiary
companion.  The times between 1+2, 2+2 and 1+3 
encounters are all comparable, suggesting that most encounters
involving hard binaries are 1+3 encounters. 

The short-period BS binaries could have formed from a
direct stellar collision that occurred within a dynamical
encounter of a hard binary and another single or (hard) binary star
(we call this Mechanism I).  If at least one of the objects going into
the encounter was a triple, then four or more stars were involved in
the interaction.  Therefore, if binaries 5078 
and 7782 were formed from this mechanism, they could 
possess triple companions with sufficiently long periods that they
would have thus far evaded detection.  This is consistent with the
requirements for both conservation of energy and angular momentum.  
Interestingly, the presence of an outer triple companion could also
contribute to hardening these BS 
binaries via Kozai cycles operating in conjunction with tidal friction
\citep{fabrycky07}.  In this case, Equation~\ref{eqn:energy-conserv}
shows that tides can contribute to making a binary's orbital energy
more negative by depositing internal energy into the component stars.

Although 2+2 encounters should also contribute to the observed BS
population, most of these should occur between a short-period binary
and a long-period binary.  This is because most encounters resulting in mergers
involve very hard binaries and binaries with long periods have 
large cross sections for collision and therefore short encounter
times.  Equation~\ref{eqn:energy-conserv2} indicates that
the minimum period of a BS binary formed during a 2+2 encounter is
usually determined by the orbital energy of the softest binary going
into the encounter.  This is because the most likely merger scenario
is one for which the hard binary is driven to coalesce by imparting
energy and angular momentum to other stars involved in the interaction 
\citep{fregeau04}.  Assuming most of 
this energy is imparted to only one of the stars causing it to escape 
the system, we can take $\Delta_m$ in
Equation~\ref{eqn:energy-conserv2} to be very small and the orbital
energy of the left-over BS 
binary will be comparable to the orbital energy of the initial wide
binary going into the encounter.  As more energy is imparted to the
star left bound to the 
merger product over the course of the interaction, its final orbital
separation effectively increases.  This in turn contributes to an increase 
in the final orbital energy of the left-over BS binary.  
Finally, from Equation~\ref{eqn:coll2+2}, the period
of the initial wide binary should be relatively long (roughly $\gtrsim
1000$ days) since the contribution from the very hard binary to the
total cross section for collision is negligible.  Otherwise, the time
required for such a 2+2 encounter to occur could exceed $\tau_{BS}$.

Now let us consider the long-period BS binaries in NGC 188.  Based on 
Equation~\ref{eqn:energy-conserv2}, we expect encounters involving
3 or more stars and only one very hard binary to typically produce
long-period BS binaries if the hard binary is driven to merge during
the encounter.  This is because the hard binary merges so that its
significant (negative) orbital energy can only be re-distributed to
the other stars by giving them positive energy.  Since the only other
orbits involved in the interaction are wide, the left-over BS binary
should also have a wide orbit.  Alternatively, BSs formed during
interactions involving more than one very hard 
binary should be left in a wide binary provided enough energy is
extracted from the orbit of the binary that merges.  In this case, a
significant 
fraction of this energy must be imparted to the other stars in order
to counter-balance the significant orbital energy of the other very
hard binary.  If not, the other short-period 
binaries are required to either merge or be ejected from the
system.  Otherwise
Equation~\ref{eqn:energy-conserv2} suggests that the left-over BS
binary should be very hard.  However, it is important to recall that we
are neglecting other non-dynamical mechanisms for energy extraction.
We will return to this last point in Section~\ref{discussion2}.  

In order to help us obtain more quantitative estimates for the
long-period BS binaries, consider two additional
mechanisms for mergers during dynamical interactions that involve both
short- and long-period orbits.  First, a
merger can occur if a sufficient amount of orbital energy is extracted
from a hard binary by other interacting stars (called Mechanism IIa).
Alternatively, a merger can occur if the encounter progresses in such
a way that the eccentricity of a hard binary becomes sufficiently
increased that the stellar radii overlap, causing the stars to
collide and merge (called Mechanism IIb).  In the case of Mechanism
IIb, most of the orbital energy of the close binary must end 
up in the form of internal and gravitational binding energy in the
merger remnant after the majority of its orbital angular momentum has been
redistributed to the other stars (and tides have extracted orbital
energy).  In the case of Mechanism IIa, however, most 
of the orbital energy of the close binary must be imparted to
one or more of the other interacting stars in the form of bulk kinetic
motion.  Consequently, one or more stars are likely to obtain a
positive total energy and escape the system.  This need not
necessarily be the case for 2+3 and 3+3 encounters provided the second
hardest binary orbit involved in the interaction has a sufficiently
negative energy.  

Regardless of the type of encounter,
Equation~\ref{eqn:energy-conserv2} shows that the extraction of
orbital energy from a hard binary in stimulating it to merge should 
increase the final orbital period of a BS binary.  To illustrate this,
we will consider 1+3 encounters since the predictions of energy
conservation are nearly identical for the other encounter types of 
interest.  Moreover, we have shown that encounters with triples
are the most likely to involve very hard binaries.
Equation~\ref{eqn:energy-conserv2} can be re-written for 1+3
encounters:  
\begin{equation}
\label{eqn:energy-conserv3}
\epsilon_{12,4} = \epsilon_{12,3} - f_{12} \times \epsilon_{12},
\end{equation}
where we have assumed that stars 1 and 2 comprise the initial hard
inner binary of the triple, star 3 is the initial outer triple
companion and star 4 is the interloping single star.  Stars 1 and 2
are assumed to merge during the encounter by exchanging energy and
angular momentum with stars 3 and 4.  We further assume that enough
energy is imparted to star 3 that it escapes the system.  We let $f_{12}$ 
represent the fraction of energy extracted from the orbital energy of
the hard inner binary of the triple in the form of bulk kinetic motion
by star 4.  Since the remaining orbital energy
of the close inner 
binary will end up in the form of internal and gravitational binding
energy in the merger remnant, we have assumed $(1 + f_{12}) \times
\epsilon_{12} \sim T_3 - \Delta_m$ in obtaining
Equation~\ref{eqn:energy-conserv3} from
Equation~\ref{eqn:energy-conserv2}.

This formation mechanism could leave the BS as a single star if 
$f_{12}$ exceeds a few percent.  As shown in
Figure~\ref{fig:sintri}, the period of a BS binary formed 
during a 1+3 encounter via Mechanism IIa ($P_{12,4}$) is only slightly
smaller than the period of the outer orbit of the triple initially
going into the encounter ($P_{12,3}$).  This assumes, however, that no
energy is exchanged between the hard inner binary and star 4
(i.e. $f_{12} = 0$).  The
predictions from energy conservation for this case are therefore
identical for Mechanism IIb.  In
general, as the amount of energy extracted from the hard inner binary
of the triple by star 4 increases, so too does the rate at
which $P_{12,4}$ increases with increasing $P_{12,3}$.  
If the amount of energy extracted is $\gtrsim$ 5\% of the orbital
energy of the initial inner binary of the triple, $P_{12,4}$ becomes a
very steeply increasing function of $P_{12,3}$.  

\begin{figure} [!h]
  \begin{center}
 \includegraphics[scale=0.5]{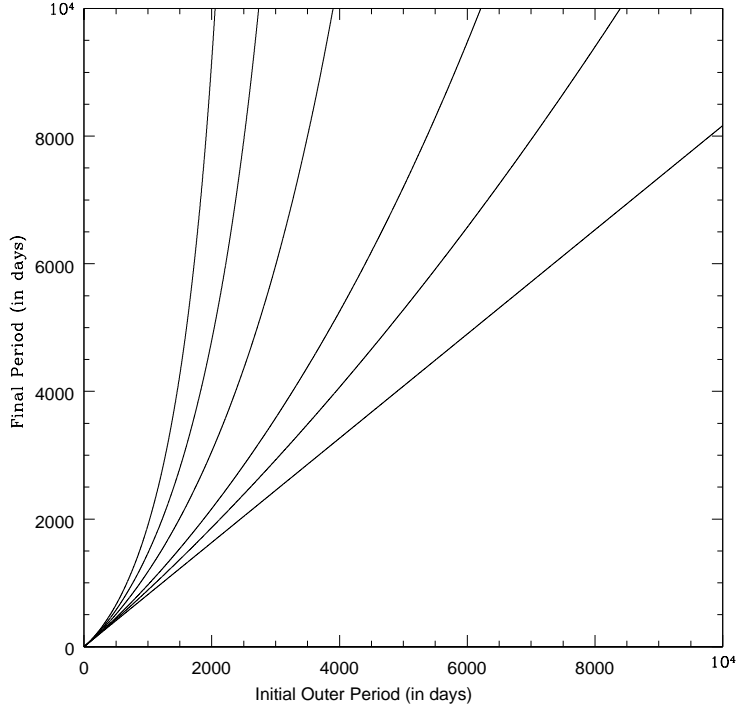}
   \caption[Plot showing the typical periods of BS binaries expected to
form during 1+3 encounters in which the hard inner binary of the
triple merges]{Plot showing the typical periods of BS binaries
  formed from 1+3 encounters in which the hard inner binary of the
  triple merges.  As described in the text, the period of the BS binary
  formed during the interaction is denoted $P_{12,4}$ and corresponds
  to the y-axis, whereas the period of the outer orbit of the triple
  initially going into the encounter is denoted $P_{12,3}$ and
  corresponds to the x-axis.  The lower straight line 
  corresponds to the case where no energy was extracted from the inner
  binary of the triple by star 4 (i.e. $f_{12}=0$).  As the amount of
  energy extracted increases, however, so too will the 
  rate at which $P_{12,4}$ increases with increasing $P_{12,3}$.
  Cases where $f_{12}=0.005$, $f_{12}=0.01$, $f_{12}=0.02$,
  $f_{12}=0.03$ and $f_{12}=0.04$ are shown as lines of increasing
  slope.
   \label{fig:sintri}}
  \end{center}
\end{figure}

Interestingly, the two general qualitative merger scenarios described above 
(Mechanisms I and II) naturally create a bi-modal period distribution
similar to the period gap observed for the BS binaries if we assume
that 1+3 encounters produced these objects.  To illustrate this, 
Figure~\ref{fig:per-num} shows a histogram of periods for 15 BS
binaries formed during 1+3 encounters via these two generic merger
scenarios.  In order to obtain this plot, we have 
used the observed period-eccentricity distribution for the regular
MS-MS binary population in NGC 188 from \citet{geller09} to obtain
estimates for the orbital energies of any binaries and/or triples
going into encounters.  Specifically, in order to obtain periods for
the outer orbits of triples undergoing encounters, we randomly sampled 
the regular period distribution, including only those 
binaries with periods satisfying $400$ days $< P < 4000$ days.  We
have shown that the initial outer orbits of triples going into 1+3
encounters provide a rough lower limit for the periods of BS binaries
formed via Mechanism II.  Therefore, any BS binaries formed in this
way could only have been identified as binaries by radial velocity
surveys if the triple going into the encounter had a period $< 4000$
days (since this corresponds to the current cut-off for detection).
All triples are taken to have a 
ratio of 30 between their inner and outer orbital semi-major axes.
This ratio has been chosen to be arbitrarily large enough that the
triples should be dynamically stable, however we will return to this
assumption in Section~\ref{discussion2}.  

We will adopt a ratio based on the observations of
\citet{mathieu09} for the fraction of outcomes that result in each of
these two possible merger scenarios.  In particular, if we assume that
the three BS 
binaries with $P < 150$ days were formed via Mechanism I whereas the
other 12 were formed via Mechanism II (either IIa or IIb since energy
conservation predicts similar periods for the left-over BS binaries),
this would suggest that Mechanism II is $\sim 4$ times more likely to
occur than Mechanism I during any given 1+3 encounter.  We will return
to this assumption in Section~\ref{discussion2}.

\begin{figure} [!h]
  \begin{center}
 \includegraphics[scale=0.5]{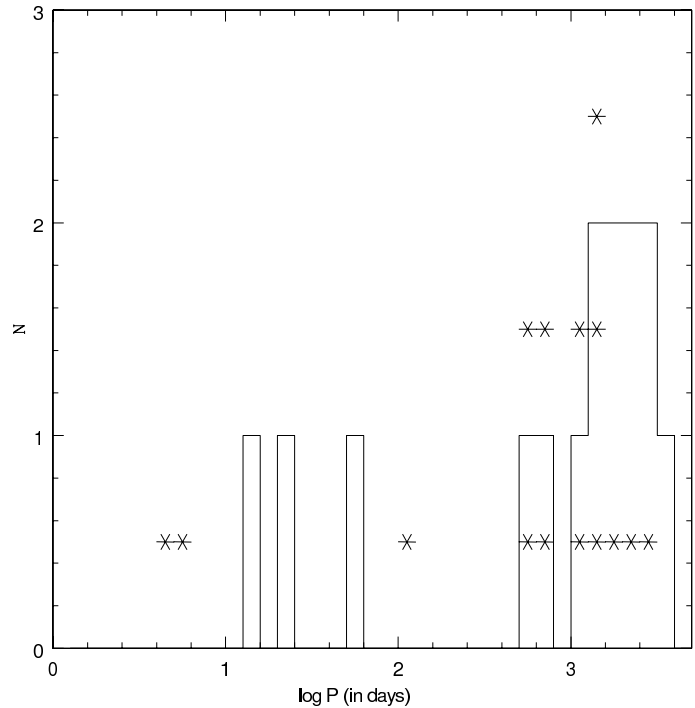}
    \caption[Histogram of the period distribution expected for BS binaries
formed during 1+3 encounters]{Histogram of the period distribution (in
  days) expected for BS binaries formed during 15 1+3 encounters.
  The parameter space assumed for the encounters is described in the
  text.  The stars show the observed BS binary period
  distribution in NGC 188 taken from \citet{mathieu09}, where each
  star represents a single BS binary.
    \label{fig:per-num}}
  \end{center}
\end{figure}

As a result of the requirement for conservation of angular
momentum, we would expect to see a large spread in the distribution of
eccentricities for BSs in long-period binaries formed from encounters
involving triples.  This is because conservation of momentum requires
that the total momentum going into the encounter must be
equal to the total momentum left over after the interaction.  
However, the total initial momentum depends not only on the
initial orbital eccentricities of any binaries or triples going into the
encounter, but also the relative orientations and trajectories of the
colliding objects.  Since the relative orientations and trajectories
are random, the final eccentricities of the BS binaries can take on a
range of values.  In other words, we cannot predict the final distribution of
momenta for BS binaries formed from dynamical encounters.  However,
the observed BS binaries in NGC 188 are observed to 
have a range of eccentricities and this is not inconsistent with a
dynamical origin involving triples.

BSs in short-period binaries formed via Mechanism I can also end up
with just about any eccentricity immediately after the encounter for
the same reasons outlined above.  However, tidal effects become increasingly significant
with decreasing orbital separation so that the hardest binaries should
typically circularize the fastest.  Although theoretical estimates for
the rate of tidal circularization are uncertain \citep{meibom05},
the circularization cut-off period is estimated to be $\sim$ 15 days in NGC
188 \citep{mathieu04}.  From this, it is entirely possible that
recently formed BS 
binaries with P $\sim$ 10 days have not yet become fully
circularized.  

\end{enumerate}

\section{Summary \& Discussion} \label{discussion2}

In this chapter, we have presented a generalized analytic prescription
for energy conservation during stellar encounters.  Our method can be
used to identify the most probable
dynamical formation scenario for an observed binary or triple
system containing one or more merger products.  We have shown
that, using the observed 
orbital parameters of the system, the allowed initial orbital
semi-major axes of any binary or triple
systems involved in its formation can be constrained.  The
initial semi-major axes of the orbits in turn provide an 
estimate for the collisional cross section and therefore the time-scale
for the encounter to occur in its host cluster.  In order to apply our
technique, repeated spectroscopic measurements of the binary or triple
system containing the merger product(s) are needed in order to obtain
its orbital solution and systemic velocity.  
However, the time-scales provided in Appendix~\ref{Appendix-A} can still
be applied if only the fraction of binary and triple stars are known, 
which can be determined either spectroscopically
\citep[e.g.][]{mathieu90, latham05} or photometrically
\citep[e.g.][]{fan96}.

As we have shown, consideration of the requirement for energy
conservation is ideal for identifying trends during
stellar encounters, whereas 
numerical scattering experiments can require hundreds or even thousands of
simulation runs before patterns will emerge.  Some of these trends include: 

\begin{itemize}

\item At least one short-period binary is usually required in a
  dynamical interaction to produce another binary having a similarly
  short-period (provided no stars are ejected with escape velocities
  $\gtrsim 100$ km/s).  This is because the orbital energy of a
  short-period binary is sufficiently negative that it tends to
  considerably outweigh the other energy terms in
  Equation~\ref{eqn:energy-conserv} for most of the encounters that
  typically occur in the cores of globular and especially open
  clusters.  This has been confirmed by \citet{hurley05}. 

\item Previous studies have found that in order for triples to remain
  stable for many dynamical times, the ratio of their inner to outer
  orbital periods must be relatively large (roughly a factor of ten or
  more) \citep[e.g.][]{mardling01}.  Based on our results, this has
  two important corollaries for stellar mergers in dense cluster
  environments hosting a significant population of triples:

\begin{enumerate}

\item  The longest-lived triples will contain very hard inner binaries
with a large $|\epsilon|$.  This is important since stellar radii are 
in general more likely to overlap and hence mergers to occur
during resonant interactions involving very hard binaries
\citep[e.g.][]{fregeau04, hurley05}.  We therefore expect stellar mergers
to be common during encounters involving stable triple systems.

\item  The longest-lived triples will contain wide outer orbits, creating
a large cross section for collision.  This suggests that the 
time-scale required for a stable triple system to encounter another
object is typically short relative to the cluster age in dense
environments.  A significant fraction of encounters
involving very hard binaries, and hence resulting in stellar mergers,
will therefore involve triples in old open clusters such as M67
and NGC 188. 

\end{enumerate}

\end{itemize}

\citet{vandenberg01} and \citet{sandquist03} suggest that
back-to-back binary-binary encounters, or even a single 3+3 encounter,
could have formed S1082.  We have
improved upon these previous studies by estimating time-scales required
for possible dynamical formation scenarios to occur.  
Since we have argued in Section~\ref{results2} that the
formation of S1082 must have involved at least 6 stars, it follows that 
only a 3+3 interaction could have reproduced the observed configuration
via a single encounter.  However, the derived 3+3 encounter time-scale
is sufficiently long that we expect very few, if any, 3+3 encounters
occurred within the lifetime of a typical merger product.  Moreover,
we have argued that the times for multiple encounters to occur are
longer than the cluster age.  Although we cannot rule out a dynamical
origin for S1082, our results suggest that it is unlikely (provided the
derived encounter time-scales were not significantly higher in the
recent past, which we will return to below).  From this, it follows that a
dynamical link between the close binary and third star is unlikely to
exist.   

On the other hand, we have so far ignored the cluster evolution,
and assumed that the currently observed cluster parameters have not
changed in the last few Gyrs.  N-body models suggest that the central
density in M67 could have been significantly higher in the recent
past.  Specifically, Figure 5 of \citet{hurley05} indicates that the
presently observed central density in M67 could have been higher
within the past Gyr by a factor of $\gtrsim 2$.  If this was indeed
the case, our previous estimates for each of the different encounter
frequencies should increase by a factor of $\sim 4$, so that a
significant number of dynamical encounters involving single, binary
and triple stars should have occurred in M67 within the last
$\tau_{BS}$ years.  It follows that a dynamical
origin for S1082 is not unlikely if the central density in M67 was 
recently larger than its presently observed value by a factor of
$\gtrsim 2$.  This also increases the 
probability that a scenario involving multiple encounters created
S1082, although we have shown that such a scenario is still likely to
have involved one or more triples.  

Based on the preceding arguments, S1082 offers an excellent example of
how observations of 
individual multiple star systems containing BSs can be used to
directly constrain the dynamical history of their host cluster.  
If a definitive dynamical link between components A and B is 
established, this would suggest that the central density in M67 
was higher in the last 1-2 Gyrs.  This is also required in order
for the cluster dynamics to have played a role in the formation of a
significant fraction of the observed BS population in M67.  Based on
the current density, the encounter time-scales are sufficiently long
that too few encounters should have occurred in the last $\tau_{BS}$
years for mergers during dynamical interactions to be a significant
contributor to BS formation.

We have obtained quantitative constraints 
for two generic channels for mergers during encounters involving
triples -- one in which a direct stellar collision occurs within a
dynamical interaction of the hard inner binary of a triple and another
single or (hard) binary star (Mechanism I) and
one in which the hard inner binary of a triple is driven to coalesce
by imparting energy and/or angular momentum to other stars involved in the
interaction (Mechanism II).  Our results
suggest that these two general merger mechanisms could contribute to a
bi-modal period distribution for BS binaries similar to that
observed in NGC 188.  These dual mechanisms predict a gap in
period, with those BS binaries formed via Mechanism I having periods of a
few to $\sim 100$ days and those formed via Mechanism II having
periods closer to $\sim 1000$ days.  Some 2+2, 2+3 and even 1+3
encounters could involve orbits with periods in this range, and
Equation~\ref{eqn:energy-conserv2} confirms that the final period of
a BS binary formed via Mechanism II will typically be determined by
that of the second hardest binary orbit.  Therefore, we might still
expect some BS binaries to have periods that fall in the gap (100 days
$\lesssim$ P $\lesssim$ 1000 days).  Our results do indeed predict one
such BS binary, as shown in Figure~\ref{fig:per-num}.  

A number of assumptions went into obtaining Figure~\ref{fig:per-num},
many of which were chosen specifically to reproduce the observed BS
binary period distribution.  Regardless, our
assumptions were chosen to reflect encounter scenarios that are the
most likely to result in mergers.  These should involve triples with
very hard inner binaries since these are the most likely to merge
during encounters.  The triples should also have outer companions on
very wide orbits since these have the largest cross sections for
collision.  From this, we have assumed a ratio of
30 between the inner and outer semi-major axes of all triples.  This
ensures that all triples are dynamically stable.  This also leads us
to assume a minimum period of $400$ days for the outer
orbits of triples so that the corresponding minimum period of their
inner orbits is not too small.  These assumptions serve to show that
encounters involving 
triples could produce both long-period and short-period BS binaries as
well as a period gap.  

Our results predict that the short-period peak in
Figure~\ref{fig:per-num} is at a slightly longer period than the
observations suggest.  If we decrease the assumed  
ratio between the inner and outer orbital separations of triples, the
short-period peak will move to even longer periods.  However, if one
or more stars were ejected with a high escape velocity or the
dissipative effects of tides are considered (which are expected to be
the most significant for encounters involving hard binaries) this
would move the short-period peak to even shorter periods.  In order to
obtain the desired agreement with the observations at the short-period
end of the BS period distribution, energy must have somehow been
dissipated or removed from the hard inner binaries of the triples
during (or even after) the encounter, or
the inner binaries must have initially been even harder than 
we have assumed in obtaining Figure~\ref{fig:per-num}.  

Our results could suggest that the hard BS binaries in
NGC 188 (binaries 5078 and 7782) may have outer triple companions,
perhaps with sufficiently long periods that they would have thus far
evaded detection.  This is consistent with the requirement for energy
conservation since the orbital energy of the outer
orbit of the triple is negligible compared to that of its inner
binary.  If binaries 5078 and 7782 do have outer triple companions, it
is also possible that Kozai oscillations combined with tidal friction
contributed to decreasing their orbital periods \citep{eggleton06}.
Finally, the BS binaries in NGC 188 are observed 
to have a wide range of eccentricities, which we have argued is
not inconsistent with a dynamical origin involving triples.  

We have assumed that Mechanism II is more likely to occur than
Mechanism I during any given 1+3 encounter.  This is a
reasonable assumption since numerical scattering experiments of 1+2
and 2+2 encounters have shown that the coalescence
of a hard binary is much more likely to occur during an encounter than
a direct collision between a hard binary and an interacting single
or binary star \citep[e.g.][]{fregeau04}.  The ratio we have
assumed for the frequencies with which Mechanisms I and II occur was
specifically chosen in order to reproduce the observed numbers of
short- and
long-period BS binaries.  The important point to take away is that the
observed BS 
binary period-eccentricity distribution offers a potential constraint
on the fraction of encounters that result in different merger
scenarios.  

Based on our results, Mechanism II must occur $\sim 4$ times more
often than Mechanism I in order to reproduce the observed BS period
distribution from 1+3 encounters (or, equivalently, 2+3 encounters
involving a very wide binary and 2+2 encounters between a 
short-period binary and a long-period binary).  This can be tested by performing
numerical scattering experiments of encounters involving triples.
Therefore, our
results highlight the need for simulations of
1+3, 2+3 and 3+3 encounters to be performed in order to better
understand their expected contributions to BS populations in open and
globular clusters.  Once a preferred encounter scenario has been
identified for an observed binary or triple containing one or more
BSs, numerical scattering experiments can be used to
further constrain the conditions under which that scenario will occur
(or to show 
that it cannot occur).  We have demonstrated that a combination of 
observational and analytic constraints can be used to isolate the
parameter space relevant to the dynamical formation of an observed
multiple star system (or population of star systems) containing
one or more merger products.  This will drastically narrow the
relevant parameter space for numerical scattering experiments.

We have improved upon the results of \citet{perets09} and
\citet{mathieu09} since we have shown that dynamical encounters
involving triples could not only be contributing to
the long-period BS binaries in NGC 188, but they could also be an
important formation
mechanism for short-period BS binaries and triples containing BSs.  We
have not ruled out mass transfer or Kozai-induced 
mergers in triples (primordial or otherwise) \citep{mathieu09,
  perets09}, or even various 
combinations of different mechanisms, as contributing formation channels
to the BS binary population in NGC 188.  For instance, a 1+3 exchange
interaction could stimulate a merger indirectly if the resulting angle
of inclination between the inner and 
outer orbits of the triple exceeds $\sim 39^{\circ}$, ultimately
allowing
the triple to evolve via the Kozai mechanism so that the eccentricity
of the inner binary increases while its period remains roughly
constant \citep{eggleton06}.  

There is evidence to suggest that mass transfer via Roche lobe
over flow could play a
role in the formation of at least some BSs.  It is difficult to account for 
the near zero eccentricities of some of the long-period BS binaries
without at least one episode of mass transfer having occurred. This is
because none of the normal MS-MS binaries with similar periods have
such small eccentricities \citep{mathieu09}.  On the other hand, it
may not be unreasonable to expect that some collision products left in
binaries undergo mass transfer since they are expected to expand
adiabatically post-collision, and will sooner or later evolve to
ascend the giant branch.  As a result of conservation
of energy and angular momentum, the mass transfer process will usually
act to 
increase the orbital periods of these binaries provided it is
conservative \citep{iben91}.  Interestingly, the cut-off period for
Roche lobe overflow is $\sim 1000$ days for low-mass stars
\citep{eggleton06}, which is in rough agreement with the long-period
peak in the observed period-eccentricity distribution of the BS binary
population in NGC 188.  Therefore, mass transfer could also be
contributing to the 
period gap observed for the BS binaries.  

According to the results of
\citet{geller09}, the number of giant-MS binaries with $P \lesssim
1000$ days is comparable to the
number of BS binaries (A. Geller, private communication).  It is unlikely that
every giant-MS binary will form a BS from mass transfer, however,
suggesting that at most a few of the long-period BS binaries in NGC
188 were formed via this mechanism.  Finally, if the outer
companion of a triple system evolves to
over-fill its Roche lobe it could transfer mass to both of the
components of the close inner binary.  This mechanism could therefore
also produce two BSs in a close binary, although it predicts the presence of
an orbiting triple companion.  For these reasons,
a better understanding of triple evolution, as well as binary
evolution in binaries containing merger products, is needed.

The dissipational effects of tides tend
to convert stars' bulk translational kinetic energies into internal or
thermal energy within the stars, leading to an increase in the total
gravitational binding energy of the stellar configuration
\citep[e.g.][]{mcmillan90}.  By
increasing the terms U$_{ii}$ in Equation~\ref{eqn:energy-conserv},
the initial orbital energies of any binaries going into an encounter
can increase accordingly in order to conserve energy.  A higher orbital energy
corresponds to a larger semi-major axis and hence cross section for
collision.  This suggests that the derived encounter time-scales can be
taken as upper limits in the limit that tidal dissipation is
negligible.  We expect tides to be particularly
effective during encounters for which the total energy is very
negative as a result of
one or more very hard binaries being involved.  

We have argued in Section~\ref{general2} that the average stellar 
mass is expected to be comparable to (but slightly less than) the mass
of the MSTO in old OCs and 
low-mass GCs.  We have also argued that most encounters will involve
stars having masses slightly larger than the average stellar mass.  We
might therefore expect to 
find that a high proportion of merger products have masses that
exceed that of the MSTO in very dynamically-evolved clusters that have
lost a large fraction of their low-mass stars.  Consequently, a larger 
number of merger products could appear sufficiently bright to
end up in the BS region of the cluster CMD in these clusters than in
their less dynamically-evolved counterparts.  This 
is consistent with the results of \citet{knigge09} and
\citet{leigh09} 
who found that the number of BSs in the cores of GCs scales
sub-linearly with the core mass.  In particular, since the cluster
relaxation time 
increases with increasing cluster mass, it is the lowest mass GCs that
should have lost the largest fraction of their low-mass stars.
Therefore, if a larger fraction of merger products do indeed end up
more massive than the MSTO in these clusters, this could be a
contributing factor to the observed sub-linear dependence on core
mass.  It is also interesting to note that, since BSs are among the
most massive cluster members and many are thought to have a binary
companion, BSs should be preferentially retained in clusters as they
evolve dynamically compared to low-mass MS stars.  This could also
contribute to the observed sub-linear dependence on core mass.  

The exchange or conversion of energies that occurs during an
encounter takes place over a finite period of time, so it is important to
specify whether or not the system has fully relaxed
post-encounter when discussing the remaining stellar configuration.
For one thing, \citet{sills01} showed that, although collision
products may be in hydrodynamic equilibrium, they are not in thermal
equilibrium upon formation and so contract on a thermal time-scale.
Simulations also suggest that most merger products should be rapid
rotators \citep{sills02, sills05}.  However, at least in the case of
blue stragglers, this is
rarely supported by the observations.  Some mechanism for
angular momentum loss must therefore be operating either during or
after the merger takes place in order to spin down the remnant.
The time-scale
considered must also be sufficiently short that subsequent dynamical
interactions are unlikely to have occurred since these could affect
the total energy and momentum of the system.  

With this last point in mind, N-body simulations considering BS
formation have shown that after they are formed, BSs are often
exchanged into other multiple star systems \citep{hurley05}.  This
suggests that for multiple star systems containing more than one BS,
the BSs could have first been formed separately or in parallel, and
then exchanged into their presently observed configuration.  For the
case of S1082, this would require at least 3 separate dynamical
interactions.  Given that the derived times between encounters are
relatively long and the fact
that the most likely formation scenario is usually that for which the
number of encounters is minimized, the current state of M67 suggests
that the probability of S1082 having formed from a scenario involving
3 encounters is low.  Conversely, the derived encounter time-scales in
NGC 188 are sufficiently short that many of the BS binaries 
could have experienced a subsequent dynamical interaction after their
formation.  BSs tend to 
be more massive than normal MS stars, contributing to an increase in
their gravitationally-focussed cross section for collision.  This
suggests that the encounter time-scale for multiple star systems
containing BSs is slightly shorter than for otherwise
identical systems composed only of normal MS stars.  This contributes
to a slight increase in the probability that a BS will experience an
exchange encounter after it is formed.  Interestingly, it could also contribute to
an increase in the probability that a close binary containing two BSs
will form during an encounter between two different multiple star
systems each containing their own BSs (\citet{mathieu09}; R. Mathieu,
private communication).  This is because it is the
heaviest stars that will experience the strongest gravitational
focussing and are therefore the most likely to experience a close
encounter, end up in a closely bound configuration, or even merge.

Most exchange interactions will involve wide binaries for which
the cross section for collision 
is large.  Since wide binaries are typically relatively soft and the
hardest binary involved in the interaction will usually determine the
orbital energy of the left-over BS binary, most exchange interactions
will leave the periods of BS binaries relatively unaffected.  This
need not be the case, of course, provided one or more stars are
ejected from the system with a very high escape velocity.  With
these last points in mind, we have assumed throughout our analysis that
all binaries and triples are dynamically hard.  This is a reasonable
assumption since the hard-soft boundary corresponds to a period of
$\sim 10^6$ days in both M67 and NGC 188, for which the cross section
for collision is sufficiently large that we do not expect such
binaries to survive for very long.  Nonetheless, considerations such
as these must be properly taken into account when isolating a
preferred formation scenario and predicting the final distribution of
energies.

We have presented an analytic technique to constrain the dynamical
origins of multiple star systems containing one or more BSs.  Our
results suggest that, in old open clusters, most dynamical
interactions resulting in mergers involve triple stars.  If most
triples are formed dynamically, this could suggest that
many stellar mergers are the culmination of a hierarchical build-up of
dynamical interactions.  
Consequently, this mechanism for BS formation should 
be properly included in future N-body simulations of cluster
evolution.  A better
understanding of the interplay between the cluster dynamics and the
internal evolution of triple systems is needed in order to better
understand the expected period distribution of BS binaries formed from
triples.  Simulations will therefore need to
track both the formation 
and destruction of triples as well as their internal evolution via
Kozai cycles, stellar and binary evolution, etc.  On the observational
front, our results highlight the need for a more detailed knowledge of
binary and especially triple populations in clusters.

\section*{Acknowledgments}

We would like to thank Bob Mathieu for many helpful comments and
suggestions.  We would also like to thank Aaron Geller, Evert
Glebbeek, Hagai Perets, David Latham, Daniel Fabrycky and Maureen van
den Berg for useful discussions.  This research has been supported by
NSERC as well as the National Science Foundation under Grant
No. PHY05-51164 to the Kavli Institute for Theoretical Physics.




\pagestyle{fancy}
\headheight 20pt
\lhead{Ph.D. Thesis --- N. Leigh }
\rhead{McMaster - Physics \& Astronomy}
\chead{}
\lfoot{}
\cfoot{\thepage}
\rfoot{}
\renewcommand{\headrulewidth}{0.1pt}
\renewcommand{\footrulewidth}{0.1pt}
%
%
%

%
%
%
%



\chapter{Stellar
  Populations in Globular Cluster Cores:  Evidence for a Peculiar
  Trend Among Red Giant Branch Stars} \label{chapter3}



\thispagestyle{fancy}
%
%
%
\section{Introduction} \label{intro3}

Studying the radial distributions of the various stellar 
populations (red giant branch, horizontal branch, main-sequence,
etc.) found in globular clusters (GCs) can provide useful hints 
regarding their dynamical histories.  As clusters evolve, they are
expected to undergo relatively rapid mass stratification as a
consequence of two-body relaxation, with the heaviest stars quickly
sinking to the central cluster regions
\citep{spitzer87}.  The shorter the relaxation time (typically
evaluated at the half-mass radius), the quicker this process occurs.  
Clusters tend towards dissolution as two-body
relaxation progresses and they lose mass due to stellar evolution and
the preferential escape of low-mass stars.  External effects like
tidal perturbations, encounters with giant molecular clouds, and
passages through the Galactic disk serve only to speed up the process
\citep[e.g.][]{baumgardt03, kupper08}.  Stellar evolution complicates
this otherwise simple picture of GC evolution, however.  Stars are
expected to change in size and lose mass as they evolve, often
dramatically, and this could significantly impact the outcomes of
future dynamical interactions with other stars.  For instance, a
typical star in a GC is expected to expand by up to a few orders of
magnitude as it ascends the red giant branch (RGB) and will shed up to
a quarter of its mass upon evolving from the tip of the RGB to the 
horizontal-branch (HB) \citep[e.g.][]{caloi08, lee94}.  

Red giant branch stars have been reported to be deficient in the cores
of some Milky Way (MW) GCs.  For instance, \citet{bailyn94} found that the
morphology of the giant branch in the dense core of 47 Tuc differs
markedly from that in the cluster outskirts.  In particular, there
appear to be fewer bright RGB stars in the core as well as an
enhanced asymptotic giant branch (AGB) sequence.  While
a similar deficiency of bright giants has been observed in the cores
of the massive GCs NGC 2808 and NGC 2419, better agreement between the
observations and theoretical luminosity functions obtained with the
Victoria-Regina isochrones was found for M5 \citep{sandquist07,
  sandquist08}.  \citet{sandquist07} speculate that the giant star 
observations in NGC 2808 could be linked to its unusually blue horizontal
branch if a fraction of the stars near the tip of the RGB experience
sufficiently enhanced mass loss that they leave the RGB early.
Alternatively, \citet{beer03} suggest that RGB stars could be depleted
in dense stellar environments as a result of collisions between red
giants and binaries. 

While stellar populations have been studied and compared on an
individual cluster basis, a statistical analysis in which their core
populations are compared over a large sample of GCs is ideal for
isolating trends in their differences.  Though a handful of studies of this
nature have been performed \citep[e.g.][]{piotto04}, we present an
alternative method by which quantitative constraints can be found for
the relative sizes of different stellar populations.  Specifically, a 
cluster-to-cluster variation in the central stellar mass function can
be looked for by comparing the core masses to the sizes of their
various stellar populations.  Since stellar evolution is the principal
factor affecting their relative numbers in the core, we expect the
size of each stellar population to scale linearly with the core mass.
If not, this could be evidence
that other factors, such as stellar dynamics, are playing an
important role.  
In this chapter, we present a comparison of the core 
RGB, main-sequence (MS) and HB populations of 56 GCs.
In particular, we use star 
counts for each stellar population to show that RGB stars are either 
over-abundant in the least massive cores or under-abundant in the most
massive cores, and that this effect is not
seen for MS stars.  We present the data in Section~\ref{data3} and our
methodology and results in Section~\ref{results3}.  In
Section~\ref{discussion3}, we discuss the implications of our results
and explore various possibilities for the source of the observed
discrepancy between RGB and MS stars.  Concluding remarks are
presented in Section~\ref{summary3}. 

\section{The Data} \label{data3}

Colour-magnitude diagrams (CMDs) taken from \citet{piotto02} are used to
obtain star counts for the RGB, HB, MS and blue straggler (BS)
populations in the cores of 56 GCs.  We apply the same selection 
criterion as outlined in Leigh, Sills \& Knigge (2007) to derive our
sample as well as to define the location of the main-sequence turn-off
(MSTO) in the (F439W-F555W)-F555W plane.  An example of this selection
criterion, applied to the 
CMD of NGC 362, is shown in Figure \ref{fig:ngc0362_labels}.  We
include all stars in the \citet{piotto02} database.  Since
\citet{piotto02} took their HST snapshots with the centre of the PC
chip aligned with the cluster centre, a portion of the cluster core
was not sampled for most GCs.  We have therefore applied a geometric
correction factor to the star counts in these clusters in order to
obtain numbers that are representative of the entire core
\citep{leigh07, leigh08}.  The total number of stars in the core is
found by summing over all stars brighter than 1 mag below the
MSTO and then multiplying by the appropriate geometric correction factor.

\begin{figure} [!h]
  \begin{center}
 \includegraphics[scale=0.5]{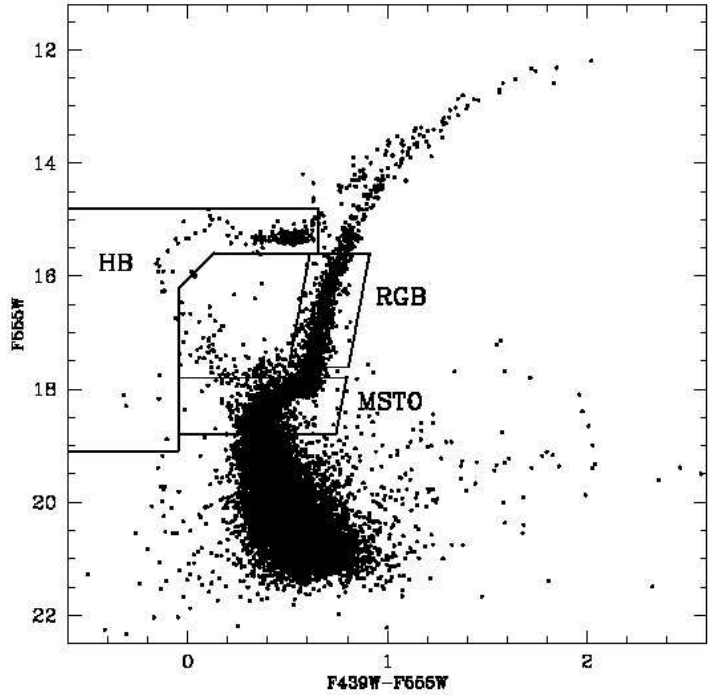}
    \caption{Colour-magnitude diagram for
  NGC 362 in the (F439W-F555W)-F555W plane.  Boundaries enclosing the
  selected RGB, HB and MSTO populations are shown.  
    \label{fig:ngc0362_labels}}
  \end{center}
\end{figure}  

Errors on the number of stars for each stellar population were
calculated using Poisson statistics.  Core radii,  
distance moduli, extinction corrections, central luminosity
densities and central surface brightnesses were taken from the Harris
Milky Way Globular Cluster catalogue \citep{harris96}.  Calibrated
apparent magnitudes in the F555W, F439W and Johnson V bands 
were taken from \citet{piotto02}.

\section{Results} \label{results3}

This chapter focuses on the core RGB, MS and HB populations of 56
GCs, comparing their numbers to the core masses.  Note that we
are focusing on the total number of stars in the core as a proxy for the
core mass instead of the total luminosity in the core in order to
avoid concerns regarding cluster-to-cluster variations in the central
stellar mass function and selection effects.  Given that a single
bright HB star can be as luminous as 100 regular MS stars, 
a small surplus of bright stars could have a dramatic impact on
the total luminosity.  
Therefore, the total number of stars in
the core is a more direct and reliable estimate for the core mass than
is the core luminosity.

Upon plotting the logarithm of the number of core RGB stars
versus the logarithm of the total number of stars in the core and
performing a weighted least-squares fit, we find a relation of
the form:
\begin{equation}
\label{eqn:power-laws}
\log (N_{RGB}) = (0.89 \pm 0.03)\log (N_{core}/10^3) + (2.04 \pm 0.02)
\end{equation}
The sub-linear slope is either indicative of a surplus of RGB stars in the
least massive cluster cores or a deficiency in the most massive cores.
Errors for lines of best fit were  
found using a bootstrap methodology in which we generated 1,000 fake
data sets by randomly sampling (with replacement) RGB counts from the
observations.  We obtained lines of best fit for each fake data set, fit a
Gaussian to the subsequent distribution and extracted its standard
deviation.  In order to avoid the additional uncertainty introduced
into our RGB number counts from trying to distinguish AGB stars from RGB
stars, as well as the difficulty in creating a selection criterion that
is consistent from cluster-to-cluster when including the brightest
portion of the RGB, stars that satisfy the RGB 
selection criterion shown in Figure \ref{fig:ngc0362_labels} are
referred to as RGB stars throughout this chapter.  Note that it is
the brightest portion of the RGB that should be the most affected by
stellar evolution effects such as mass-loss.  If we
extend our selection criterion to include the entire RGB, however, our
results remain unchanged.

Interestingly, MS plus sub-giant branch stars (hereafter
collectively referred to as MSTO stars, the selection criterion for
which is shown in Figure 1) show a more linear relationship than do RGB
stars and appear to dominate the central star counts.  If we count
only those stars having a F555W mag within half a magnitude above and
below the turn-off, we obtain a relation of the form:
\begin{equation}
\log (N_{MSTO}) = (1.02 \pm 0.01)\log (N_{core}/10^3) + (2.66 \pm 0.01)
\end{equation}
A nearly identical fit is found when counting only those
stars having a F555W mag between the turn-off and one
magnitude fainter than the turn-off. 

We also tried plotting the logarithm of the number of core
helium-burning stars (labeled HB in Figure~\ref{fig:ngc0362_labels})
versus the logarithm of the number of stars in the core, yielding a
relation of the form:
\begin{equation}
\log (N_{HB}) = (0.91 \pm 0.10)\log (N_{core}/10^3) + (1.58 \pm 
0.05)
\end{equation}   
Note the large uncertainty associated with the fit, indicating that
the slope is consistent with both those of the RGB and MSTO samples.
We will discuss this stellar population further in
Section~\ref{discussion3}.

The number of MSTO, RGB and HB stars are shown as a function of the
total number of stars in the core in Figure \ref{fig:N_vs_ncore}.
Interestingly, the blue stragglers in our sample also scale
sub-linearly with core mass, albeit more dramatically, obeying a
relation of the form N$_{BS}$ $\sim$ M$_{core}^{0.38 \pm 0.04}$
\citep{knigge09}.  Note that N$_{core}$ can be used interchangeably
with M$_{core}$.  In this case, we obtain a fit of:
\begin{equation}
\label{eqn:power-law-bs}
\log (N_{BS}) = (0.47 \pm 0.06)\log (N_{core}/10^3) + (1.22 \pm 0.02)
\end{equation}

\begin{figure} [!h]
  \begin{center}
 \includegraphics[scale=0.5]{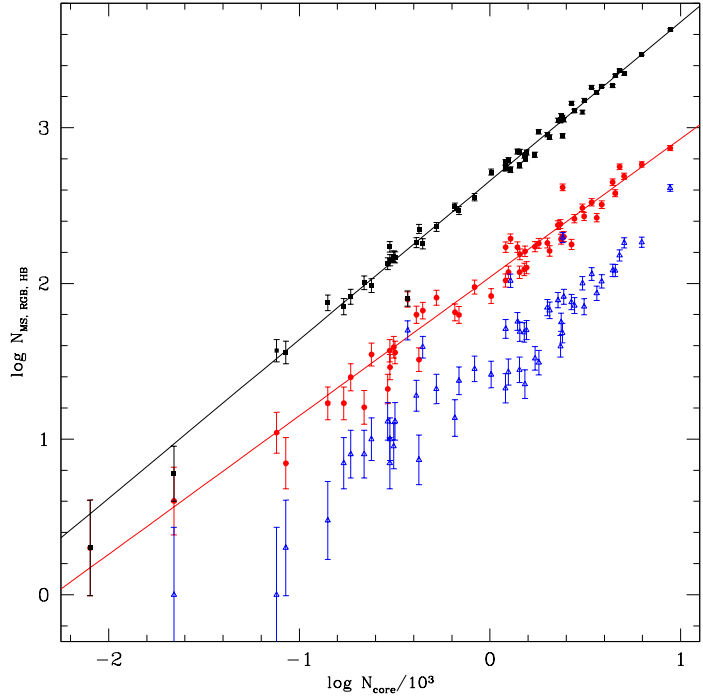}
    \caption[N$_{core}$ versus N$_{MS,RGB,HB}$]{The number of RGB
  (red circles), MSTO (black squares) and HB (blue triangles) stars
  found within the cluster core plotted versus the total number of
  stars in the core brighter than 1 mag below the MSTO, along with the 
  corresponding lines of best fit for the RGB and MSTO samples.
    \label{fig:N_vs_ncore}}
  \end{center}
\end{figure}

In an effort to explore the influence of selection effects, we re-did our
plots having removed from our sample clusters denser than log $\rho$
$>$ 10$^5$ L$_{\odot}$ pc$^{-3}$ since we are the most likely to be
under-counting stars in the most crowded cluster cores where blending
of the stellar light is the most severe.  This cut also removes from
our sample the post-core collapse (PCC) clusters for which the core
radii are poorly defined since King models are known to provide a poor
fit to the observed surface brightness profiles in these clusters.
Similarly, we applied a cut in the central surface brightness,
removing from our sample clusters satisfying $\Sigma_0$ $<$ 15.1 V mag
arcsec$^{-2}$.  Finally, since clusters having both high surface
brightnesses and small cores are the most likely to suffer from
selection effects, we also tried adding to 
the aforementioned cut in $\Sigma_0$ a cut in the angular core radius,
removing clusters with r$_c$ $\le$ 0.05'.  In all cases, the
sub-linear power-law index reported for the RGB remains unchanged to
within one standard deviation of our original result.  Selection 
effects do not appear to be the source of the observed sub-linearity,
though it is clear that its effects must properly be accounted for in
future studies.

In order to assess the effects of age-related cluster-to-cluster
variations in the stellar mass function, as well as to test our
assumption that the number of stars in the core provides a reliable
estimate for the core mass, we have obtained MSTO 
masses for most of the GCs in our sample.  We fit theoretical
isochrones provided in \citet{pols98} to the cluster CMDs, using
the bluest point along the MS of a given isochrone as a proxy for the
MSTO mass.  Isochrones were calculated
using the metallicities of \citet{piotto02} and cluster ages were
taken from \citet{deangeli05} using the Zinn \& West (1984)
metallicity scale.  Core masses were estimated by multiplying the mass
corresponding to the MSTO (m$_{MSTO}$) by the number of stars in the
core brighter than 1 mag below the turn-off.  This is a reasonable
assumption given the very small dispersion in the ages of MW
GCs \citep{deangeli05} and the fact that we are only considering stars
brighter than 1 mag below the TO.  Consequently, the range of stellar
masses upon which we are basing our 
number counts is very small.  Our results remain entirely unchanged
upon using M$_{core}$ $\sim$ N$_{core}$m$_{MSTO}$ as a proxy for the
core mass instead of pure number counts.

In order to further check the sensitivity of our results to our estimate for
the core masses, we re-did all plots shown in
Figure~\ref{fig:N_vs_ncore} using various approximations for the total core
luminosity instead of pure number counts.  Core luminosities are
calculated in the Johnson V band directly from the stellar fluxes
which are summed over all stars in the core and then multiplied by the
appropriate geometric correction factor.  We also adopted L$_{core}$ =
$\frac{4}{3}\pi$r$_c^3$$\rho_0$, where $\rho_0$ is the
central luminosity density in L$_{\odot}$ pc$^{-3}$ taken from
\citet{harris96}.  Additionally, since the number of core RGB stars is
in reality a projected quantity, we tried plotting N$_{RGB}$ versus
L$_{core}$ = $\pi$r$_c^2$$\Sigma_0$, where 
$\Sigma_0$ is the central surface brightness in L$_{\odot}$ 
pc$^{-2}$, so that we are consistently comparing two projected
quantities.  Finally, we can adopt slightly more realistic
estimates for the total core luminosity by integrating over King
density profiles.  We fit single-mass King models calculated using the
method of \citet{sigurdsson95} to the surface brightness profiles of the
majority of the clusters in our sample using the concentration
parameters of \citet{mclaughlin05} and the central luminosity
densities of \citet{harris96}.  We then integrated the derived
luminosity density profiles numerically in order to estimate the total
stellar light contained within the core.  After removing clusters
labelled as post-core collapse in \citet{harris96} for which King
models are known to provide a poor fit, we once again compared the
integrated core luminosities to the number of RGB stars in the core.
For all four of these estimates for the total core luminosity, we
find that our fundamental results remain unchanged, with the power-law
index for RGB stars remaining sub-linear at the 3-$\sigma$ level.
Therefore, we conclude that our result is robust to changes in choices
of cluster and population parameters.  

\section{Discussion} \label{discussion3}

We have shown that the number of RGB stars in globular cluster cores
does not directly trace the total stellar population in those cores.  In
particular, the number of RGB (but not MSTO) stars 
scales sub-linearly with core mass at the 3-$\sigma$ level.  Given
that the MS lifetime is expected to be a factor of 10-100 
longer than that of the RGB sample \citep{iben91}, the ratio
N$_{MSTO}$/N$_{RGB}$ indicates that the relative sizes of these stellar
populations are in better agreement with the expectations of stellar evolution
theory in the most massive cores.  This suggests that our results are
consistent with a surplus of RGB stars in the least massive cores.  We 
discuss below some of the key considerations in understanding the
evolution of GC cores and the stars that populate them in an effort to
explain our result.

\subsection{Stellar evolution}

Could this trend be a reflection of a stellar evolution process?  The
evolution and distribution of stellar populations can be thought 
of as the sum of many single stellar evolution tracks, which depend
only on a star's mass and composition.  Since there is no relation
between a cluster's mass and its metallicity \citep{harris96} and the
dispersion in the relative ages of MW GCs is quite small
\citep{deangeli05}, there is no reason to expect the RGB 
lifetime to depend on the cluster mass.  On the other hand, recent
studies suggest that 
the chemical self-enrichment of GCs during their early evolutionary
stages could help to explain many of the population differences
observed among them \citep[e.g.][]{caloi07}.  In particular, many of
the most massive GCs are thought to be enriched in helium and this is
expected to reduce the time scale for stellar evolution
\citep[e.g.][]{romano07}.  While this scenario predicts a deficiency
of RGB stars in the most massive cores, it would contribute to
depressing the slope of the RGB sample relative to that of the MSTO
population in Figure~\ref{fig:N_vs_ncore}.  

\subsection{Single star dynamics} \label{dynamics3}

Two-body relaxation is the principal driving force behind the dynamical
evolution of present-day GCs, slowly steering them towards a state of
increased mass stratification as predominantly  
massive stars fall into the core and typically low-mass stars are
ejected via dynamical encounters.  The relaxation time increases
with the cluster mass \citep{spitzer87} and the variance in the relative ages
(and hence MSTO masses) of MW GCs is quite small
\citep{deangeli05}.  Therefore, it is the
least massive clusters that should show signs of being the
most dynamically evolved.  This assumes that cluster-to-cluster
variations in the initial mass function and the degree of
initial mass segregation are small.  In general, however,
proportionately fewer massive stars should have had sufficient time to
migrate into the most massive cores, while 
fewer low-mass stars should have been ejected out.  While qualitatively
correct, this effect should contribute little to the observed
difference between the core RGB and MSTO populations since RGB stars
are only slightly more massive \citep[e.g.][]{demarchi07}.

Stars expand considerably as they ascend the RGB.  Both the increase
in collisional cross-section and the change 
in the average stellar density could have an important bearing on the
outcomes of dynamical interactions involving RGB stars.  Indeed,
\citet{bailyn94} suggests that interactions between giants and other
cluster members in the core could strip the outer envelope of the
giant before it has a chance to fully ascend the RGB.  Since our
adopted RGB selection criterion does not include the brightest giants, 
we are only considering giants that are larger than MSTO stars
by a factor of $\sim$ 10 \citep{iben91}.  This small degree of
expansion will have only a minor effect on the collision rate.  Any
scenario that relies on dynamical encounters to explain a depletion of
RGB stars should be operating in very dense cores.  Our results are
consistent with some of the densest clusters in our sample having a
surplus of giants, however.  

\subsection{Binary effects}

Stripping of the envelopes of large stars could also be mediated by a
binary companion as the expanding giant overfills its Roche lobe
\citep{bailyn94}.  While this process should preferentially
occur in the centres of clusters where binaries will congregate as
a result of mass segregation, two-body relaxation progresses more
slowly in the most massive clusters.  Binaries should therefore sink 
to the cluster core more quickly in the least massive clusters,
contributing to an increase in the core binary fraction at a rate that
decreases with increasing cluster mass.  Observational evidence has
been found in support of this, most notably by \citet{sollima07} and
\citet{milone08} who found an anti-correlation between the cluster 
mass and the core binary fraction.  Any mechanism for RGB
depletion that relies on binary stars should therefore operate
more efficiently in the least massive cores where the binary
fraction is expected to be the highest.  Our results are consistent
with a surplus of giants in the least massive cores,
however.  This therefore argues against a binary mass-transfer origin
for RGB depletion in massive GC cores.  For similar reasons, it seems
unlikely that our result can be explained by collisions between RGB
stars and binaries.  If, on the other hand, 
RGB stars are more commonly found in binaries than are MS stars,
perhaps as a result of their larger cross-sections for tidal capture, 
binary stars could still be contributing to the observed trend.  Note
that in the cluster outskirts where the velocity dispersion has
dropped considerably from its central value, individual encounters are
more likely to result in tidal capture.  Since mass segregation should deliver
binaries to the core faster in the least massive clusters, a
larger fraction of their RGB stars could have hitched
a ride to the core as a binary companion.   However, both 
the average half-mass relaxation time of MW GCs and the RGB lifetime
tend to be on the order of a Gyr 
\citep{harris96, iben91}.  This does not leave much time for giants to be
captured into binaries and subsequently fall into the core before
evolving away from the RGB.

\subsection{Core helium-burning stars}

The fit for the HB sample is consistent with those of the RGB and MSTO
samples at the 3-sigma level so that we are unable to draw any
reliable conclusions for this stellar population.  The high
uncertainty stems from a number of outlying clusters.  
Selection effects and contamination from the Galaxy are likely to be
playing a role in this, in addition to our formulaic selection
criterion which may not be as suitable to the varying morphology of
the HB as it is to other stellar populations.  That is, the creation of
a purely photometry-based cluster-independent selection criterion may
not be possible for HB stars.  Given that stellar evolution effects
are expected to be the 
most dramatic at the end of the RGB lifetime, an interplay with the
cluster dynamics could also be contributing.  In particular, if the
central relaxation time is shorter than the HB lifetime, 
significant numbers of HB stars could be ejected from the core via
dynamical encounters as a result of having lost around a quarter of
their mass upon evolving off the tip of the RGB.  Moreover, since
stars expand considerably as they ascend the RGB, many of
the dynamical arguments presented in Section~\ref{dynamics3} may more
strongly affect the size of the HB sample if they are the direct
evolutionary descendants of RGB stars.  Since at most a handful of
studies have been performed comparing the radial HB and RGB
distributions in GCs \citep[e.g.][]{iannicola09}, more data is needed
before any firm constraints can be placed on the source of the 
poor fit found for the HB sample.

\subsection{An evolutionary link with blue stragglers?}

The addition of a small number of extra RGB stars to every cluster is one
way to account for the observed sub-linear dependence on core
mass since the fractional increase in the size of
the RGB population will be substantially larger in the least massive
cores.  In log-log space, the result is a reduction of the
slope.  Since blue stragglers will evolve
into RGB and eventually HB stars \citep{sills09}, evolved BSs could be
the cause of a surplus of RGB (and possibly  
core helium-burning) stars in these clusters.  This scenario also
predicts that MSTO stars should scale slightly more linearly with core
mass since there should be a smaller contribution from evolved BSs, as
we have shown.  Given the fits for the RGB and BS  
samples presented in Section~\ref{results3} and their corresponding
uncertainties, we find that the addition of evolved BSs to the RGB
populations could inflate the slope enough that the dependence on core
mass becomes linear.  Upon subtracting the BS sample from the RGB
sample, we find that the new fit is consistent with being linear:
\begin{equation}
\log (N_{RGB}) = (0.94 \pm 0.04)\log (N_{core}/10^3) + (1.97 \pm 0.02)
\end{equation}
The slope becomes larger if we have under-estimated the
number of BSs, perhaps as a result of selection effects, our adopted
selection criterion or a larger population size having existed in the
past.

\section{Summary} \label{summary3}

In this chapter, we have performed a cluster-to-cluster comparison
between the number of core RGB, MSTO \& HB stars 
and the total core mass.  We have introduced a technique for comparing
stellar populations in clusters that is well suited to studies of both
cluster and stellar evolution, in addition to the interplay thereof.
Using a sample of 56 GCs taken from Piotto et al.'s 2002 HST database,
we find a sub-linear scaling for RGB stars at the 3-$\sigma$ level,
whereas the relation is linear for MSTO stars.  While the preferential
self-enrichment of massive GCs, two-body relaxation, and evolved BSs
could all be contributing to the observed sub-linear dependence,
further studies with an emphasis on selection effects are needed in
order to better constrain the source of this curious observational result.

\section*{Acknowledgments}

We would like to thank an anonymous referee for a 
number of helpful suggestions, as well as Bill Harris, 
David Chernoff, Barbara Lanzoni and Francesco Ferraro for useful
discussions.  This research has been supported by NSERC as well as the
National Science Foundation under Grant No. PHY05-51164 to the Kavli
Institute for Theoretical Physics.





\pagestyle{fancy}
\headheight 20pt
\lhead{Ph.D. Thesis --- N. Leigh }
\rhead{McMaster - Physics \& Astronomy}
\chead{}
\lfoot{}
\cfoot{\thepage}
\rfoot{}
\renewcommand{\headrulewidth}{0.1pt}
\renewcommand{\footrulewidth}{0.1pt}




\chapter{Dissecting the
  Colour-Magnitude Diagram:  A Homogeneous Catalogue of Stellar
  Populations in
  Globular Clusters} \label{chapter4}
%
\thispagestyle{fancy}

\section{Introduction} \label{intro4}
Colour-magnitude diagrams (CMDs) are one of the most important tools
available to astronomers for studying stellar evolution, stellar
populations and star clusters.  And yet, there remain several
features found in CMDs whose origins are still a mystery.
Examples include horizontal branch (HB) morphology, the presence of
extended horizontal branch (EHB) stars, and blue stragglers (BSs) 
\citep[e.g.][]{sandage53, zinn96,
  peterson03, dotter10}.  Previous studies have shown that the
observed differences in the HBs of Milky Way globular clusters (GCs)
are related to metallicity \citep{sandage60}, however at least one
additional parameter is required to explain the spread in their
colours.  Many cluster properties have been suggested as possible
Second and Third Parameters, including age, central density and
cluster luminosity, although no definitive candidates have been
identified \citep[e.g.][]{rood73, fusi93}.  An 
explanation to account for the existence of BSs has proved equally
elusive.  Many BS formation mechanisms have been proposed,
including stellar collisions \citep[e.g.][]{leonard89, sills99} and
binary mass-transfer \citep{mccrea64, mathieu09}.  However, no clear
evidence has yet emerged in favour of a dominant formation
mechanism.    

In short, we still do 
not understand how many of the physical processes operating within
star clusters should affect the appearance of CMDs
\citep[e.g.][]{fusi92, buonanno97, ferraro99, beccari06}.  In general,
the importance of these processes can be constrained by looking for
correlations between particular features in CMDs and cluster properties
that serve as proxies for different effects.
For example, the central density can be used as a rough proxy for the
frequency with which close dynamical encounters occur.  Similarly, the
cluster mass can be used as a proxy for the rate of two-body relaxation.  
Once the relevant effects are accounted for, CMDs can continue to
provide an ideal tool to further our 
understanding of stellar evolution, stellar populations and star
clusters. 

It is now clear that an important interplay
occurs in clusters between stellar dynamics and stellar evolution.
For example, dynamical models have shown that 
star clusters expand in response to mass-loss driven by stellar
evolution, particularly during their early evolutionary phases when 
massive stars are still present \citep[e.g.][]{chernoff90, portegieszwart98,
  gieles10}.  Mass-loss resulting from stellar evolution has also been
proposed to cause horizontal branch stars to exhibit more
extended radial distributions relative to 
red giant branch and main-sequence turn-off (MSTO) stars in
globular clusters having short 
central relaxation times relative to the average HB lifetime
\citep[e.g.][]{sigurdsson95, leigh09}.  This can be understood as
follows.  Red giant branch (RGB) stars
should be more mass segregated than other stellar populations since
they are among the most massive stars in GCs.  HB stars, on the other
hand, are much less massive since RGB stars experience significant
mass loss upon evolving into HB stars.  Consequently, two-body
relaxation should act to 
re-distribute HB stars to wider orbits within the cluster potential
relative to RGB and MSTO stars \citep{spitzer75}, provided the average
HB lifetime is shorter than 
the central relaxation time.  Studies have shown that the radial
distributions of the HB populations in some GCs could differ from
those of other stellar populations.  For
instance, \citet{saviane98} 
presented evidence that blue HB stars could be more centrally
concentrated than red HB and sub-giant branch stars in the GC NGC
1851.  Conversely, \citet{cohen97} showed that blue HB stars could be
centrally depleted relative to other stellar types in the GC NGC
6205.  To date, no clear evidence has been found linking the spatial
distributions of HB stars to any global cluster properties.  

Peculiar trends have also been reported for the radial distributions
of RGB stars.  
For example, a deficiency of bright red giants has been observed 
in the GC NGC 1851 \citep[e.g.][]{iannicola09}.  \citet{sandquist07} 
discussed the possibility that this deficiency could be the result of
strong mass loss on the RGB.  
Alternatively, some authors have suggested that dynamical effects could
deplete red giants.  For instance, 
giants could experience collisions more frequently than other stellar
populations due to their larger cross-sections for collision
\citep{beers04}.

One important example of the interplay that occurs in clusters between
stellar evolution and stellar dynamics can be found in the study
of blue stragglers.  Found commonly in both open and globular clusters, BSs are
thought to be produced by the addition of hydrogen to the cores of
low-mass main-sequence (MS) stars, and therefore appear as an
extension of the MSTO in cluster CMDs \citep{sandage53}.  This can
occur via multiple channels, most of which involve the mergers of
low-mass MS stars since a significant amount of mass is typically 
required to reproduce the observed locations of BSs in CMDs
\citep[e.g.][]{sills99}.  Stars in close binaries can merge if enough
orbital angular momentum is lost, which can be mediated by dynamical
interactions with other stars, magnetized stellar winds, tidal
dissipation or even an outer triple companion
\citep[e.g.][]{leonard92, li06, perets09, dervisoglu10}.
Alternatively, MS stars can collide directly, although this is 
also thought to usually be mediated by multiple star systems
\citep[e.g.][]{leonard89, fregeau04, leigh10}.  Finally, BSs have also
been hypothesized to form by 
mass-transfer from an evolving primary onto a normal MS companion
via Roche lobe overflow \citep{mccrea64}.  

Whatever the dominant BS
formation mechanism(s) operating in dense star clusters, it is now
thought to somehow involve multiple star systems.  This was shown 
to be the case in even the dense cores of GCs \citep{knigge09} where
collisions between single stars are thought to occur frequently
\citep{leonard89}.  In \citet{knigge09}, we showed that the numbers of
BSs in the cores 
of a large sample of GCs correlate with the core masses.  We 
argued that our results are consistent with what is expected if BSs
are descended from binary stars.  \citet{mathieu09} also showed
that at least $76\%$ of the BSs in the old open cluster NGC 188 have 
binary companions.  Although the nature of these companions remains
unknown, it is clear that binaries played a role in the
formation of these BSs.  Dynamical
interactions occur frequently enough in dense clusters that they
are also expected to be at least partly responsible for the observed
properties of BSs.  It follows
that the current properties of BS populations
should reflect the dynamical histories of their host clusters.  As a
result, BSs could provide an indirect means of probing
the physical processes that drive star cluster evolution.

In this chapter, we present a homogeneous catalogue for red giant
branch, main-sequence turn-off, horizontal branch and blue
straggler stars in a sample of 35 Milky Way (MW) GCs
taken from the Advanced Camera for Surveys
(ACS) Survey for Globular Clusters \citep{sarajedini07}.  With this
catalogue, we investigate two important issues related to stellar
populations in GCs.  First, we test the observational correlation
found for BSs presented in \citet{knigge09} by re-doing the study with
newer and more accurate photometry.  The larger spatial coverage
considered in our new sample 
offers an important additional constraint for the origin of this correlation.  
Second, we perform the same statistical comparison for RGB, HB and
MSTO stars in 
order to study their radial distributions.  This will allow us to test
some of the results and hypotheses introduced in \citet{leigh09},
where we first presented this
technique for studying stellar populations.  In
particular, we found evidence for a surplus of RGB stars in
low-mass GC cores relative to MSTO stars.  However,
we concluded that the study needed to be re-done with better
photometry.  The ACS data are of sufficiently high quality to address
this issue.

In Section~\ref{method4}, we present our selection criteria to
determine the numbers of BS, RGB, HB and MSTO stars located in the central
cluster regions.  The spatial coverage of the photometry extends out
to several core radii from the cluster centre for most of the clusters
in our sample.  For these clusters, we have obtained number counts within
several circles centred on the cluster centres provided in
\citet{goldsbury10} 
for various multiples of the core radius.  This catalogue is presented in
Section~\ref{results4}.  In this section, we also present a comparison
between the sizes of the different stellar populations and the total
stellar masses contained within each circle and annulus.  Finally,
we discuss our results for both BSs and the other 
stellar populations in Section~\ref{discussion4}.   

\section{Method} \label{method4}

In this section, we present our sample of CMDs and define our
selection criteria for each of the different stellar populations.  We
also discuss the spatial coverage offered by the ACS sample, and
describe how we obtain estimates for several different fractions of
the total cluster mass from King models.

\subsection{The Data} \label{data4}

The data used in this study consists of a sample of 35 MW GCs taken
from the ACS Survey for Globular Clusters \citep{sarajedini07}.\footnote[1]{The
data can be found at http://www.astro.ufl.edu/~ata/public\_hstgc/.}  The
ACS Survey provides unprecedented deep photometry in the F606W ($\sim$
V) and F814W ($\sim$ I) filters 
that is nearly complete down to $\sim 0.2$ M$_{\odot}$.  In other
words, the 
CMDs extend reliably from the HB all the way down to about 7
magnitudes below the MSTO.  We have confirmed that
the photometry is nearly complete above at least 0.5
magnitudes below the MSTO for every cluster in our sample.  This was
done using the results of artificial star tests taken from  
\citet{anderson08}, and confirms that the photometric quality of
the stellar population catalogue presented in this chapter is very
high.\footnote[2]{Artificial star tests were obtained directly from Ata
Sarajedini via private communication.}  We have also considered
foreground contamination by field stars, and it is negligible.

Each cluster was centred in the ACS field, which
extends out to several core radii from the cluster 
centre in most clusters and, in a few cases, beyond even 15 core
radii.  Coordinates for the cluster centres were taken from 
\citet{goldsbury10}.  These authors found their centres by fitting
a series of ellipses to the density distributions within the inner 2'
of the cluster centre, and computing an average value.  The core
radii were taken from \citet{harris96}.

\subsection{Stellar Population Selection Criteria} \label{criteria4}

In order to select the number of stars belonging to each stellar
population, we define a series of lines in the (F606W-F814W)-F814W
plane that act as boundaries enclosing each of the different stellar
populations.  
To do this, we fit theoretical isochrones taken from \citet{dotter07}
to the CMDs of every cluster in our sample.  Each isochrone
was generated using the metallicity and age of the cluster, and fit to
its CMD using the corresponding distance modulus and extinction
provided in \citet{dotter10}.  
The MSTO was then defined using our isochrone fits by selecting the
bluest point along the MS.  This acts as our primary point of
reference for defining the boundaries in the CMD for the different
stellar populations.  Consequently, the selection criteria
provided in this chapter are a significant improvement upon the criteria
presented in \citet{leigh07}, and our new catalogue for the
different stellar populations is highly homogeneous.

Two additional points of reference must also
be defined in order for our selection criteria to be applied 
consistently from cluster-to-cluster.  First, the selection criteria 
for the HB are determined by fitting a line through the approximate
mid-point of the points that populate it in the CMD.  This
line is then used to define upper and lower boundaries for the
HB.  Theoretical isochrones become highly uncertain at the HB, so it
is necessary to specify this additional criterion by eye.  Second, the
lower boundary of the RGB is defined for each cluster as 
the point along its isochrone corresponding to a helium core mass of
0.08 M$_{\odot}$.  We do not include RGB
stars brighter than the HB since the tilt 
of the upper RGB varies significantly from cluster-to-cluster, 
presenting a considerable challenge for the consistency of our
selection criteria.  Moreover, the distinction between RGB and
asymptotic giant branch stars in the CMD is often ambiguous.

Example selection criteria for each of the different stellar
populations are shown in Figure~\ref{fig:fig1}.  Formal definitions
for the boundaries in the cluster CMD 
that define the BS, RGB, HB and MSTO populations are provided in
Appendix~\ref{Appendix-B}.  We note that the sizes of our selection
boxes have been chosen to accommodate the photometric errors, which
contribute to broadening the various evolutionary sequences in the
CMD.  With superior photometry, the sizes of our selection
boxes could therefore be reduced.  This would further 
decrease contamination from field stars in our samples.

\begin{figure} [!h]
  \begin{center}
 \includegraphics[scale=0.5]{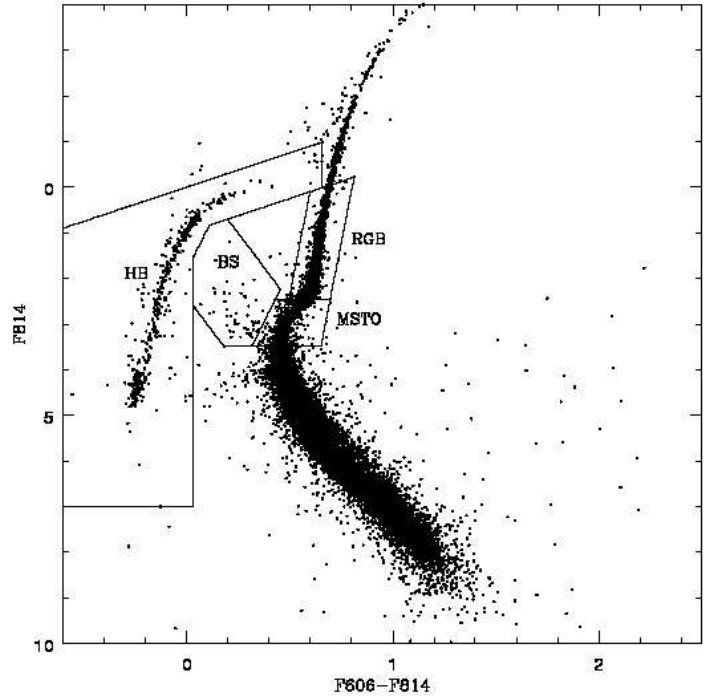}
\caption[Plot showing the parameter space in the (F606W-F814W)-F814W
plane defining each of the different stellar populations for the GC
NGC 6205]{Colour-magnitude diagram for the Milky Way globular cluster
  NGC 6205.  Boundaries enclosing the parameter space in the 
  (F606W-F814W)-F814W plane that define each of the different stellar
  populations are indicated with solid lines, as described in the
  text.  Absolute 
  magnitudes are shown, converted from apparent magnitudes using the
  distance moduli and extinctions provided in \citet{dotter10}.
  Labels for blue straggler, red giant branch, horizontal branch and
  main-sequence turn-off stars are indicated.  Stars with large
  photometric errors have been omitted from this plot.
\label{fig:fig1}}
\end{center}
\end{figure}

\subsection{Spatial Coverage} \label{spatial4}

The ACS field of view extends out to several core radii from the
cluster centre for nearly every cluster in our sample.  Consequently,
we have obtained estimates for the number 
of stars contained within four different circles centred on the central
cluster coordinates provided in \citet{goldsbury10}.  We list
these numbers only for clusters for which the indicated circle is
completely sampled by the field of view.  
The radii of the circles were taken to be integer multiples of the
core radius, and we focus our attention on the inner four core radii
since the field of view extends beyond this for only a handful of the
clusters in our sample.  An example
of this is shown in Figure~\ref{fig:fig2}.  The numbers we
list are cumulative, so that entries for each circle include
stars contained within all smaller circles.  

\begin{figure} [!h]
  \begin{center}
 \includegraphics[scale=0.5]{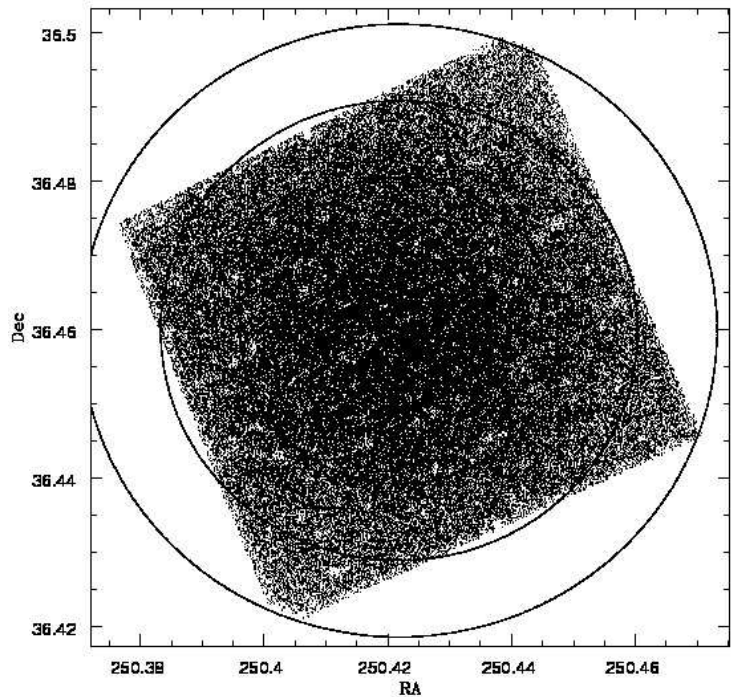}
\caption[Plot showing the RA and Dec coordinates for all stars in NGC
6205]{RA and Dec coordinates for all stars in the GC NGC 6205.
  Circles corresponding to one, two, three and four core radii are 
  shown.  
\label{fig:fig2}}
\end{center}
\end{figure}

\subsection{King Models} \label{king4}

In order to obtain accurate estimates for the total stellar mass
contained within each circle, we generated 
single-mass King models calculated using the method of
\citet{sigurdsson95} to obtain luminosity density profiles
for the 
majority of the clusters in our sample.  The profiles were obtained using
the concentration parameters of \citet{mclaughlin05} and the central
luminosity densities of \citet{harris96} for each cluster in
\citet{mclaughlin05} that overlaps with our sample.  We then
integrated the derived 
luminosity density profiles numerically in order to estimate the total
stellar light contained within each circle.  After removing clusters
with high concentration parameters \citep{harris96} for which King
models are known to provide a poor fit, we multiplied the total
stellar light by a mass-to-light ratio of 2 in order to obtain
estimates for the total stellar mass contained within each circle.
Calculating the total stellar mass contained within each
circle from King models requires a number of assumptions that we will
discuss fully in Section~\ref{discussion4}.

\section{Results} \label{results4}

In this section, we present our catalogue along with the results of
our comparisons between the 
sizes of the different stellar populations and the total stellar mass
contained within each circle and annulus.

\subsection{Catalogue}  \label{catalogue4}

The numbers of BS, RGB, HB and MSTO stars found within several
different circles are shown for all clusters in
Table~\ref{table:catalogue}, along with the total number of stars 
with magnitudes brighter than 0.5 mag below the MSTO.  Number counts
are only shown whenever the spatial coverage is complete within the
indicated circle.

\clearpage

\begin{sidewaystable}
\tiny
\centering
\caption{Stellar Population Catalogue
  \label{table:catalogue}}
\begin{tabular}{|p{1cm}@{}|p{1.5cm}@{}|p{1.7cm}@{}|@{}c@{}|@{}c@{}|@{}c@{}|@{}c@{}|@{}c@{}|@{}c@{}|@{}c@{}|@{}c@{}|@{}c@{}|@{}c@{}|@{}c@{}|@{}c@{}|c@{}|c@{}|c@{}|c@{}|c@{}|c@{}|c@{}|c@{}|}
\hline
Cluster ID  &  Alternate ID  &  Core Radius (in arcmin)  & \multicolumn{4}{|c|}{N$_{BS}$}           & \multicolumn{4}{|c|}{N$_{HB}$}           & \multicolumn{4}{|c|}{N$_{RGB}$}          & \multicolumn{4}{|c|}{N$_{MSTO}$}         & \multicolumn{4}{|c|}{N$_{TOT}$}          \\
\hline
            &              &                           & $< r_c$ & $< 2r_c$ & $< 3r_c$ & $< 4r_c$ & $< r_c$ & $< 2r_c$ & $< 3r_c$ & $< 4r_c$ & $< r_c$ & $< 2r_c$ & $< 3r_c$ & $< 4r_c$ & $< r_c$ & $< 2r_c$ & $< 3r_c$ & $< 4r_c$ & $< r_c$ & $< 2r_c$ & $< 3r_c$ & $< 4r_c$ \\
\hline
 104 &    47 Tuc    & 0.36 & 62 & 100 & 120 & 128 & 172 & 344 & 486 & 615 &  397 &  944 & 1454 & 1798 & 2190 & 5004 & 7300 & 9080 &   4874 & 11430 & 16985 & 21183 \\
1261 &              & 0.35 & 56 &  79 &  95 & 104 &  73 & 170 & 216 & 250 &  241 &  481 &  664 &  755 & 1102 & 2268 & 2953 & 3369 &   2713 &  5576 &  7347 &  8429 \\
1851 &              & 0.09 & 34 &  58 &  74 &  90 &  33 & 107 & 161 & 213 &   93 &  223 &  307 &  385 &  178 &  692 & 1128 & 1524 &    417 &  1418 &  2421 &  3400 \\
2298 &              & 0.31 & 27 &  32 &  37 &  38 &  16 &  41 &  56 &  63 &   61 &  120 &  158 &  186 &  208 &  429 &  568 &  662 &    549 &  1117 &  1490 &  1753 \\
3201 &              & 1.30 & 40 &  -- &  -- &  -- &  43 &  -- &  -- &  -- &  160 &   -- &   -- &   -- &  635 &   -- &   -- &   -- &   1691 &    -- &    -- &    -- \\
4147 &              & 0.09 & 16 &  26 &  30 &  34 &   7 &  18 &  35 &  44 &   23 &   61 &   93 &  120 &   89 &  206 &  316 &  400 &    234 &   569 &   844 &  1064 \\
4590 &      M 68    & 0.58 & 29 &  59 &  -- &  -- &  33 &  66 &  -- &  -- &  152 &  269 &   -- &   -- &  480 &  977 &   -- &   -- &   1321 &  2623 &    -- &    -- \\
5024 &      M 53    & 0.35 & 57 & 103 & 133 & 149 & 114 & 235 & 333 & 387 &  293 &  704 & 1059 & 1260 & 1215 & 2864 & 4106 & 4891 &   3118 &  7504 & 10730 & 12827 \\
5139 & $\Omega$ Cen & 2.37 & 49 &  87 &  -- &  -- & 408 & 762 &  -- &  -- & 1441 & 2592 &   -- &   -- & 4643 & 8637 &   -- &   -- &  12652 & 23178 &    -- &    -- \\
5272 &       M 3    & 0.37 & 74 & 111 & 127 & 135 & 153 & 311 & 379 & 413 &  496 &  995 & 1277 & 1387 & 1909 & 3828 & 5052 & 5512 &   4971 & 10020 & 13195 & 14429 \\
5286 &              & 0.28 & 82 & 120 & 138 & 144 & 218 & 413 & 530 & 599 &  442 &  970 & 1308 & 1535 & 1723 & 3666 & 4983 & 5876 &   4016 &  8934 & 12448 & 14826 \\
5466 &              & 1.43 & 30 &  -- &  -- &  -- &  37 &  -- &  -- &  -- &  123 &   -- &   -- &   -- &  487 &   -- &   -- &   -- &   1276 &    -- &    -- &    -- \\
5904 &       M 5    & 0.44 & 37 &  57 &  64 &  68 &  97 & 212 & 291 & 338 &  233 &  516 &  729 &  885 &  997 & 2260 & 3190 & 3843 &   2483 &  5700 &  8123 &  9846 \\
5927 &              & 0.42 & 28 &  71 &  93 & 122 &  91 & 207 & 294 & 358 &  188 &  513 &  748 &  922 & 1214 & 3043 & 4528 & 5667 &   2619 &  6714 & 10108 & 12688 \\
5986 &              & 0.47 & 57 &  88 &  -- &  -- & 220 & 386 &  -- &  -- &  614 & 1136 &   -- &   -- & 2359 & 4549 &   -- &   -- &   5756 & 11255 &    -- &    -- \\
6093 &      M 80    & 0.15 & 79 & 114 & 133 & 135 &  94 & 199 & 269 & 331 &  252 &  543 &  773 &  984 & 1045 & 2176 & 3090 & 3790 &   2008 &  4627 &  6840 &  8637 \\
6101 &              & 0.97 & 26 &  -- &  -- &  -- &  68 &  -- &  -- &  -- &  173 &   -- &   -- &   -- &  681 &   -- &   -- &   -- &   1798 &    -- &    -- &    -- \\
6121 &       M 4    & 1.16 & 11 &  18 &  -- &  -- &  21 &  46 &  -- &  -- &   52 &  126 &   -- &   -- &  243 &  574 &   -- &   -- &    553 &  1350 &    -- &    -- \\
6171 &     M 107    & 0.56 & 19 &  43 &  54 &  -- &  16 &  37 &  56 &  -- &   63 &  153 &  223 &   -- &  264 &  667 &  933 &   -- &    677 &  1688 &  2414 &    -- \\
6205 &      M 13    & 0.62 & 41 &  58 &  -- &  -- & 207 & 416 &  -- &  -- &  527 & 1162 &   -- &   -- & 1960 & 4250 &   -- &   -- &   5015 & 10973 &    -- &    -- \\
6218 &      M 12    & 0.79 & 28 &  50 &  -- &  -- &  32 &  68 &  -- &  -- &  114 &  245 &   -- &   -- &  447 & 1118 &   -- &   -- &   1127 &  2680 &    -- &    -- \\
6254 &      M 10    & 0.77 & 36 &  52 &  -- &  -- &  93 & 169 &  -- &  -- &  257 &  540 &   -- &   -- &  955 & 1985 &   -- &   -- &   2483 &  5165 &    -- &    -- \\
6304 &              & 0.21 & 19 &  36 &  51 &  67 &  27 &  65 &  95 & 112 &   82 &  207 &  313 &  397 &  453 & 1112 & 1657 & 2143 &    994 &  2584 &  3864 &  5008 \\
6341 &      M 92    & 0.26 & 41 &  73 &  84 &  91 &  60 & 126 & 177 & 217 &  140 &  367 &  540 &  684 &  543 & 1290 & 1896 & 2376 &   1409 &  3341 &  4943 &  6252 \\
6362 &              & 1.13 & 35 &  -- &  -- &  -- &  61 &  -- &  -- &  -- &  165 &   -- &   -- &   -- &  716 &   -- &   -- &   -- &   1844 &    -- &    -- &    -- \\
6535 &              & 0.36 &  7 &  11 &  12 &  12 &   8 &  18 &  24 &  26 &   21 &   42 &   56 &   72 &   54 &  107 &  174 &  227 &    165 &   338 &   493 &   629 \\
6584 &              & 0.26 & 36 &  54 &  63 &  -- &  52 &  95 & 135 &  -- &  217 &  386 &  482 &   -- &  788 & 1499 & 1863 &   -- &   2023 &  3810 &  4830 &    -- \\
6637 &      M 69    & 0.33 & 50 &  85 &  96 & 106 &  80 & 148 & 204 & 239 &  200 &  443 &  592 &  702 & 1067 & 2257 & 3063 & 3605 &   2413 &  5209 &  7129 &  8414 \\
6652 &              & 0.10 & 16 &  19 &  24 &  27 &  10 &  21 &  34 &  40 &   32 &   61 &   87 &  122 &  127 &  272 &  417 &  536 &    286 &   619 &   919 &  1218 \\
6723 &              & 0.83 & 39 &  -- &  -- &  -- & 113 &  -- &  -- &  -- &  354 &   -- &   -- &   -- & 1594 &   -- &   -- &   -- &   3777 &    -- &    -- &    -- \\
6779 &      M 56    & 0.44 & 21 &  41 &  48 &  49 &  44 &  99 & 133 & 158 &  128 &  302 &  435 &  528 &  411 &  993 & 1495 & 1875 &   1126 &  2679 &  3982 &  4912 \\
6838 &      M 71    & 0.63 & 17 &  45 &  -- &  -- &  10 &  30 &  -- &  -- &   36 &   95 &   -- &   -- &  144 &  385 &   -- &   -- &    355 &   960 &    -- &    -- \\
6934 &              & 0.22 & 35 &  54 &  57 &  60 &  50 & 100 & 137 & 163 &  150 &  322 &  431 &  508 &  612 & 1240 & 1681 & 1974 &   1528 &  3208 &  4308 &  5088 \\
6981 &      M 72    & 0.46 & 31 &  49 &  56 &  -- &  52 &  78 & 103 &  -- &  140 &  285 &  354 &   -- &  652 & 1272 & 1596 &   -- &   1594 &  3159 &  4000 &    -- \\
7089 &       M 2    & 0.32 & 83 & 129 & 143 & 150 & 277 & 551 & 729 & 838 &  535 & 1205 & 1652 & 1960 & 2264 & 4832 & 6603 & 7795 &   5394 & 12038 & 16669 & 19851 \\
\hline
\end{tabular}
\end{sidewaystable}

\clearpage

\subsection{Population Statistics} \label{statistics4}

How can we use our catalogue to learn which, if any, 
cluster properties affect the appearance of CMDs?
One way to accomplish this is by plotting the size of a
given stellar population in a particular circle versus the total stellar
mass contained within it.  From this, lines of best fit can be found
that provide equations relating the size of each stellar population to 
the total stellar mass contained within each circle.  As described
below, this is ideal for probing the effects of the cluster dynamics
on the appearance of CMDs.  

The rate of two-body relaxation for a cluster can be approximated
using the half-mass relaxation time \citep{spitzer87}:
\begin{equation}
\label{eqn:t-rh4}
t_{rh} = 1.7 \times 10^5[r_h(pc)]^{3/2}N^{1/2}[m/M_{\odot}]^{-1/2} years,
\end{equation}
where $r_h$ is the half-mass radius, $N$ is the total number of stars
within $r_h$ and $m$ is the average stellar mass.  
The half-mass radii of MW GCs are remarkably similar independent of mass,
and simulations have shown that $r_h$ changes by a factor of at most a few
over the course of a cluster's lifetime \citep{murray09, henon73}.  
The GCs that comprise our sample show a range of masses spanning roughly 3
orders of magnitude, and have comparably old ages \citep{deangeli05}.
Therefore, Equation~\ref{eqn:t-rh4} suggests that the degree of
dynamical evolution (due to two-body relaxation) experienced by a
cluster is primarily determined by the total cluster mass for the 
GCs in our sample.  In particular, more massive clusters are less
dynamically evolved, and vice versa.  Consequently, if the size of a
given stellar population is affected by two-body relaxation, the
effects should be the most pronounced in the least massive clusters in
our sample.  
Additionally, the rate of (direct) stellar collisions increases 
with increasing cluster mass \citep[e.g.][]{davies04}.  This suggests
that, if a given stellar population is affected by collisions, the
effects should be the most pronounced in the most massive clusters in
our sample.  Therefore, by comparing the size of each stellar population
to the total stellar mass contained within a given circle, the 
effects of the cluster dynamics can be quantified.  
This technique also ensures a normalized and consistent comparison
since it accounts for cluster-to-cluster
differences in the fractional area sampled by the ACS field of view.
That is, we are consistently comparing the same
structural area for each cluster.  The validity and implications of
all of these assumptions will be discussed further in
Section~\ref{discussion4}. 

Plots showing the number of stars belonging to each stellar population
as a function of the total stellar mass contained within each circle
are shown in Figure~\ref{fig:Mshell_vs_Npop_2x2_cum}.  Uncertainties
for the number of stars belonging to each stellar population were
calculated using Poisson statistics.  We also plot 
in Figure~\ref{fig:Mshell_vs_Npop_2x2_noncum} the number of stars
belonging to each stellar population as a 
function of the total stellar mass contained in each annulus outside
the core.  That is, we
considered the populations for each annulus individually, as opposed
to considering every star with a 
distance from the cluster centre smaller than the radius of the
outer-most circles.  Recall that we have neglected 
clusters for which
our theoretical King models provide a poor description of the true
density distributions.  This was the case for clusters
in our sample having a high concentration parameter, most of
which are labelled as post-core collapse in \citet{harris96}.

\begin{figure} [!h]
  \begin{center}
 \includegraphics[scale=0.5]{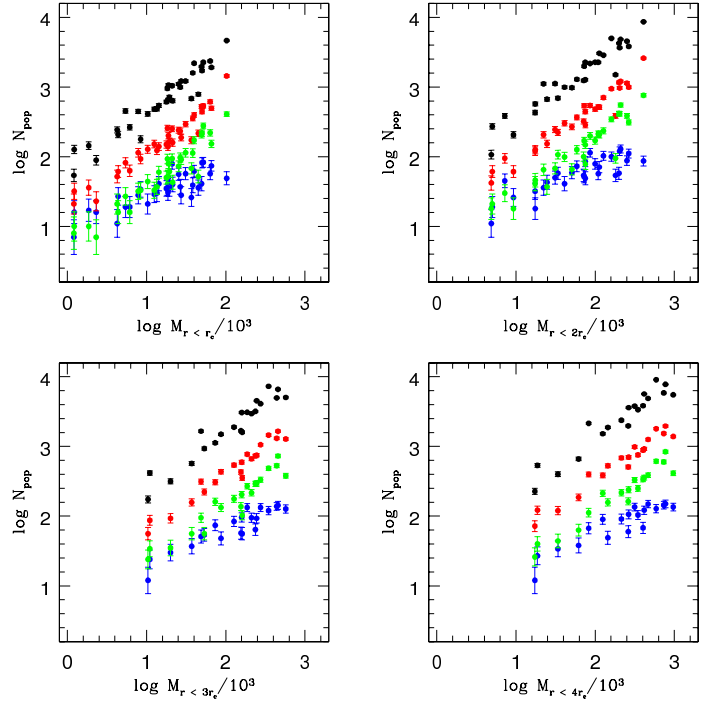}
\caption[Plot showing for each circle the logarithm of the number of
stars belonging to each stellar population as a function of the
logarithm of the total stellar mass]{The logarithm of the number of
  stars belonging to each stellar population is shown for each circle
  as a function of the logarithm of the total 
stellar mass.  From left to right and top to bottom, each frame
corresponds to number counts contained within a circle having a
radius of $r_c$, $2r_c$, $3r_c$ and $4r_c$.  Blue corresponds to blue
stragglers, red to red giant branch stars, green to horizontal branch
stars and black to main-sequence turn-off stars.
Estimates for the total
stellar mass contained within each circle were found using single-mass King
models, as described in Section~\ref{king4}.  
\label{fig:Mshell_vs_Npop_2x2_cum}}
\end{center}
\end{figure}

\begin{figure} [!h]
  \begin{center}
 \includegraphics[scale=0.5]{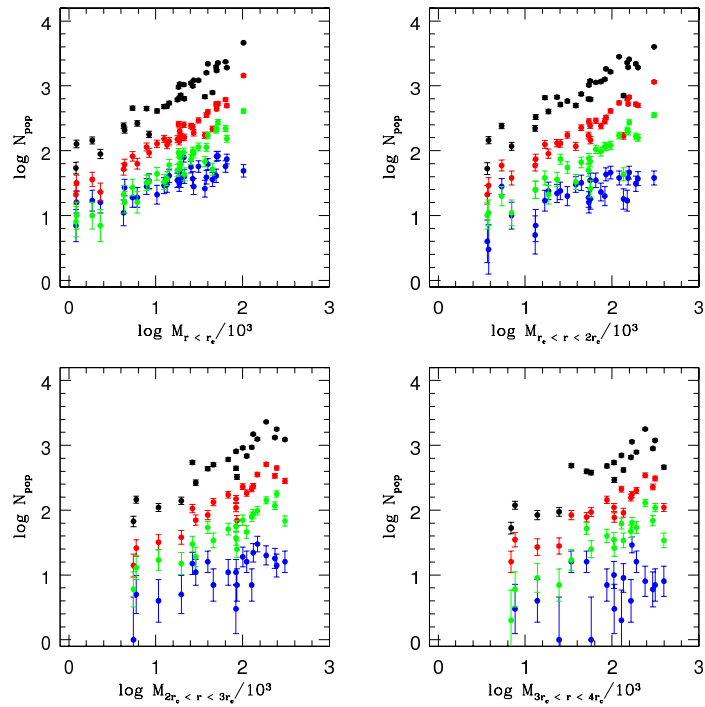}
\caption[Plot showing for each annulus the logarithm of the number of
stars belonging to each stellar population as a function of the
logarithm of the total stellar mass]{The logarithm of the number of
  stars belonging to each stellar population is shown for each annulus
  as a function of the logarithm of the total
stellar mass.  The annulus and colour corresponding to each inset and
stellar population, respectively, are the same as in
Figure~\ref{fig:Mshell_vs_Npop_2x2_cum}.
\label{fig:Mshell_vs_Npop_2x2_noncum}}
\end{center}
\end{figure}

\clearpage

\begin{sidewaystable}
\tiny
\centering
\caption{Lines of Best Fit for log(M$_{circle}$/10$^3$) Versus log(N$_{pop}$)
  \label{table:Mshell_vs_Npop_cum}}
\begin{tabular}{|l|p{4cm}|p{4cm}|p{4cm}|p{4cm}|}
\hline
Circle      &        BS        &        RGB         &         HB         &       MSTO      \\
\hline
$<$ r$_c$   & log(N$_{BS}$) = (0.39 $\pm$ 0.05)log(M$_{< r_c}$/10$^3$) + (1.22 $\pm$ 0.05)  &    log(N$_{RGB}$) = (0.95 $\pm$ 0.11)log(M$_{< r_c}$/10$^3$) + (1.36 $\pm$ 0.11)    & log(N$_{HB}$) = (0.95 $\pm$ 0.06)log(M$_{< r_c}$/10$^3$) + (0.92 $\pm$ 0.06)   & log(N$_{MSTO}$) = (0.90 $\pm$ 0.07)log(M$_{< r_c}$/10$^3$) + (2.03 $\pm$ 0.07)    \\
$<$ 2r$_c$  & log(N$_{BS}$) = (0.36 $\pm$ 0.05)log(M$_{< 2r_c}$/10$^3$) + (1.26 $\pm$ 0.08)  &    log(N$_{RGB}$) = (0.87 $\pm$ 0.08)log(M$_{< 2r_c}$/10$^3$) + (1.27 $\pm$ 0.12)    & log(N$_{HB}$) = (0.85 $\pm$ 0.06)log(M$_{< 2r_c}$/10$^3$) + (0.84 $\pm$ 0.09)   & log(N$_{MSTO}$) = (0.82 $\pm$ 0.06)log(M$_{< 2r_c}$/10$^3$) + (1.98 $\pm$ 0.09)    \\
$<$ 3r$_c$  & log(N$_{BS}$) = (0.47 $\pm$ 0.04)log(M$_{< 3r_c}$/10$^3$) + (1.02 $\pm$ 0.08)  &    log(N$_{RGB}$) = (0.80 $\pm$ 0.06)log(M$_{< 3r_c}$/10$^3$) + (1.26 $\pm$ 0.12)    & log(N$_{HB}$) = (0.79 $\pm$ 0.08)log(M$_{< 3r_c}$/10$^3$) + (0.83 $\pm$ 0.14)   & log(N$_{MSTO}$) = (0.82 $\pm$ 0.11)log(M$_{< 3r_c}$/10$^3$) + (1.86 $\pm$ 0.20)    \\
$<$ 4r$_c$  & log(N$_{BS}$) = (0.45 $\pm$ 0.05)log(M$_{< 4r_c}$/10$^3$) + (1.01 $\pm$ 0.12)  &    log(N$_{RGB}$) = (0.75 $\pm$ 0.07)log(M$_{< 4r_c}$/10$^3$) + (1.28 $\pm$ 0.15)    & log(N$_{HB}$) = (0.75 $\pm$ 0.09)log(M$_{< 4r_c}$/10$^3$) + (0.83 $\pm$ 0.19)   & log(N$_{MSTO}$) = (0.78 $\pm$ 0.12)log(M$_{< 4r_c}$/10$^3$) + (1.86 $\pm$ 0.25)    \\
\hline
\end{tabular}
\end{sidewaystable}

\begin{sidewaystable}
\tiny
\centering
\caption{Lines of Best Fit for log(M$_{annulus}$/10$^3$) Versus log(N$_{pop}$)
  \label{table:Mshell_vs_Npop_noncum}}
\begin{tabular}{|l|p{4.1cm}|p{4.1cm}|p{4.1cm}|p{4.1cm}|}
\hline
Annulus                 &         BS        &        RGB        &        HB       &       MSTO      \\
\hline
 r$_c$ $<$ r $<$ 2r$_c$ & log(N$_{BS}$) = (0.27 $\pm$ 0.08)log(M$_{r_c < r < 2r_c}$/10$^3$) + (1.04 $\pm$ 0.13)  &    log(N$_{RGB}$) = (0.80 $\pm$ 0.06)log(M$_{r_c < r < 2r_c}$/10$^3$) + (1.22 $\pm$ 0.08)    & log(N$_{HB}$) = (0.77 $\pm$ 0.06)log(M$_{r_c < r < 2r_c}$/10$^3$) + (0.79 $\pm$ 0.09)   & log(N$_{MSTO}$) = (0.76 $\pm$ 0.05)log(M$_{r_c < r < 2r_c}$/10$^3$) + (1.91 $\pm$ 0.08)    \\
2r$_c$ $<$ r $<$ 3r$_c$ & log(N$_{BS}$) = (0.39 $\pm$ 0.09)log(M$_{2r_c < r < 3r_c}$/10$^3$) + (0.52 $\pm$ 0.16)  &    log(N$_{RGB}$) = (0.79 $\pm$ 0.11)log(M$_{2r_c < r < 3r_c}$/10$^3$) + (0.97 $\pm$ 0.17)    & log(N$_{HB}$) = (0.68 $\pm$ 0.10)log(M$_{2r_c < r < 3r_c}$/10$^3$) + (0.67 $\pm$ 0.16) & log(N$_{MSTO}$) = (0.78 $\pm$ 0.12)log(M$_{2r_c < r < 3r_c}$/10$^3$) + (1.64 $\pm$ 0.20)    \\
3r$_c$ $<$ r $<$ 4r$_c$ & log(N$_{BS}$) = (0.15 $\pm$ 0.21)log(M$_{3r_c < r < 4r_c}$/10$^3$) + (0.77 $\pm$ 0.35)  &    log(N$_{RGB}$) = (0.63 $\pm$ 0.11)log(M$_{3r_c < r < 4r_c}$/10$^3$) + (1.04 $\pm$ 0.18)    & log(N$_{HB}$) = (0.69 $\pm$ 0.18)log(M$_{3r_c < r < 4r_c}$/10$^3$) + (0.45 $\pm$ 0.31) & log(N$_{MSTO}$) = (0.69 $\pm$ 0.16)log(M$_{3r_c < r < 4r_c}$/10$^3$) + (1.60 $\pm$ 0.27)    \\
\hline
\end{tabular}
\end{sidewaystable}

\clearpage

We performed a weighted least-squares fit for every relation in
Figure~\ref{fig:Mshell_vs_Npop_2x2_cum} and
Figure~\ref{fig:Mshell_vs_Npop_2x2_noncum}.  Slopes and y-intercepts
for these lines are shown in Table~\ref{table:Mshell_vs_Npop_cum} and
Table~\ref{table:Mshell_vs_Npop_noncum}, respectively.  Uncertainties for the
slopes and y-intercepts were 
found using a bootstrap methodology in which we generated 1,000 fake
data sets by randomly sampling (with replacement) number counts from the
observations.  We obtained lines of best fit for each fake data set, fit a
Gaussian to the subsequent distribution and extracted its standard
deviation.  

As shown in Table~\ref{table:Mshell_vs_Npop_cum}, the power-law index
is sub-linear for BSs within the core at much better than the
$3-\sigma$ confidence level, and it is consistent with the slope 
obtained in our earlier analysis presented in \citet{knigge09}.  The
slopes for the BSs are also sub-linear 
at better than the $3-\sigma$ confidence level for all circles outside the
core.  This is also the case for all annuli outside the core, as shown
in Table~\ref{table:Mshell_vs_Npop_noncum}.  Note, however, that the
uncertainties for the BS slopes are very 
large for all annuli outside the core, whereas this is not always the case
for corresponding circles outside the core.  This is the result of the
fact that the number of BSs drops off rapidly outside
the core in several clusters so that the corresponding Poisson
uncertainties, which are given by the square-root of the number of
BSs, are significant.  The rapid decline of BS numbers with 
increasing distance from the cluster centre in these clusters has also
contributed to an 
increased degree of scatter in the relations for annuli outside the core
relative to the corresponding relations for circles outside the core.

The slopes are consistent with being linear for all other stellar
populations in the core within their respective $3-\sigma$ confidence
intervals.  This agrees with the results of our earlier analysis
presented in \citet{leigh09} when we performed the comparison using
the total core masses.  The slopes are also consistent with being 
linear for all circles outside 
the core for both HB and MSTO stars.  The power-law indices are sub-linear 
at the $3-\sigma$ confidence level only for RGB stars, and this is
only the case for circles outside the core.  The power-law index is
nearly unity for the core RGB population, 
yet the associated uncertainty is very large.  Upon closer inspection,
the distribution of power-law indices obtained from our bootstrap
analysis for RGB stars in the core is strongly bi-modal, with
comparably-sized peaks centred at $\sim 0.82$ and $\sim 1.0$.  This
bi-modality is most likely an artifact of our bootstrap analysis
caused by a chance 
alignment of data points in the log M$_{core}$-log N$_{RGB}$ plane.  
Upon performing the comparison for only those stars found within 
particular annuli, our results suggest that the slopes are consistent
with being sub-linear for all stellar 
populations at the $3-\sigma$ confidence level in only the annulus
immediately outside the core (i.e. $r_c < r < 2r_c$).  The slopes are
consistent with being linear for RGB, HB and MSTO stars in all other
annuli.

We also tried 
performing the same comparisons using the total number of stars in each
circle and annulus as a proxy for the total stellar mass.  In this
case, the slopes are 
sub-linear for BSs within all circles and annuli at the
$3-\sigma$ confidence level.  Once again, the uncertainties are very
large for all annuli outside the core, whereas this is not the case
for corresponding circles outside the core.  The slopes are consistent
with being linear within the $1-\sigma$ confidence
interval for all other stellar populations in all circles and annuli.
Our results are therefore inconsistent with those presented in
\citet{leigh09} for the core RGB populations, in which we found that
RGB numbers scale sub-linearly with the number of stars in the core at
the $3-\sigma$ confidence level.  We will discuss the implications of
these new results in Section~\ref{discussion4}.

\subsection{Blue Stragglers and Single-Single Collisions} \label{collisions4}

As a check of our previous results reported in \citet{knigge09}, we
also looked for a correlation between the observed number of BSs in
the cluster core and the number predicted from single-single (1+1)
collisions.  The results of this comparison are shown in
Figure~\ref{fig:Ncoll_vs_Nbs}.  We define the predicted
number of BSs formed from 1+1 collisions as $N_{1+1} =
\tau_{BS}/\tau_{1+1}$, where $\tau_{BS}$ is the average BS lifetime
and $\tau_{1+1}$ is the average time between 1+1 collisions in the
cluster core.  We adopt the same
definition for $\tau_{1+1}$ as used in \citet{knigge09}, and assume
$\tau_{BS} = 1.5$ Gyrs as well as an average stellar mass and radius of
$0.5$ M$_{\odot}$ and $0.5$ R$_{\odot}$, respectively.  We also adopt
a constant mass-to-light ratio of $M/L = 2$ for all clusters.  Central
luminosity densities and velocity dispersions were taken from 
\citet{harris96} and \citet{webbink85}, respectively.

Upon performing a weighted line of best fit for every cluster in our
sample that overlaps with the catalogue of \citet{webbink85}, we find
a power-law index of $0.15 \pm 0.03$ (the uncertainty
was found using the bootstrap methodology described in
Section~\ref{statistics4}).  For the subset of dense 
clusters having a central luminosity density satisfying log
$\rho_0 > 4$, we find a power-law index of $0.36 \pm 0.14$.  As before,
we find no significant correlation with collision rate, even for the 
subset of dense clusters.  Although we do find a weak dependence of BS
numbers on collision rate for the entire sample, this is not
unexpected since the collision rate and the core mass are themselves
correlated, and our results suggest that there exists a strong
correlation between BS numbers and the core masses.

\begin{figure} [!h]
  \begin{center}
 \includegraphics[scale=0.5]{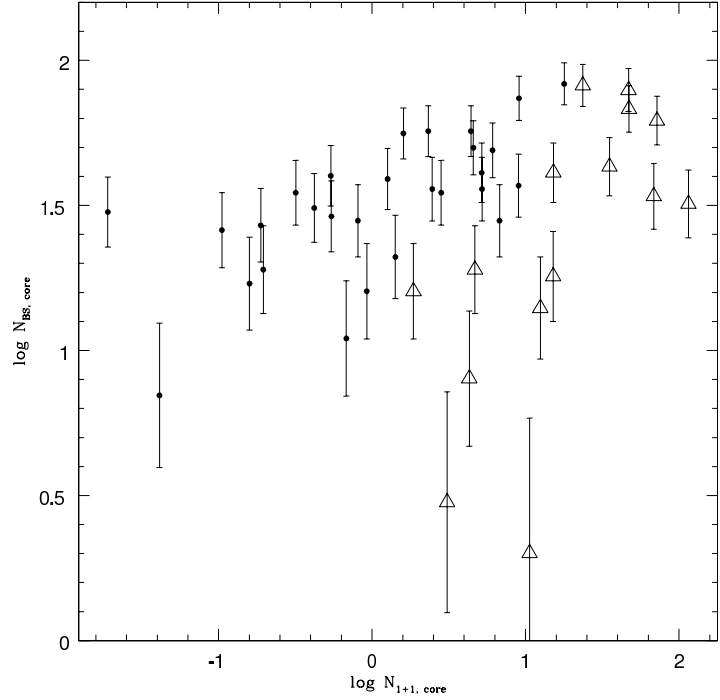}
\caption[Plot showing the logarithm of the number of BSs predicted to
have formed in the core from single-single collisions versus the logarithm of
the observed number of BSs in the core.]{The logarithm of the number
  of BSs predicted to have formed in the core from single-single collisions
  N$_{1+1}$ 
  versus the logarithm of the observed number of BSs in the core N$_{BS}$.
  Filled circles correspond to clusters
having central luminosity densities satisfying log $\rho_0 < 4$, 
whereas open triangles correspond to dense clusters for which log
$\rho_0 > 4$.  The adopted definition for N$_{1+1}$ has been provided
in the text.
\label{fig:Ncoll_vs_Nbs}}
\end{center}
\end{figure}

\section{Summary \& Discussion} \label{discussion4}

In this chapter, we have presented a catalogue for BS, RGB,
HB and MSTO stars in a sample of 35 GCs.  Our catalogue provides
number counts for each stellar population within several different
circles centred on the cluster centre.  The radii of the circles were
taken to be integer multiples of the
core radius, and we have focussed on the inner four core radii
since the field of view extends beyond this for only a handful of the
clusters in our sample.  Particular consideration was
given to our selection criteria for the 
different stellar populations in order to ensure that they were
applied consistently from cluster-to-cluster.  In particular, we have
improved upon our previous selection criteria \citep{leigh07} by
fitting theoretical isochrones to the cluster CMDs.  This provides an
unambiguous definition for the location of the MSTO, which acts as the 
primary point of reference for the application of our selection
criteria.  As a result, our new catalogue is highly homogeneous.

We have used our catalogue to quantify the dependence
of the size of each stellar population on the total stellar mass
enclosed within the same radius.  As described in
Section~\ref{statistics4}, this provides a means of quantifying the
effects, if any, had by the cluster dynamics in shaping the appearance 
of CMDs above the MSTO.  Below, we summarize the implications of our
results for each of the different stellar populations.
 
\subsection{Blue Stragglers} \label{BSs4}

We have confirmed our previous result that the numbers of BSs in the
cores of GCs scale sub-linearly with the core masses
\citep{knigge09}.  That is, we find proportionately larger BS
populations in low-mass GCs.  
There exist several possibilities that could explain the origin of
this sub-linear dependence.  First, we previously suggested 
that this could be an artifact of an anti-correlation between the binary
fraction and the cluster (or core) mass \citep{knigge09}.  This
assertion stems from the fact 
that, if BSs have a binary origin, we expect their numbers to scale
with the core mass as $N_{BS} \sim f_bM_{core}$, where $f_b$ is the binary
fraction in the core.  As before, we find that $N_{BS} \sim
M_{core}^{0.4-0.5}$.  Our result could therefore be explained if $f_b \sim
M_{core}^{-(0.5-0.6)}$.  Second, we suggested in \citet{leigh09} that
the fact that the least massive GCs in our sample should be
more dynamically evolved than their more massive
counterparts could be contributing to the observed sub-linearity for
BSs.  In particular, the low-mass clusters in our sample should have
experienced a significant depletion of their very low-mass stars as a
result of stellar evaporation induced by two-body relaxation 
\citep[e.g.][]{spitzer87, heggie03, demarchi10}.  In turn, this could contribute
to a higher fraction of merger products having masses that exceed that
of the MSTO in low-mass GCs.  As a result, more merger products would 
appear brighter and bluer than the MSTO in these clusters' CMDs,
leading to more merger products being identified as BSs.  Finally,
mass segregation could also be contributing to the observed sub-linear
dependence for BSs.  Again, this is the result of the fact that the
rate of two-body relaxation, and therefore dynamical friction, is in
general the 
fastest in low-mass clusters.  BSs are among the most massive stars in
GCs, so they should rapidly migrate into the core via 
dynamical friction in clusters for which the half-mass relaxation time
is shorter than the average BS lifetime.  It follows that proportionately
more BSs could have drifted into the core via dynamical friction in
low-mass GCs.  This would also contribute to the observed sub-linear
dependence of BS numbers on the core masses. 

This last hypothesis can be tested by comparing our scaling
relations for progressively larger circles outside the core.  If mass
segregation is indeed the cause of the observed sub-linear dependence
of BS numbers on the core masses, then we might expect the power-law
index to systematically increase as we consider progressively larger
circles.  That is, we could be including more BSs that have not yet migrated
into the core via dynamical friction, particularly in the most massive
clusters in our sample.  However, our results suggest 
that the power-law index remains roughly constant for all circles.  
This is the case for both the comparison to the total stellar masses as
well as to the total number of stars contained in each circle.  
The fact that these relations are comparably sub-linear within all
circles can be interpreted as evidence that mass segregation is not
the dominant effect contributing to the 
observed sub-linear dependence of BS numbers on the total stellar
masses (or the total number of stars).  We note that for many of the
clusters in our sample, the spatial coverage is comparable to or
exceeds the half-mass radius.  This is a sufficiently large fraction
of the total cluster area for our comparison to be sensitive to the
effects of mass segregation.  

On the other hand, several GCs are known
to show evidence for a bi-modal BS radial distribution
\citep[e.g.][]{mapelli06, lanzoni07}.  That is, in
these clusters the number of BSs is highest in the central cluster
regions and decreases with increasing distance from the cluster centre
until a second rise in BS numbers occurs in the cluster outskirts.
This secondary outer peak has been shown to occur at a distance 
from the cluster centre that exceeds 20 core radii in several
cases.  Consequently, 
the spatial coverage provided by the ACS data likely does not extend
sufficiently far in most clusters to 
detect any bi-modality in the BS radial distribution.  Nonetheless,
if applied to a 
statistically-significant sample for which the spatial coverage
is complete out to the tidal radius, the technique we have
presented in this chapter could provide a powerful constraint for the
origin of the 
bi-modal BS radial distribution observed in several MW GCs by
addressing the role played by mass segregation.

We also tried to correlate the number of BSs observed in 
the cluster core with the number predicted from single-single 
collisions.  As in \citet{knigge09}, we find that BS numbers depend
strongly on
core mass, but not on collision rate.  This also proved to be the case
for the subset of 
dense clusters satisfying log $\rho_0 > 4$.  Our previous
interpretation 
that our results provide strong evidence for a binary, as opposed to a
collisional, origin for BSs in GCs therefore remains the same.

\subsection{Red Giant Branch Stars} \label{RGBs4}

The technique used in this chapter to compare the sizes of
the different stellar populations was first presented in
\citet{leigh09}.  In that study, we introduced our method 
and applied it to a sample of 56 GCs taken from \citet{piotto02} in
order to study their RGB populations.  Our results 
were consistent with a sub-linear dependence of RGB numbers on the
core masses.  In particular, we found evidence for a surplus of RGB
stars relative to MSTO stars in the cores of low-mass GCs.  We
considered several possible causes for this result, 
but concluded that our analysis should ideally be repeated with
superior photometry in order to properly assess effects such as 
completeness.  Given the high-quality of the ACS data, we are now in
a position to reassess our previous result for RGB stars.

Upon applying our technique to the ACS sample, we find that the
numbers of RGB stars scale linearly with the core masses
to within one standard deviation.  This is also the case for our
comparison to the total number of stars in the core.  This 
suggests that we should reject our previous conclusions for this stellar
population reported in \citet{leigh09}.  Specifically, if we take a
strict $3-\sigma$ limit as our 
criterion for whether or not the slopes are sub-linear at a
statistically significant level, then the core RGB slope reported in
this chapter is consistent with being linear whereas this was not the case for
the core RGB slope reported in \citet{leigh09}.  However, if we take a
more stringent criterion for statistical significance,
then there is no inconsistency between our new and old results for RGB
stars, and both slopes are consistent with being linear.  

We also tried comparing the RGB catalogue presented in this chapter with
the one presented in \citet{leigh09}.  This showed that the
old RGB numbers are slightly 
deficient relative to the new numbers at the high-mass end.  Although this
difference is not sufficiently large to completely account for the
difference in slopes (for the comparison with the core mass) found
between our new and old RGB catalogues, it works in the right
direction and is likely a contributing factor.  If our uncertainties
are also factored in, then our new and old slopes agree to within one
standard deviation (due mainly to the large uncertainty for the new
slope).  The source of the disagreement between our old and new RGB
catalogues is unclear, and we cannot say whether or not 
incompleteness (in the old data set) is the culprit.  The results of
our artificial star tests have at least confirmed that incompleteness
is not an issue for our new catalogue, however it could certainly have
contributed to the lower RGB numbers reported in \citet{leigh09}.
Indeed, the central cluster density tends to be 
higher in more massive clusters, which should negatively affect
completeness.  It is not clear, however, why this would have affected
RGB stars more than MSTO stars in the old data set.  
Given all of these considerations, we feel that our new results show
that this issue needs to be looked at in more detail before any firm
conclusions can be drawn.  

The evidence in favour of RGB numbers scaling linearly with the core masses 
is interesting.  For one thing, it suggests that two-body relaxation 
does not significantly affect RGB population size relative to other
stellar populations of comparable mass in even the dense central
regions of GCs.  This is not surprising, since two-body 
relaxation is a long-range effect for which the stellar radius plays a
negligible role.   Second, it suggests that collisions do not
significantly deplete RGB stars relative to other stellar populations
despite their much larger radii.  This is because the collision rate
increases with increasing cluster mass, so we would expect RGB stars
to appear preferentially depleted in massive clusters if they are
significantly affected by collisions \citep[e.g.][]{beers04, davies04}.  
Third, it suggests that the sub-linear relation found for BSs does not 
contribute to a sub-linear relation for RGB stars despite the fact that
BSs should eventually evolve to occupy our RGB selection box, as
discussed in \citet{leigh09}.  This is likely the result of the
relatively small sizes of the BS populations in our sample when
compared to the numbers of RGB stars since the rate at which evolved
BSs ascend the RGB is thought to be comparable to the RGB lifetimes of
regular MSTO 
stars \citep{sills09}.  Alternatively, this result could, at least in
part, be explained if
a smaller fraction of BSs end up sufficiently bright and blue to
be identified as BSs in the CMDs of massive GCs.  In other words, it
could be that a larger fraction of BSs are 
hidden along the MS in massive clusters, as discussed in
Section~\ref{BSs4}.  In this case, the
contributions to RGB populations from evolved BSs could be comparable
in all clusters, in which case a linear relationship between RGB
numbers and the core masses would be expected.  Finally, evolved BSs
would be expected to have a negligible impact on RGB population size
if the average BS lifetime is considerably longer than the lifetimes
of RGB stars.  This effect is difficult to quantify, however, given
that BS lifetimes are poorly constrained 
in the literature \citep[e.g.][]{sandquist97, sills01}.

\subsection{Horizontal Branch Stars} \label{HBs4}

Our results suggest that HB numbers scale linearly with the core
masses.  This can be interpreted as evidence that two-body relaxation
does not significantly affect the radial distributions of HB 
stars in GCs relative to the other stellar populations above the
MSTO.  One reason to perhaps expect that two-body relaxation should
affect the spatial distributions of HB stars stems from the fact 
that RGB stars are among the most massive stars in clusters, and they
undergo significant mass loss upon evolving into HB stars.
Consequently, the progenitors of HB stars should be heavily mass
segregated.  HB stars themselves, however, have relatively low-masses
so that two-body relaxation and 
strong dynamical encounters should act to re-distribute them to wider
orbits within the cluster potential.  The HB 
lifetime is roughly constant at $10^8$ years \citep{iben91} and it is
comparable to or exceeds the core relaxation times for most of the
low-mass clusters in our sample \citep{harris96}.  Therefore, we might
expect the HB populations in these 
clusters to exhibit more extended radial profiles relative to 
more massive clusters.  This
would contribute to a sub-linear relationship between the numbers of
HB stars and the core masses.  Our uncertainties are sufficiently
large that this possibility cannot be entirely ruled out, however our
results are consistent with a linear relationship between HB numbers
and the core masses.  

\subsection{Additional Considerations} \label{final4}

Recent observations have revealed the presence of
multiple stellar populations in a number of MW GCs
\citep[e.g.][]{pancino03}.  The majority of these cases have
been reported in very massive clusters.  Moreover, their existence is
thought to be related to the chemical properties of GCs, in
particular an observed anti-correlation between their sodium and oxygen 
abundances.  In turn, these chemical signatures have been argued to be
linked to the cluster metallicity, mass and age \citep{carretta10}.  

We identified clusters in our sample currently known to host multiple stellar
populations, but none of these were clear outliers in
our plots.  Consequently, the effects had on our results by multiple stellar
populations remains unclear.  It is certainly possible that multiple
stellar populations have contributed to the uncertainties for
the weighted lines of best fit performed for the relations in 
Figure~\ref{fig:Mshell_vs_Npop_2x2_cum} and
Figure~\ref{fig:Mshell_vs_Npop_2x2_noncum}.  It is difficult to
quantify the possible severity of this effect, however, given the
limited evidence linking multiple stellar populations to cluster
properties.  

Although the uncertainties are sufficiently large that the slopes are
consistent with being linear at the $3-\sigma$ confidence level for
all stellar populations when 
performing the comparisons with the total stellar mass, the reported
slopes are typically less than unity within the $1-\sigma$, and often 
even the $2-\sigma$, confidence interval.  This does not appear 
to result from the fact that we have 
obtained our estimates for the total stellar masses 
by numerically integrating 3-dimensional density distributions and are
comparing to number counts, which are projected quantities.  To
address this, we also tried obtaining the total stellar masses by
numerically integrating 2-dimensional surface brightness profiles so
that we are consistently comparing only projected quantities.  Despite
this, our results remain unchanged and the new slopes agree with the 
old ones to within 
one standard deviation for all stellar populations.  Another
possibility to account for this trend that is perhaps worth 
considering is a systematic dependence of the mass-to-light
ratios of clusters on their total mass.  There are two ways this could
have affected our analysis.  First, stellar remnants have been shown to 
affect the dynamical evolution of clusters, and therefore the sizes of
their cores \citep[e.g.][]{lee91, trenti10}.  It follows that, if the
number of stellar remnants depends on the cluster mass, then this
could contribute to an additional underlying dependence of 
the core radius on the cluster mass.  This could perhaps
arise as a result of the fact that the ratio of the rate of stellar
evolution to the rate of dynamical evolution is larger in more massive
clusters, since in general the rate of two-body relaxation decreases with
increasing cluster mass whereas the rate of stellar evolution is
independent of the cluster mass.  Coupled with their deeper gravitational
potential wells, this could contribute to more massive clusters
retaining more stellar remnants.  Second, variations in the
mass-to-light ratios of clusters can also occur as a result of 
changes in the average stellar mass (not including stellar remnants)
\citep{kruijssen09}.  That is, we can approximate the total stellar
mass contained in the core as:
\begin{equation}
\label{eqn:M_core4}
M_{core} \sim \frac{4}{3}{\pi}\frac{M}{L}\rho_0r_c^3 \sim mN_{core},
\end{equation}
where $M/L$ is the mass-to-light ratio, $\rho_0$ is the central 
luminosity density, $m$ is the average stellar mass and $N_{core}$
is the total number of stars in the core.  Based on our results,
$M_{core} \propto N_{core}^{0.9}$, where we have used the total number
of stars in the core with magnitudes brighter than 0.5 mag below the
MSTO as a proxy for $N_{core}$.  This could suggest that $m \propto
N_{core}^{-0.1} \propto M_{core}^{-0.1}$.  In other words, the
average stellar mass in the core decreases weakly with increasing core
mass.  This could in part be due to the fact that more massive clusters 
should be less dynamically evolved than their less massive
counterparts, and should therefore be less depleted of their low-mass
stars due to stellar evaporation induced by two-body relaxation
\citep[e.g.][]{ambartsumian38, spitzer58, henon60, demarchi10}.
Similarly, mass 
segregation should also tend to operate more rapidly in low-mass
clusters, which acts to migrate preferentially massive stars into the
core \citep{spitzer69, spitzer71, farouki82, shara95, king95,
  meylan97}.  
Alternatively, differences in the stellar mass function in the 
core could result from variations in the degree of primordial mass
segregation, or even variations in the initial stellar mass function.

We have assumed throughout our analysis that the core mass is a
suitable proxy for the total cluster mass.  We have checked that these
two quantities are 
indeed correlated, however this does not tell the whole story since we
are also using the total cluster mass as a proxy for the degree of
dynamical evolution.  The central concentration 
parameter, defined as the logarithm of the ratio of the tidal to core
radii, describes the degree to which a cluster is centrally
concentrated.  Previous studies have shown that there exists a weak
correlation between the concentration parameter and the total cluster
mass \citep[e.g.][]{djorgovski94, mclaughlin00}.  In order to better use our
technique to reliably probe the effects of the cluster
dynamics on the sizes and radial distributions of the different
stellar populations, the concentration parameter should ideally be
accounted for when applying our normalization technique in future
studies.  It is not yet clear how the concentration parameter can be
properly absorbed into the normalization, however its effect on our
analysis should be small given the weak dependence on cluster mass.  

The assumption that the degree of dynamical evolution experienced by a
given cluster depends only on its mass is also incorrect.  Two-body 
relaxation has been shown to dominate cluster evolution for a
significant fraction of the lives of old MW GCs
\citep[e.g.][]{gieles11}, however other effects can also play a
significant role.  For example, stellar evolution is known to affect
the dynamical evolution of star 
clusters, although its primary role is played during their early
evolutionary phases 
\citep[e.g.][]{applegate86, chernoff90, fukushige95}.  Tidal effects
from the Galaxy have also been shown to play an important role in
deciding the dynamical fates of clusters by increasing the rate of
mass loss across the 
tidal boundary \citep[e.g.][]{heggie03}.  Consequently, clusters with
small perigalacticon distances should appear more dynamically evolved
than their total mass alone would suggest.  This effect can 
be significant, and has likely contributed to increasing the uncertainties
found for the comparisons to the total stellar mass.  Therefore, tidal
effects from the Galaxy should also ideally 
be absorbed into our normalization technique in future studies.  This
can be done by using the perigalacticon distances of clusters as a
rough proxy for the degree to which tides from the Galaxy should have
affected their internal dynamical evolution \citep{gieles11}.  

Interestingly, tides could also
help to explain why the uncertainties for the comparisons to the total
number of stars in each circle are considerably smaller than for the
comparisons to the total stellar mass.  We have used 
the number of stars with magnitudes brighter than 0.5 mag below the
MSTO as a proxy for the total number of stars.  Consequently, we are
comparing stars within a very narrow mass range, so that all 
populations of interest should have been comparably affected by
two-body relaxation (except, perhaps, for HB stars) independent of
tidal effects from 
the Galaxy.  In other words, tides should affect all stars above the
MSTO more or less equally, and this is consistent with our results.
It is also worth mentioning here that our King models consider only a
single stellar 
mass.  This assumption is not strictly true and could also be
contributing to increasing the uncertainties found for the comparisons
to the total stellar masses.

An additional concern is that we do not know if the clusters in our
sample are currently in a phase of core contraction or expansion.  
This has a direct bearing on the recent history of the stellar density
in the core, and therefore the degree to which stars in the core
should have been affected by close dynamical interactions.  These
effects are independent of two-body relaxation and occur on a
time-scale that is typically much shorter than the half-mass relaxation
time \citep{heggie03}.  The effects could be significant
in clusters that were recently in a 
phase of core-collapse but have since rebounded back out of this 
highly concentrated state.  This could occur, for example, as a result
of binary 
formation induced by 3-body interactions combined with their subsequent
hardening via additional encounters \citep{hut83, heggie03}.  In general,
binaries play an important role in the dynamical evolution of
clusters, and could have affected our results in a number of ways.
This is a difficult issue to address even qualitatively 
given how little is currently known about the binary populations in
globular clusters.  Theoretical models suggest, however, that the
time-scale for core contraction is often longer than a Hubble time, and
that this evolutionary phase will only come to an end once the central
density becomes sufficiently high for hardening encounters involving
binaries to halt the process \citep{fregeau09}.  It
follows that the cores of most MW GCs are expected to currently be in
a phase of core contraction.  This process is ultimately driven by
two-body relaxation, so that our assumption that the total
cluster mass provides a suitable proxy for the degree of dynamical
evolution is still valid.

In summary, our results suggest that effects related to the cluster
dynamics do not significantly affect 
the relative sizes of the different stellar populations above the
MSTO.  This is the case for 
at least RGB, HB and MSTO stars.  BSs, on the other hand, show
evidence for a sub-linear dependence of population size on the total
stellar mass contained within the same radius.  Whether or not the
cluster dynamics is responsible for this sub-linearity is still not
clear.  Notwithstanding, our results have provided 
evidence that mass segregation is not the dominant cause for this
result, although it will be necessary to redo the comparison performed
in this study with a larger spatial coverage in order
to fully address this question.  Further insight into the origin of
the sub-linearity found for BSs will be provided by reliable
binary fractions for the clusters in our sample, which are forthcoming
(Sarajedini 2010, private communication).

\section*{Acknowledgments}

We would like to thank Ata Sarajedini, Aaron Dotter and Roger Cohen
for providing the data on which this study was based and for their
extensive support in its analysis.  We would also like to thank Evert
Glebbeek for useful discussions.  This research has been
supported by NSERC and OGS.




\pagestyle{fancy}
\headheight 20pt
\lhead{Ph.D. Thesis --- N. Leigh }
\rhead{McMaster - Physics \& Astronomy}
\chead{}
\lfoot{}
\cfoot{\thepage}
\rfoot{}
\renewcommand{\headrulewidth}{0.1pt}
\renewcommand{\footrulewidth}{0.1pt}




\chapter{An Analytic Model for
  Blue Straggler Formation in Globular Clusters} \label{chapter5}
%
\thispagestyle{fancy}

\section{Introduction} \label{intro5}

Commonly found in both open and globular clusters (GCs), blue
stragglers (BSs) appear as an
extension of the main-sequence (MS) in cluster colour-magnitude
diagrams (CMDs), occupying the region that is just brighter and bluer
than the main-sequence turn-off (MSTO) 
\citep{sandage53}.  BSs are
thought to be produced via the addition of hydrogen to 
low-mass MS stars \citep[e.g.][]{sills01, lombardi02}.  This can
occur via multiple channels, most of which involve the mergers of
low-mass MS stars since a significant amount of mass is typically
required to reproduce the observed locations of BSs in CMDs
\citep[e.g.][]{sills99}.  Stars in close binaries can merge if enough
orbital angular momentum is lost, which can be mediated by dynamical
interactions with other stars, magnetized stellar winds, tidal
dissipation or even an outer triple companion
\citep[e.g.][]{leonard92, li06, perets09, dervisoglu10}.
Alternatively, MS stars can collide directly, although this is
also thought to usually be mediated by multiple star systems
\citep[e.g.][]{leonard89, leonard95, fregeau04, leigh11b}.  First proposed by
\citet{mccrea64}, BSs have also been hypothesized to form by
mass-transfer from an evolving primary onto a normal MS companion
via Roche lobe overflow.

Despite numerous formation
mechanisms having been proposed, a satisfactory explanation to account
for the presence of BSs in star clusters eludes us still.  Whatever the
dominant BS 
formation mechanism(s) operating in dense clusters, it is now
thought to somehow involve multiple star systems.  This was shown
to be the case in even the dense cores of GCs \citep{leigh07, leigh08,
  knigge09} where
collisions between single stars are thought to occur frequently
\citep{leonard89}.  In \citet{knigge09}, we showed that the numbers of
BSs in the cores
of a large sample of GCs correlate with the core masses.  We
argued that our results are consistent with what is expected if BSs
are descended from binary stars since this would imply a dependence of
the form $N_{BS} \sim f_bM_{core}$, where $N_{BS}$ is the number of
BSs in the core, $f_b$ is the binary fraction in the core and
$M_{core}$ is the total stellar mass contained within the core.
\citet{mathieu09} also showed 
that at least $76\%$ of the BSs in the old open cluster NGC 188 have
binary companions.  Although the nature of these companions remains
unknown, it is clear that binaries played a role in the
formation of these BSs.  

Blue stragglers are typically concentrated in the dense cores of
globular clusters where the high 
stellar densities should result in a higher rate of stellar encounters 
\citep[e.g.][]{leonard89}.  Whether or not this fact is
directly related to BS formation remains unclear, since mass segregation 
also acts to migrate BSs (or their progenitors) into the core 
\citep[e.g.][]{saviane98, guhathakurta98}.  Additionally, numerous BSs
have been observed in more sparsely populated open clusters
\citep[e.g.][]{andrievsky00} and the fields of GCs where
collisions are much less likely to occur and mass-transfer within
binary systems is thought to be a more likely formation scenario
\citep[e.g.][]{mapelli04}.  

Several studies have provided evidence
that BSs show a bimodal spatial distribution in some GCs
\citep{ferraro97, ferraro99, lanzoni07}.  In these clusters, the BS
numbers are the highest in the central cluster 
regions and decrease with increasing distance from the cluster centre
until a second rise occurs in the cluster outskirts.  
This drop in BS numbers at intermediate cluster radii is often
referred to as the ``zone of avoidance''.  Some authors have suggested
that it is the result of two separate formation mechanisms
occurring in the inner and outer regions of the cluster, with
mass-transfer in primordial binaries dominating in the latter and
stellar collisions dominating in the former
\citep{ferraro04, mapelli06}.  Conversely, mass segregation could also
give rise to a ``zone of avoidance'' for BSs if the time-scale for dynamical
friction exceeds the average BS lifetime in only the outskirts of GCs that
exhibit this radial trend \citep[e.g.][]{leigh11a}. 

Dynamical interactions occur frequently enough in dense clusters that
they are expected to be at least partly responsible for the observed
properties of BSs \citep[e.g.][]{stryker93, leigh11b}.  It follows
that the current properties of BS populations
should reflect the dynamical histories of their host clusters.  As a
result, BSs could provide an indirect means of probing
the physical processes that drive star cluster evolution
\citep[e.g.][]{heggie03, hurley05, leigh11b}.

In this chapter, our
goal is to constrain the dominant BS formation mechanism(s) operating
in the dense cores of GCs by analyzing the principal processes
thought to influence their production.  To this end, we
use an analytic treatment to obtain predictions for the number of BSs
expected to be found within one core radius of the cluster centre at
the current cluster age.  Predicted numbers for the core are calculated
for a range of free parameters, and then compared to the
observed numbers in order to find the
best-fitting model parameters.  In this way, we are able to quantify
the degree to which each of the considered formation mechanisms
should contribute to the total predicted numbers in order to
best reproduce the observations.  

In Section~\ref{data5}, we describe the BS catalogue used for comparison
to our model predictions.  In 
Section~\ref{method5}, we present our analytic model for BS 
formation as well as the statistical technique we have developed to
compare its predictions to the observations.  
These predictions are then compared to the observations in
Section~\ref{results5} for a range of model parameters.  In
Section~\ref{discussion5}, we discuss the implications of our results
for BS formation, as well as the role played by the cluster dynamics
in shaping the current properties of BS populations.

\section{The Data} \label{data5}

The data used in this study was taken from \citet{leigh11a}.  In that 
chapter, we presented a catalogue for blue straggler, red giant branch
(RGB), horizontal branch (HB) and main-sequence turn-off stars
obtained from the colour-magnitude diagrams of 35 Milky Way GCs
taken 
from the ACS Survey for Globular Clusters \citep{sarajedini07}.  The
ACS Survey provides unprecedented deep photometry in the F606W ($\sim$
V) and F814W ($\sim$ I) filters
that extends reliably from the HB all the way down to about 7
magnitudes below the MSTO.  The clusters in our sample span a range of
total masses (by nearly 3 orders of magnitude) and central
concentrations \citep{harris96}.  We have confirmed that 
the photometry is nearly complete in the BS region of the CMD for
every cluster in our sample.  This was 
done using the results of artificial star tests taken from
\citet{anderson08}.

Each cluster was centred in the ACS field, which 
extends out to several core radii from the cluster
centre in most of the clusters in our sample.  Only the core
populations provided in \citet{leigh11a} are used in this chapter.
We have taken estimates for the core radii and central luminosity
densities for the clusters in our sample from
\citet{harris96}, whereas central velocity 
dispersions were taken from \citet{webbink85}.  Estimates for the
total stellar mass contained within the core were obtained from
single-mass King models, as described in \citet{leigh11a}.  All of the
clusters in our sample were chosen to be non-post-core collapse, and
have surface brightness profiles that provide good fits to our King
models. 

\section{Method} \label{method5}

In this section, we present our model and outline our
assumptions.  We also present the statistical technique used to
compare the 
observed number counts to our model predictions in order to identify
the best-fitting model parameters. 

\subsection{Model} \label{model5}

Consider a GC core that is home to N$_{BS,0}$ BSs at some time t $=$
t$_0$.  At a specified time in the future, the number of BSs in the
core can be approximated by: 
\begin{equation}
\label{eqn:number-bss}
N_{BS} = N_{BS,0} + N_{coll} + N_{bin} + N_{in} - N_{out} - N_{ev},
\end{equation}
where N$_{coll}$ is the number of BSs formed from collisions during
single-single (1+1), single-binary (1+2) and binary-binary (2+2)
encounters, N$_{bin}$ is the number formed from 
binary evolution (either partial mass-transfer between the binary
components or their complete coalescence), N$_{in}$ is the number 
of BSs that migrate into the core due to dynamical friction, N$_{out}$
is the number that migrate out of the core via kicks experienced during
dynamical encounters, and N$_{ev}$ is the number of BSs that have
evolved away from being brighter and bluer than the MSTO in the
cluster CMD due to stellar evolution.  

We adopt an average stellar mass of $m = 0.65 M_{\odot}$ and an
average BS mass of $m_{BS} = 2m = 1.3 M_{\odot}$.  The mass of a BS
can provide a rough guide to its lifetime, although a range of
lifetimes are still possible for any given mass.  
For instance, \citet{sandquist97} showed that a
1.3 M$_{\odot}$ blue straggler will have a lifetime of around $0.78$
Gyrs in unmixed models, or 1.57 Gyrs in 
completely mixed models.  Combined with the results of \citet{sills01},
\citet{lombardi02} and \citet{glebbeek08}, we expect a lifetime in the
range 1-5 Gyrs for a 1.3 M$_{\odot}$ BS.  As a first approximation, we
choose a likely value of $\tau_{BS} = 1.5$ Gyrs for the average BS
lifetime \citep[e.g.][]{sills01}.  The effects had on our
results by changing our assumption for the average BS lifetime will be
explored in Section~\ref{results5} and discussed in
Section~\ref{discussion5}. 

We consider only the last $\tau_{BS}$ years.  This is because we are
comparing our model predictions to current
observations of BS populations, so that we are only concerned with
those BSs formed within the last few Gyrs.  Any BSs formed before this
would have evolved away from being brighter and bluer than the MSTO by
the current cluster age.  Consequently, we set N$_{BS,0}$ =
N$_{ev}$ in Equation~\ref{eqn:number-bss}.  We further assume that
all central cluster parameters have not changed in the last
$\tau_{BS}$ years, including the central velocity dispersion, the
central luminosity density, the core radius and the core binary
fraction.  It follows that the rate of BS formation is constant for
the time-scale of interest.  This time-scale is
comparable to the half-mass relaxation time but much longer than the
central relaxation time for the majority of the clusters in our sample
\citep{harris96}.  This suggests that core
parameters such as the central density and the core radius will
typically change in a time $\tau_{BS}$ since the time-scale on
which these parameters vary is the central relaxation
time \citep{heggie03}.  Therefore, our assumption of constant rates
and cluster parameters is not strictly correct, however it provides a
suitable starting point for our model.  We will discuss the
implications of our assumption of time-independent cluster properties
and rates in Section~\ref{discussion5}.

In the following sections, we discuss each of the remaining terms in
Equation~\ref{eqn:number-bss}.

\subsubsection{Stellar Collisions} \label{collisions5}

We can approximate the number of BSs formed in the last $\tau_{BS}$
years from collisions during dynamical encounters as:  
\begin{equation}
\label{eqn:number-coll}
N_{coll} = f_{1+1}N_{1+1} + f_{1+2}N_{1+2} + f_{2+2}N_{2+2},
\end{equation}
where N$_{1+1}$, N$_{1+2}$ and N$_{2+2}$ are the number of
single-single, single-binary and binary-binary encounters,
respectively.  The terms f$_{1+1}$, f$_{1+2}$ and f$_{2+2}$ are the
fraction of 1+1, 1+2 and 2+2 encounters, respectively, that will
produce a BS in the last $\tau_{BS}$ years.  We treat these three
variables as free parameters since we do not know what fraction of
collision products will produce BSs (i.e. stars with an appropriate
combination of colour and brightness to end up in the BS region of the
CMD), nor do we know what fraction of 1+2 and 2+2 encounters will
result in a stellar collision.  Numerical scattering experiments have
been performed to study the outcomes of 1+2 and 2+2 encounters
\citep[e.g.][]{hut83, mcmillan86, fregeau04}, however a large fraction
of the relevant parameter space has yet to be explored.

In terms of the core radius $r_c$ (in parsecs), the central number
density $n_0$ (in pc$^{-3}$), the root-mean-square velocity $v_{m}$
(in km s$^{-1}$), the average stellar mass $m$ (in M$_{\odot}$) and
the average stellar radius $R$ (in R$_{\odot}$), the mean time-scale
between single-single collisions in the core of a GC is
\citep{leonard89}:
\begin{equation}
\begin{gathered}
\label{eqn:coll1+15}
\tau_{1+1} = 1.1 \times 10^{10}(1-f_b)^{-2} \Big(\frac{1 pc}{r_c}
\Big)^3 \Big(\frac{10^3 pc^{-3}}{n_0} \Big)^2 \\
 \Big(\frac{v_{m}}{5
  km/s} \Big) \Big(\frac{0.5 M_{\odot}}{m} \Big) \Big(\frac{0.5
  R_{\odot}}{R} \Big)\mbox{ years}
\end{gathered}
\end{equation}
The additional factor (1-f$_b$)$^{-2}$ comes from the fact that we
are only considering interactions between single stars and the
central number density of single stars is given by (1-f$_b$)n$_0$,
where f$_b$ is the binary fraction in the core (i.e. the fraction of
objects that are binaries).  For our chosen mass, we assume a
corresponding average stellar radius using the relation M/M$_{\odot}$
= R/R$_{\odot}$ \citep{iben91}.
The number of 1+1 collisions expected to have occurred in the last
$\tau_{BS}$ years is then approximated by:
\begin{equation}
\label{eqn:N-1+1}
N_{1+1} = \frac{\tau_{BS}}{\tau_{1+1}}.
\end{equation}

The rate of collisions
between single stars and binaries, as well as between two
binary pairs, can be roughly approximated in the same way
as for single-single encounters \citep{leonard89, sigurdsson93,
  bacon96, fregeau04}.  We adopt the time-scales derived in
\citet{leigh11b} for the average times between 1+2 and 2+2 encounters.
These are:
\begin{equation}
\begin{gathered}
\label{eqn:coll1+25}
\tau_{1+2} = 3.4 \times 10^7f_b^{-1}(1-f_b)^{-1}\Big(\frac{1 pc}{r_c}
\Big)^3 \Big(\frac{10^3 pc^{-3}}{n_0} \Big)^2 \\
 \Big(\frac{v_{m}}{5
  km/s} \Big) \Big(\frac{0.5 M_{\odot}}{m} \Big) \Big(\frac{1 AU}{a}
\Big)\mbox{ years}
\end{gathered}
\end{equation}
and
\begin{equation}
\begin{gathered}
\label{eqn:coll2+25}
\tau_{2+2} = 1.3 \times 10^7f_b^{-2}\Big(\frac{1 pc}{r_c}
\Big)^3 \Big(\frac{10^3 pc^{-3}}{n_0} \Big)^2 \\ 
\Big(\frac{v_{m}}{5
  km/s} \Big) \Big(\frac{0.5 M_{\odot}}{m} \Big) \Big(\frac{1 AU}{a}
\Big)\mbox{ years},
\end{gathered}
\end{equation}
where $a$ is the average binary semi-major axis in the core in AU and
we have assumed that the average binary mass is equal to twice the
average single star mass.  
The numbers of 1+2 and 2+2 encounters expected to have occurred in the 
last $\tau_{BS}$ years are given by, respectively:
\begin{equation}
\label{eqn:N-1+2}
N_{1+2} = \frac{\tau_{BS}}{\tau_{1+2}}
\end{equation}
and
\begin{equation}
\label{eqn:N-2+2}
N_{2+2} = \frac{\tau_{BS}}{\tau_{2+2}}.
\end{equation}

The outcomes of 1+2 and 2+2 encounters will ultimately contribute to
the evolution of the binary fraction in the core.  How and with what
frequency binary 
hardening/softening as well as capture, exchange and ionization
interactions affect the binary fraction in the dense cores of GCs is
currently a subject of debate \citep[e.g.][]{ivanova05,
  hurley07}.  
Observations are also lacking for binary fractions in the dense cores of
GCs, however rough constraints suggest that they range from a few to a
few tens of a percent \citep[e.g.][]{rubenstein97, cool02, sollima08,
  davis08}.  The 
situation is even worse for the distribution of binary orbital
parameters observed in dense stellar environments.  Our best
constraints come from radial velocity surveys of moderately dense open
clusters \citep{latham05, geller09}, however whether or not the
properties of the binary populations in these clusters should differ
significantly from those in the much denser cores of GCs is unclear.
As an initial assumption, we assume a time-independent core binary
fraction of 10\% for all clusters, and an average semi-major axis
of 2 AU.  This semi-major axis corresponds roughly to the hard-soft
boundary for most of the clusters in our sample, defined by setting
the average binary orbital energy equal to the kinetic energy of an
average star in the cluster \citep[e.g.][]{heggie03}.  We 
treat both the core binary 
fraction and the average binary semi-major axis as free parameters,
and explore a range of possibilities using the available observations
as a guide for realistic values.  We will return to these assumptions 
in Section~\ref{discussion5}.

\subsubsection{Binary Star Evolution} \label{mergers5}

Although we do not know the rate of BS formation from binary star
evolution, we expect a general relation of the form
N$_{bin}$ = $\tau_{BS}$/$\tau_{mt}$ for the number of BSs produced
from binary mergers in the last $\tau_{BS}$ years, where $\tau_{mt}$
is the average time between BS formation events due to binary star
evolution.  We can express the number of BSs formed from binary star
evolution in the last $\tau_{BS}$ years as:
\begin{equation}
\label{eqn:N-bin}
N_{bin} = f_{mt}f_bN_{core},
\end{equation}
where  N$_{core}$ is the total number of
objects (i.e. single and binary stars) in the core and f$_{mt}$ is
the fraction of binary stars that evolved internally to form
BSs within the last $\tau_{BS}$ years.  We treat f$_{mt}$ as a free
parameter since it is likely to depend on the mass-ratio, period and
eccentricity distributions characteristic of the binary populations of
evolved GC cores, for which data is scarce at best.  

\subsubsection{Migration Into and Out of the Core}
\label{segregation5}

Due to the relatively small sizes of the BS populations
considered, the migration of BSs into or out of the core is an
important consideration when calculating the predicted numbers.  In
other words, we are dealing with relatively small
number statistics and every blue straggler counts.  In order to
approximate the number of stars in the core as a function of time,
two competing effects need to be taken into account:  (1) mass
stratification/segregation (or, equivalently, dynamical friction) and
(2) kicks experienced during dynamical interactions.  
These effects are accounted for with the variables N$_{in}$ and
N$_{out}$ in Equation~\ref{eqn:number-bss}, respectively.  

Blue stragglers are among the most massive stars in clusters
\citep[e.g.][]{shara97, vandenberg01, mathieu09}, so they should
typically be heavily mass segregated relative to other stellar
populations \citep[e.g.][]{spitzer69, shara95, king95}.  The
time-scale for this process to occur 
can be approximated using the dynamical friction time-scale \citep{binney87}:
\begin{equation}
\label{eqn:t-dyn}
\tau_{dyn} = \frac{3}{4ln{\Lambda}G^2(2{\pi})^{1/2}}\frac{\sigma(r)^3}{m_{BS}\rho(r)},
\end{equation}  
where $\sigma(r)$ and $\rho(r)$ are, respectively, the velocity
dispersion and stellar mass density at the given distance from the
cluster centre $r$.  Both $\sigma(r)$ and $\rho(r)$ are found from
single-mass King models \citep{sigurdsson93}, which are fit to each
cluster using the 
concentration parameters provided in \citet{mclaughlin05}.  
The Coulomb logarithm is denoted by $\Lambda$, and we adopt
a value of ln$\Lambda \sim 10$ throughout this chapter 
\citep[e.g.][]{spitzer87, heggie03}.  If $\tau_{dyn} >
\tau_{BS}$ at a given distance from the 
cluster centre, then any BSs formed at this radius in the last
$\tau_{BS}$ years will not have had sufficient time to migrate into
the core by the current 
cluster age.  The maximum radius r$_{max}$ at which BSs can
have formed in the 
last $\tau_{BS}$ years and still have time to migrate into the core
via dynamical friction is given by setting $\tau_{dyn} = \tau_{BS}$.  
Therefore, N$_{in}$ depends only on the number of BSs formed in the last
$\tau_{BS}$ years at a distance from the cluster centre smaller than
r$_{max}$.  

In order to estimate the contribution to N$_{BS}$ in
Equation~\ref{eqn:number-bss} from BSs formed outside the core, we
calculate the number of BSs formed in radial shells between the
cluster centre and r$_{max}$.  Each shell is taken to be one core
radius thick, and we calculate the contribution from 
each formation mechanism in every shell.  This is done by assuming a
constant (average) density and velocity dispersion in each shell.
Specifically, we estimated the density and velocity 
dispersion at the half-way point in each shell using our single-mass King models,
and used these to set average values.  The number of BSs expected to
have migrated into the core within the last $\tau_{BS}$ years can be
written:
\begin{equation}
\begin{gathered}
\label{eqn:N-in}
N_{in} = \sum_{i=2}^N \Big( f_{1+1}N_{(1+1),i} + f_{1+2}N_{(1+2),i} +
f_{2+2}N_{(2+2),i} \\ 
+ f_{mt}N_{(bin),i} \Big) \times \Big(1 -
\frac{\tau_{(dyn),i}}{\tau_{BS}}\Big),
\end{gathered}
\end{equation}
where $i=1$ refers to the core, $i=2$ refers to the shell immediately 
outside the core, etc. and N is taken to be the integer nearest to
r$_{max}$/r$_c$.  We let the terms with N$_{(1+1),i}$, N$_{(1+2),i}$,
N$_{(2+2),i}$ and 
N$_{(bin),i}$ represent the number of BSs formed in shell $i$ from
single-single collisions, single-binary collisions, binary-binary
collisions and binary star evolution, respectively.  The time-scale
for dynamical friction in shell $i$ is denoted by $\tau_{(dyn),i}$, and
the factor (1 - $\tau_{(dyn),i}$/$\tau_{BS}$) is included to account for
the fact that we are assuming a constant formation rate for BSs, so
that not every BS formed in shells outside the core will have
sufficient time to fall in by the current cluster age.

It is typically the least massive stars
that are ejected from 1+2 and 2+2 interactions as single stars
\citep[e.g.][]{sigurdsson93}.  Combined with conservation of momentum,
this suggests that BSs are the least likely to be ejected from
dynamical encounters with velocities higher than the central velocity
dispersion due to their large masses.  This has been confirmed by
several studies of numerical scattering experiments 
\citep[e.g.][]{hut83, fregeau04}.  Based on this, we 
expect that N$_{out}$ should be 
very small and so, as a first approximation, we take N$_{out} = 0$.
However, we also explore the effects of a 
non-zero N$_{out}$ by assigning a kick velocity to all BSs at birth.
If dynamical interactions play a role in BS formation, we 
might naturally expect BSs to be imparted a recoil velocity at birth
(or shortly before) due to 
momentum conservation.  We will return to this assumption in
Section~\ref{results5}. 

\subsection{Statistical Comparison with Observations} \label{statistics5}

Our model contains 4 free parameters, which correspond to the fraction of
outcomes that produce a blue straggler for each formation mechanism
(1+1 collisions, 1+2 collisions, 2+2 collisions, and binary star
evolution).  These are the $f$ values
described in the previous section: $f_{1+1}, f_{1+2}, f_{2+2},
f_{mt}$.  We assume that these values are constant throughout each
cluster, and are also constant between clusters.  By fitting the
predictions of our model to the observational data, we can determine
best-fit values for each of these $f$ parameters, and therefore make
predictions about which blue straggler formation processes are
more important. 

In order to determine the best values for these $f$ parameters, we 
need an appropriate statistical test.  For this, we follow the approach
of \citet{verbunt08}.  The number of BSs seen in the core of
a globular cluster can be described by Poisson statistics.  In
particular, the probability of observing $N$ sources when $\mu$ are
expected is:
\begin{equation}
\label{eqn:stats5}
P(N,\mu) = \frac{\mu^N}{N!}e^{-\mu}
\end{equation}
We can calculate a probability for each cluster, and then calculate an 
overall probability $P^{\prime}$ for the model by multiplying the
individual $P$ values.  We can then vary the $f$ values to maximize
this value. 

In practice, these $P$ values are typically of order ten percent per
cluster, and with 35 clusters, the value of $P^{\prime}$ quickly
becomes extremely small.  Therefore we chose to work with a modified
version of this value:  the deviance of our model to the saturated
model.  A saturated model is one in which the observed number of
sources is exactly equal to the expected number in each cluster.  In
other words, this is the best that we can possibly do.  However,
because of the nature of Poisson statistics, the probability $P$ of
such a model (calculated by setting $N=\mu$ in
Equation~\ref{eqn:stats5}) is not equal to 1, but in fact has some
smaller value.  For the numbers of blue stragglers in our clusters,
the $P$ values for the saturated model run from 0.044 to 0.149, and
the value of $P^{\prime}$ is $2.08 \times 10^{-41}$.

The deviance of any model from the saturated model is given by
\begin{equation}
\label{eqn:deviance}
D = 2.0(\ln(P^{\prime}_{saturated}) - \ln(P^{\prime}_{model}))
\end{equation}
The model which minimizes this quantity will be our best-fit
model.  Ideally, the scaled deviance ($D/(N_{data}-N_{parameters}$)
should be equal to 1 for a best fit.  Given the simplicity of our
model, we expect that our values will not provide this kind of
agreement, and we simply look for the model which provides the minimum
of the scaled deviance.

\section{Results} \label{results5}

In this section, we present the results of comparing our model
predictions to the observations.  After presenting the results for 
a constant core binary fraction for all clusters, we explore the
implications of adopting a core binary fraction that depends on the
cluster luminosity, as reported in \citet{sollima07} and
\citet{milone08}. 

\subsection{Initial Assumptions} \label{initial}

The predictions of our model for our initial choice of assumptions are
shown in Figure~\ref{fig:model_best}.  These numbers are plotted
against the total stellar mass in the core along with the 
number of BSs observed in the core (blue triangles).  We plot both the
total number of BSs predicted to have formed within $r_{max}$ in the
last $\tau_{BS}$ years (N$_{BS}$ in Equation~\ref{eqn:number-bss};
green circles), as well as the total number formed only in the core
(N$_{coll}$ + N$_{bin}$; red circles).  Upon comparing N$_{BS}$ to the
observed number of BSs in the core, 
the best-fitting model parameters predict that most BSs are formed
from binary star evolution, with a small contribution from 2+2
collisions being needed in order to obtain the best possible match to
the observations.  The ideal contribution from 2+2 collisions
constitutes at 
most a few percent of the predicted total for most of the clusters in
our sample.  Even for our best-fitting model
parameters, our initial choice of assumptions predicts too few BSs in
clusters with small core masses. 

\begin{figure} [!h]
  \begin{center}
 \includegraphics[scale=0.5]{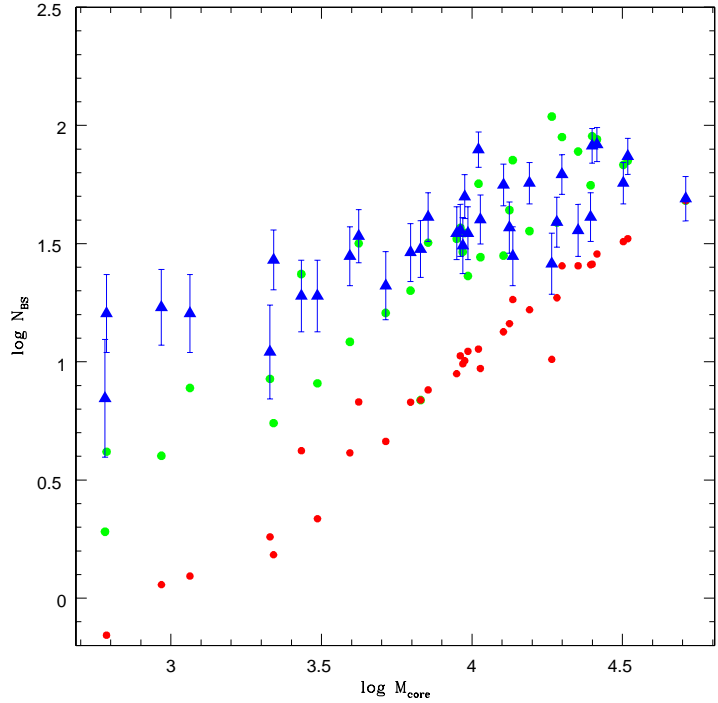}
\caption[The predicted number of BSs plotted versus the total stellar
mass in the core for the best-fitting model parameters found for our
initial choice of assumptions]{
The predicted number of BSs plotted versus the total stellar
mass in the core for the best-fitting model parameters found for our
initial choice of assumptions.  The filled blue triangles correspond to
  the observed numbers, the 
  filled green circles to the number of BSs predicted to have formed within
  r$_{max}$, and the filled red circles to the number predicted to
  have formed in only the core.  The best-fitting model parameters
  used to calculate the predicted numbers are f$_{1+1} = 0$, f$_{1+2} = 0$,
  f$_{2+2} = 7.3 \times 10^{-3}$, and f$_{mt} = 1.7 \times 10^{-3}$.  
  Estimates for the total stellar masses in the core were obtained
  from single-mass King models, as described in \citet{leigh11a}.
  Error bars have been indicated for the observed numbers using 
  Poisson statistics.
\label{fig:model_best}}
\end{center}
\end{figure}

\subsection{Binary Fraction} \label{results-binary}

We tried changing our assumption of a constant f$_b$ for all
clusters to one for which the core binary fraction varies with the
total cluster magnitude.  First, we adopted a dependence of the
form:
\begin{equation}
\label{eqn:sollima07} 
f_b = 0.13M_V + 1.07,
\end{equation}
where M$_V$ is the total cluster V magnitude.  This 
relation comes from fitting a line of best-fit to the observations of 
\citet{sollima07}, who studied the binary fractions in a sample
of 13 low-density GCs (we calculated an average of
columns 3 and 4 in their Table 3 and used these binary fractions to
obtain Equation~\ref{eqn:sollima07}).  In order to prevent negative
binary fractions, 
we impose a minimum binary fraction of f$_b^{min} = 0.01$.  In other
words, we set f$_b =$ f$_b^{min}$ if Equation~\ref{eqn:sollima07}
gives a binary fraction less than f$_b^{min}$.  As before, we adopt an
average semi-major axis of 2 AU.  The results of this comparison are
presented in Figure~\ref{fig:model_rmax_sollima_best}.  As in
Figure~\ref{fig:model_best}, both the numbers of 
observed (blue triangles) and predicted (green circles) BSs in the
core are plotted 
versus the total stellar mass in the core.  Once again, the predicted
numbers include all BSs formed within r$_{max}$ in the last
$\tau_{BS}$ years.  The best-fitting model parameters for this
comparison suggest that both single-single collisions and binary star
evolution are significant contributors to BS formation.  Single-single
collisions contribute up to several tens of a percent of the predicted
total in several clusters.  We obtain a 
deviance in this case that is significantly larger than that obtained
for the best-fitting model parameters assuming a constant f$_b$ of
$10\%$ for all clusters.  Equation~\ref{eqn:sollima07} gives higher
binary fractions in low-mass
clusters relative to our initial assumption of a constant f$_b$.  This
increases the number of BSs formed from binary star evolution and
improves the agreement between our model predictions and the
observations in low-mass clusters.  This is consistent with the
results of \citet{sollima08}.  However, adopting
Equation~\ref{eqn:sollima07} for f$_b$ also causes our model to
under-predict the number of BSs in several high-mass clusters.

\begin{figure} [!h]
  \begin{center}
 \includegraphics[scale=0.5]{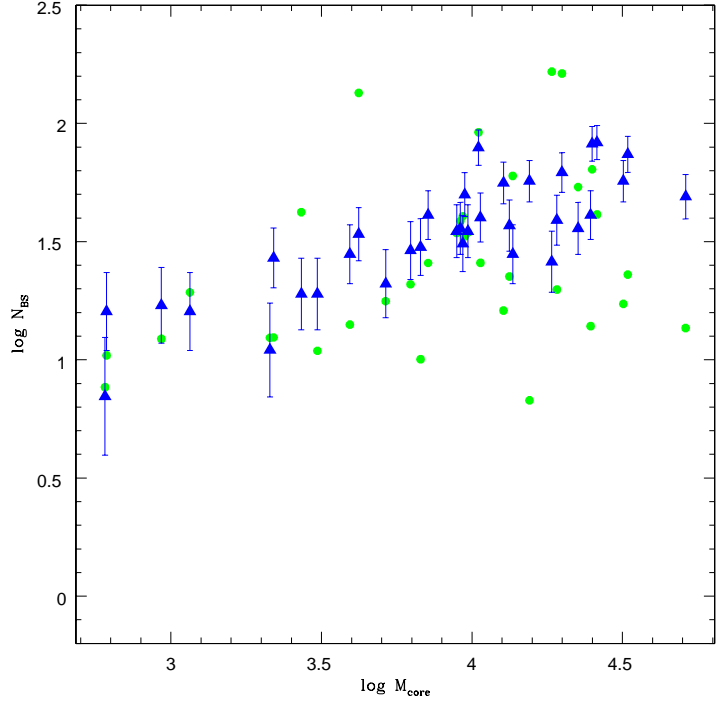}
\caption[The predicted number of BSs plotted versus the total stellar
mass in the core for the best-fitting model parameters found using the
binary fractions of \citet{sollima07} with f$_b^{min} = 0.01$]{The
  number of BSs predicted in the cluster core using the binary
  fractions of \citet{sollima07} with f$_b^{min} = 0.01$ plotted
  against the total stellar mass contained within the core.  The
  colours used to indicate the observed and predicted numbers are the
  same as in Figure~\ref{fig:model_best}.  The predicted numbers correspond to
  the best-fitting model parameters, which are f$_{1+1} = 0.41$,
  f$_{1+2} = 0$, f$_{2+2} = 0$, and f$_{mt} = 1.5 \times 10^{-3}$.
\label{fig:model_rmax_sollima_best}}
\end{center}
\end{figure}

The best fit to the observations is found
by adopting the relation for f$_b$ provided in
Equation~\ref{eqn:sollima07} and setting f$_b^{min} = 0.1$ (however we
note that a comparably good agreement is found with a slightly lower
f$_b^{min} = 0.05$).  This
improves the agreement between our model predictions and the
observations by increasing the number of BSs formed from binary star
evolution in massive clusters.  The result is a good agreement between
our model predictions and the observations in both low- and high-mass
cores, as shown in
Figure~\ref{fig:model_best_rmax_sollima_10min_tBS15}.  In this case,
the best-fitting model parameters yield the lowest deviance of any of
the assumptions so far considered.  
These best-fitting values suggest that most BSs are formed 
from binary star evolution, with a small contribution from 2+2
collisions being needed in order to obtain the best possible match to
the observations.  Similarly to what was found for our initial
assumptions, the ideal contribution from 2+2 collisions
constitutes at most a few percent of the predicted total for most of
the clusters in our sample.  
On the other hand, if we change our imposed minimum binary fraction to
f$_b^{min} = 0.05$ we find that a non-negligible (i.e. up to a few
tens of a percent) contribution from
single-single collisions is needed in several clusters to obtain the best possible
agreement with the observations (which is very nearly as good as was
found using f$_b^{min} = 0.1$).  All of this shows that, although
binary star evolution consistently dominates BS formation in our
best-fitting models, at least some contribution from collisions
(whether it be 1+1 or 2+2 collisions, or some combination of 1+1, 1+2
and 2+2 collisions) also
consistently improves the agreement with the observations.  Moreover,
it is interesting to note that an improved agreement with the
observations could alternatively be obtained if we keep f$_b^{min} =
0.01$ but all or some of
f$_{1+1}$, f$_{1+2}$ and f$_{2+2}$ increase with increasing cluster
mass.  This would also serve to improve the agreement at the high-mass
end.  We will return to this in Section~\ref{discussion5}.

\begin{figure} [!h]
  \begin{center}
 \includegraphics[scale=0.5]{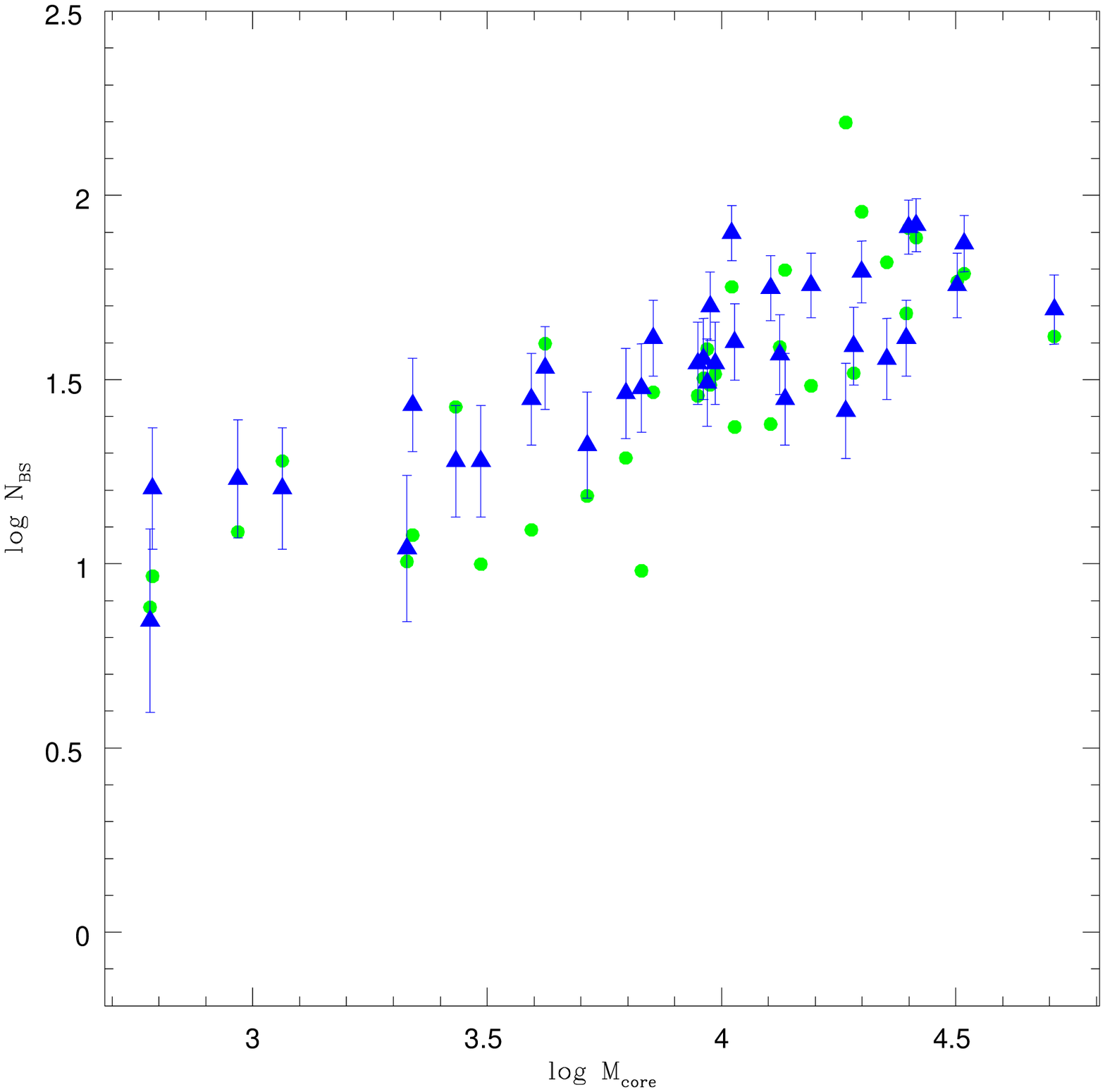}
\caption[The predicted number of BSs plotted versus the total stellar
mass in the core for the best-fitting model parameters found using the binary
fractions of \citet{sollima07} with f$_b^{min} = 0.1$]{The number of BSs
  predicted in the cluster core using the binary fractions of
  \citet{sollima07} with f$_b^{min} = 0.1$ plotted against the total
  stellar mass contained within the core.  The
  colours used to indicate the observed and predicted numbers are the
  same as in Figure~\ref{fig:model_best}.  The predicted numbers correspond to
  the best-fitting model parameters, which are f$_{1+1} = 0$,
  f$_{1+2} = 0$, f$_{2+2} = 3.6 \times 10^{-3}$, and f$_{mt} = 1.4
  \times 10^{-3}$.
\label{fig:model_best_rmax_sollima_10min_tBS15}}
\end{center}
\end{figure}

Finally, we also tried adopting the observed dependence of f$_b$ on
M$_V$ reported in \citet{milone08}, who also found evidence for an
anti-correlation between the core binary fraction and the total
cluster mass.  Despite this change, we consistently find that our
results are the same as found when using the empirical binary fraction
relation provided in Equation~\ref{eqn:sollima07}.

\subsection{Average BS Lifetime} \label{lifetime}

We also tried changing our assumption for the average BS lifetime.  We
explored a range of plausible lifetimes based on values found
throughout the literature.  Specifically, we explored the range 0.5-5
Gyrs.  We find that at 
the low end of this range, our model fits become increasingly poor.
This is because lower values for $\tau_{BS}$ correspond to smaller
values for r$_{max}$ and decrease the term (1 -
t$_{(dyn),i}$/$\tau_{BS}$) in Equation~\ref{eqn:N-in}.  This reduces the
contribution to the total predicted 
numbers from BSs formed outside the core that fall in via dynamical 
friction.  Conversely, our model fits
improve for $\tau_{BS} > 1.5$ Gyrs since this corresponds to a larger
contribution to N$_{BS}$ from N$_{in}$.  It is important to note,
however, that this same effect can be had by increasing the 
number of BSs formed outside the core, since this would also
serve to increase N$_{in}$ in Equation~\ref{eqn:number-bss}.  This
can be accomplished by, for instance, increasing the binary fraction
outside the core (which would increase the
number of BSs formed from binary star evolution outside the core that
migrate in due to dynamical friction) relative to inside the core.
This seems unlikely, however, given that observations of low-density
globular clusters and open clusters suggest that their binary
fractions tend to drop off rapidly outside the core
\citep[e.g.][]{sollima07}.  We 
will return to this issue in Section~\ref{discussion5}.

The best possible match to the observations is
found by adopting an average BS lifetime of 5 Gyrs along with the
relation for f$_b$ provided in Equation~\ref{eqn:sollima07} with 
f$_b^{min} = 0.1$.  The predictions of our model 
are shown in Figure~\ref{fig:model_best_rmax_sollima_10min_tBS50} for
these best-fitting model parameters.  As shown, the agreement
between our model predictions and the observed numbers is excellent.
The best agreement is found by adopting an average BS lifetime of 5
Gyrs, however the agreement is comparably excellent down to 
slightly less than $\tau_{BS} \sim 3$ Gyrs.  Although increasing
$\tau_{BS}$ does contribute to improving the 
agreement between our model predictions and the observations, the
effect is minor compared to the improvement that 
can be found by changing our assumption for the binary fraction.
This is apparent upon comparing 
Figure~\ref{fig:model_best_rmax_sollima_10min_tBS50} to 
Figure~\ref{fig:model_best_rmax_sollima_10min_tBS15}.  

\begin{figure} [!h]
  \begin{center}
 \includegraphics[scale=0.5]{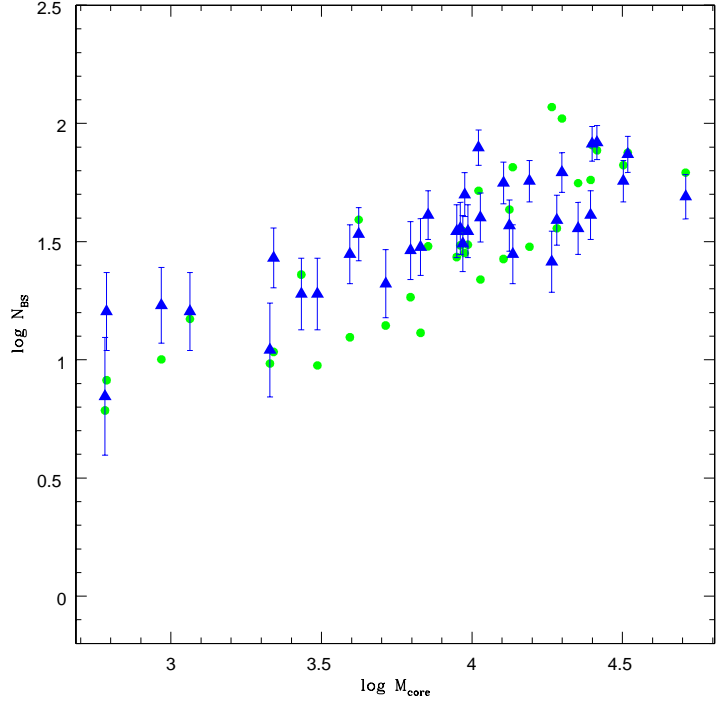}
\caption[The predicted number of BSs plotted versus the total stellar
mass in the core for the best-fitting model parameters found using an
average BS lifetime of 5 Gyrs and the relation for the cluster binary
fraction provided in Equation~\ref{eqn:sollima07} with f$_b^{min} =
0.1$]{The predicted 
  number of BSs plotted versus the total stellar mass in the core for
  the best-fitting model parameters found using an average BS lifetime
  of 5 Gyrs and the relation for the cluster binary fraction provided
  in Equation~\ref{eqn:sollima07} with f$_b^{min} = 0.1$.  The
  colours used to indicate the observed and predicted numbers are the
  same as in Figure~\ref{fig:model_best}.  The best-fitting model parameters
  used to calculate the predicted numbers are f$_{1+1} = 0$, f$_{1+2}
  = 0$, f$_{2+2} = 1.6 \times 10^{-3}$, and f$_{mt} = 9.9 \times
  10^{-4}$.  The agreement with the observations is excellent for these
  best-fitting values.
\label{fig:model_best_rmax_sollima_10min_tBS50}}
\end{center}
\end{figure}

\subsection{Migration} \label{results-migration}

In order to explore the sensitivity of our results to our assumption
for r$_{max}$, we also tried setting N$_{in}$ equal to the total
number of BSs expected to form within 10 r$_c$ for all clusters.  For
comparison, for an average BS lifetime of $\tau_{BS} = 1.5$ Gyrs,
r$_{max}$ ranges from 2 - 15 r$_c$ for the clusters in our
sample.  Despite implementing this change, our results remained the
same.  This is because the largest contribution to the total number of
BSs comes from those BSs formed in the first few shells
immediately outside the core that migrate in due to dynamical
friction.

Several GCs have been reported to show evidence for a decrease in
their binary fractions with increasing distance from the cluster
centre \citep[e.g.][]{sollima07, davis08}.  This effect is often
significant, with binary fractions decreasing by up to a factor of a
few within only a few core radii from the cluster centre.  Based on
this, our assumption that f$_b$ is independent of the distance
from the cluster centre likely results in an over-estimate of the true
binary fraction at large cluster radii.  In order to
quantify the possible implications of this for our results, we tried
setting f$_b$ $= 0$ for all shells outside the core.  Although this
assumption is certainly an under-estimate for the true binary fraction
outside the core, our results remain the same (albeit the agreement
with the observations is considerably worse than for most of our previous
model assumptions).  Once again, the
best-fitting model parameters suggest that most BSs are formed from
binary star evolution, with a non-negligible (i.e. up to a few tens of
a percent in some clusters) contribution from
binary-binary collisions.  Our results indicate that, if the
binary fraction is negligible outside the core, then the contribution
from BSs that migrate into the core due to dynamical friction is also
negligible.  This is because the time between 1+1 collisions increases
rapidly outside the core, and every other BS formation mechanism requires
binary stars to operate.  

We also explored the effects of assuming a non-zero value for
N$_{out}$ in Equation~\ref{eqn:number-bss} by imparting a constant
kick velocity to all
BSs at birth.  Using conservation of energy, we calculated the
cluster radius to which BSs should be kicked upon formation, and used
the time-scale for dynamical friction at the kick radius to calculate
the fraction of BSs expected to migrate back into the core within
a time $\tau_{BS}$.  Regardless of our assumption for the kick
velocity, this did not improve the deviance for any of our
best-fitting model parameters.

\subsection{Average Binary Semi-Major Axis} \label{avg-a}

We investigated the dependence of our results on our assumption for
the average binary semi-major axis.  However, this had a very small
effect on our results.  This is because only N$_{1+2}$ and N$_{2+2}$
depend on the average semi-major axis, and neither of these terms
dominated BS production regardless of our model
assumptions.  Only f$_{2+2}$ is
non-zero for our best-fitting models however, as before, it consistently
suggests that far fewer BSs should be formed from 2+2 collisions than from
binary star evolution. 

\section{Summary \& Discussion} \label{discussion5}

In this chapter, we have presented an analytic model to investigate BS
formation in globular clusters.  Our model predicts the number of BSs
in the cluster core at the current cluster age by estimating the
number that should have either formed 
there from stellar collisions and binary star evolution, or migrated
in via dynamical friction after forming outside the core.  We have
compared the results of our model predictions for a variety of input
parameters to observed BS numbers in 35 GCs taken from the catalogue of
\citet{leigh11a}.   

What has our model told us about BS formation in dense cluster
environments?  
The agreement between the predictions of our model and the
observations is excellent if we assume that:

\begin{itemize}

\item Binary star evolution dominates BS formation, however at least
  some contribution from 2+2 collisions (most of which
  occur in the core) must also be included in the total predicted
  numbers.  Although it is clear that including a contribution from
  dynamical encounters gives the best possible match to the
  observations, it is not clear how exactly this is accomplished in
  real star clusters.  Does the cluster dynamics increase
  BS numbers via direct collisions?  Or do dynamical interactions
  somehow modify primordial binaries to initiate more mass-transfer
  events?  We will return to this point below.

\item The binary fraction in the core is inversely
  correlated with the total cluster luminosity, similar to the
  empirical relations found by \citet{sollima07} and \citet{milone08}.
  We also require a minimum core binary fraction of $5-10\%$.  The
  inverse dependence of f$_b$ on the total cluster mass
  contributes to a better agreement with the observations at the
  low-mass end of the distribution of cluster masses, whereas 
  the imposed condition that f$_b^{min} = 0.05 - 0.1$ contributes to
  improving the agreement at the high-mass end.

\item BSs formed outside the core that migrate in by the
  current cluster age contribute to the total predicted numbers.

\item The average BS lifetime is roughly a few ($\sim$ 3-5) Gyrs, since this
  increases the fraction of BSs formed outside the core that will have
  sufficient time to migrate in due to dynamical friction.

\end{itemize}

Our model can only provide a
reasonable fit to the observations for all cluster masses if we assume
that the cluster binary fraction is inversely proportional to the total
cluster mass.  It is interesting to consider the possibility that such
an inverse proportionality could 
arise as a result of the fact that the rate of two-body relaxation is
also inversely proportional to the cluster mass \citep{spitzer87}.
Consequently, since binaries tend to be the most massive objects in
GCs, they should quickly migrate into the core in low-mass clusters,
contributing to an increase in the core binary fraction over
time \citep{fregeau09}.  This process should operate on a considerably
longer time-scale in very massive GCs since the time-scale for
two-body relaxation is very long.  Mass segregation could then 
contribute to the observed sub-linear dependence of BS numbers
on the core masses by acting to preferentially migrate the binary star
progenitors of BSs into the cores of low-mass clusters.  This is one
example of how a direct link could arise between the observed
properties of BS 
populations and the dynamical histories of their host clusters.
Although this scenario is interesting to consider, we 
cannot rule out the possibility that an anti-correlation between the
core binary fraction and the total cluster mass could be a primordial
property characteristic of GCs at birth.

When interpreting our results, it is important to bear in mind that
binary star evolution and dynamical interactions involving binaries
may not always contribute to BS formation independently.  For example,
dynamics could play an important role in changing the distribution of
binary orbital parameters so that mass-transfer occurs more commonly
in some clusters.  
One way to perhaps compensate for this effect would be to include 
a factor of $1/a$ (where $a$ is the average binary semi-major axis) in
Equation~\ref{eqn:N-bin}.  This would serve to account for the fact
that we might naively expect clusters populated by more close binaries
to be 
more likely to have a larger fraction of their binary populations
undergo mass-transfer.  This does not, however, guarantee that more
BSs will form since our poor understanding of binary star evolution
prevents us from being able to predict the outcomes of these
mass-transfer events, and whether or not they will form BSs.
Moreover, little is known about the distribution of orbital
parameters characteristic of the binary populations in globular
clusters, and how they are typically modified by the cluster
dynamics.  For 
these reasons, the interpretation of our results must be done with
care in order to ensure that reliable conclusions can be drawn.

In general, our results suggest that binary stars play a crucial role
in BS formation in dense GCs.  
In order to obtain the best possible agreement with the observations,
an enhancement in BS formation from dynamical encounters is required
in at least some clusters relative to what is expected by assuming a
simple population of binaries evolving in isolation.  It is not clear
from our results, however, how exactly this occurs in real star
clusters.  Dynamics could enhance BS
formation directly by causing stellar collisions, or this could
also occur indirectly if the cluster dynamics somehow induces episodes of
mass-transfer by reducing the orbital 
separations of binaries.  But in which clusters is
BS formation the most strongly influenced by the cluster dynamics?
Unfortunately, no clear trends have emerged from our analysis that
provide a straight-forward answer to this question.  However, our
results are consistent with dynamical interactions 
playing a more significant role in more massive clusters (although
this does not imply that the cluster dynamics does not also contribute
in low-mass clusters).  This could 
be due to the fact that more massive clusters also tend to have higher
central densities \citep[e.g.][]{djorgovski94}, and therefore
higher collision rates.  This picture is, broadly speaking, roughly
consistent with 
the results of \citet{davies04}.  These authors considered the observed
dependence (or lack thereof) of BS numbers on the total cluster masses
presented in \citet{piotto04}, and suggested that primordial binary
evolution and stellar collisions dominate BS production in low- and
high-mass clusters, respectively.

Our model neglects the dynamical evolution of GCs and the resulting
changes to their global properties, including the central density,
velocity dispersion, core radius and binary fraction.  As a young
cluster evolves, dynamical processes like mass segregation and
stellar evaporation tend to result in a smaller, denser core.  Within
a matter of a few half-mass relaxation times, a gravothermal
instability has set in and the collapse ensues on a time-scale
determined by the rate of heat flow out of the core
\citep[e.g.][]{spitzer87}.  We are focussing 
on the last $\tau_{BS}$ years of cluster 
evolution, a sufficiently late period in the lives of most GCs
that gravothermal collapse will have long since taken over as the
primary driving force affecting the stellar concentration in the
core.  Most of the GCs in our sample should currently be in a phase
of core contraction \citep{fregeau09, gieles11}, and their central
densities and core radii should have been steadily
decreasing over the last $\tau_{BS}$ years.  Therefore, by using
the currently observed central cluster parameters and assuming that
they remained constant over the last
$\tau_{BS}$ years, we have effectively calculated upper limits for
the encounter rates.  This could suggest that we have over-estimated
the importance of dynamical interactions for BS formation.  On the
other hand, some theoretical
models of GC evolution suggest that the hard binary fraction in the
core of a dense stellar system will generally increase
with time \citep[e.g.][]{hurley05, fregeau09}.  This can be understood as an
imbalance between the migration of
binaries into the core via mass segregation and the destruction of
binaries in the core via both dynamical encounters and their internal
evolution.  This could suggest that our estimate for the number of BSs
formed from binary star evolution should also be taken as an upper
limit.  The key point is that GC evolution can act to
increase the number of BSs in the cluster core via several different
channels.  The effects we have discussed should typically be small,
however, since 
$\tau_{BS}$ is much shorter than the cluster age \citep{deangeli05}
and any changes to global cluster properties that occur during this
time will often be small.

Our model adopts the same values for all free parameters in all
clusters.  In particular, this is the case for several global cluster
properties, including the average stellar mass, the average BS 
mass, the average BS lifetime, and the average 
binary semi-major axis.  With the exception of the average stellar
mass, there is no conclusive observational or theoretical
evidence to indicate that these parameters should differ from
cluster-to-cluster, although we cannot rule out this possibility.  
For instance, the distribution of binary orbital parameters could
depend on cluster properties 
like the total mass, density or velocity dispersion
\citep[e.g.][]{sigurdsson93}.  In particular, the central velocity
dispersion should be higher in more massive GCs
\citep[e.g.][]{djorgovski94}, which should correspond to a smaller
binary orbital separation for the hard-soft boundary.  
This could contribute to massive GCs tending to have smaller average
binary orbital separations since soft binaries should not survive
for long in the dense cores of GCs \citep[e.g.][]{heggie03}.  In turn,
this could 
affect the occurrence of mass-transfer events, or of 
mergers during 1+2 and 2+2 encounters.  This last point follows from
the fact that numerical scattering experiments have shown that the 
probability of mergers occurring during 1+2 and 2+2 interactions
increases with decreasing binary orbital separation
\citep[e.g.][]{fregeau04}.  Both the average stellar mass and the
average BS mass (and hence lifetime) could also depend on the total 
cluster mass, as discussed in \citet{leigh09} and \citet{leigh11a}.

We have also neglected to consider the importance of triples for BS
formation throughout our analysis \citep[e.g.][]{perets09} since we
are unaware of any 
observations of triples in GCs in the literature.  Interestingly,
however, our results for binary star evolution can be generalized to
include the internal evolution of triples since they should have the
same functional dependence on the core mass (i.e. N$_{te} \propto
f_tM_{core}$, where N$_{te}$ is the number of BSs formed from triple
star evolution and f$_{t}$ is the fraction of objects that are
triples). 

Finally, our model 
assumes that several parameters remain constant as a function of
the distance from the cluster centre, including the binary fraction
and the average 
semi-major axis.  However, observations of GCs suggest that their
binary fractions could fall off rapidly outside the core
\citep[e.g.][]{sollima07, davis08}.  Our results suggest that, if the
binary fraction is negligible outside the core, then the contribution
from BSs that migrate into the core due to dynamical friction is also
negligible.  This is because the time between 1+1 collisions increases
rapidly outside the core, and every other BS formation mechanism requires
binary stars to operate.  On the other hand, the presently observed
binary fraction outside the core could be low as a result of binaries
having previously migrated into the core due to dynamical friction
\citep[e.g.][]{fregeau09}.  If these are the binary star progenitors
of the BSs currently populating the core, then dynamical friction
remains an important effect in determining the number of BSs
currently populating the core. 

Despite all of these 
simplifying assumptions, we have shown that our model can reproduce the 
observations with remarkable accuracy.  Notwithstanding, the effects
we have discussed 
could be contributing to cluster-to-cluster differences in the
observed BS numbers.  Our model provides a well-suited resource for
addressing the role played by these effects, however future
observations will be needed in order to obtain the desired constraints
(e.g. binary fractions, distributions of binary orbital parameters, etc.).   

\section*{Acknowledgments}

We would like to thank Ata Sarajedini, Aaron Dotter and Roger Cohen
for providing the observations to which we compared our model
predictions, as well as for providing a great deal of guidance in 
analyzing the data.  We would also like to thank Evert Glebbeek, Bob Mathieu
and Aaron Geller for useful discussions.  This research has been
supported by NSERC and OGS. 




\pagestyle{fancy}
\headheight 20pt
\lhead{Ph.D. Thesis --- N. Leigh }
\rhead{McMaster - Physics \& Astronomy}
\chead{}
\lfoot{}
\cfoot{\thepage}
\rfoot{}
\renewcommand{\headrulewidth}{0.1pt}
\renewcommand{\footrulewidth}{0.1pt}

\chapter{A New Quantitative Method for Comparing the Stellar Mass
  Functions in a Large Sample of Star Clusters:  Evidence for a
  Universal Initial Mass Function in Massive Star Clusters in the
  Early Universe} \label{chapter6}
\thispagestyle{fancy}






%

\section{Introduction} \label{intro6}
It is now known that most, if not all, of the stars in our Galaxy were
born in star clusters \citep[e.g.][]{lada95, lada03, mckee07}.  And yet, 
there remain several key details of the star formation process that are still
not understood.  Part of the 
problem lies in the fact that populations of young stars are typically
hidden by a dense veil of optically-thick gas and dust.  This
prevents the escape of most of the light produced by infant stars, and
often renders these regions difficult to observe \citep[e.g.][]{grenier05,
  lada07}.  Most of 
these clusters are sparsely populated and are of relatively low mass (M
$\lesssim$ 10$^4$ M$_{\odot}$) \citep[e.g.][]{lada85}.  They are also
very young since these clusters are unlikely to survive for more than
1 Gyr \citep[e.g.][]{portegieszwart10}.    
At the other end of the mass spectrum, most massive star clusters (M
$\gtrsim$ 10$^4$ M$_{\odot}$) in our Galaxy tend to be at least a few
Gyrs old, and 
in many cases are nearly as old as the Universe itself
\citep[e.g.][]{harris96, deangeli05}.  
Unfortunately, the conditions present at the
time of their formation have been largely erased
\citep[e.g.][]{hurley05, murray09}.  This presents a 
considerable challenge for studying star formation in the regime of
cluster masses and metallicities that characterize Milky Way globular 
clusters.  These old star clusters contain the fossil record of a very
early episode of star formation in the Universe, and are the only
means of studying it locally.

One of the primary
observational tests for star formation theories is the stellar initial 
mass function (IMF).  Current observational evidence suggests that the
IMF is very similar in several different regions of our Galaxy,
including the disk and young star clusters
\citep[e.g.][]{elmegreen99}, however this is still being debated
throughout the literature \citep[e.g.][]{scalo98}.  The arguably
``standard'' IMF to come from these observations is that of
\citet{kroupa01}, who fit a three-part 
power-law with breaks at 0.08 M$_{\odot}$ and 0.5 M$_{\odot}$.  This
can be expressed as:
\begin{equation}
\label{eqn:kroupa}
\frac{dN}{dlm} = {\beta}m^{-\alpha},
\end{equation}
where
$\alpha$ is a constant given by $\alpha = 2.3$ for 0.5 $<$
$m$/M$_{\odot}$ $<$ 50, $\alpha = 1.3$ for 0.08 $<$
$m$/M$_{\odot}$ $<$ 0.5, and $\alpha = 0.3$ for 0.01 $<$
$m$/M$_{\odot}$ $<$ 0.08, and $\beta$ is a constant determined by the
total cluster mass.  By considering the mass function up to only
$\sim$ 1 M$_{\odot}$, \citet{miller79} found a good fit to the
observed mass distribution using a log-normal functional form:
\begin{equation}
\label{eqn:miller}
\frac{dN}{dlnm} \propto exp\Big[-\frac{(lnm-lnm_c)^2}{2\sigma^2}\big],
\end{equation}
where m$_c$ $\sim$ 0.2 M$_{\odot}$ and $\sigma$ $\sim$ 0.55
\citep{chabrier05}.  

Different star formation theories tend to predict different IMFs.  These
tend to vary with the properties of the gas clouds from which the
stars are born \citep[e.g.][]{elmegreen01, bonnell07}.  Given the
sensitive nature of the observations, a
large sample of IMFs spanning the entire range of cluster properties
exhibited by star clusters in the Milky Way, 
including total mass and chemical composition, has yet to be
compiled.  
This is a sorely needed step in order to advance our understanding
of star formation by providing direct comparisons for theoretical 
predictions.  This is especially true of massive (M $\gtrsim$ 10$^4$
M$_{\odot}$), metal-poor star clusters since we are particularly
lacking observations of IMFs in this regime of cluster masses and
metallicites, especially in our own Galaxy \citep[e.g.][]{mckee07,
  portegieszwart10}.   

For the very first time, the ACS Survey for Globular Clusters has
provided photometry that is nearly reliable all the way down to the
hydrogen burning limit in a large sample of Milky Way globular
clusters.  This offers a 
large sample of current stellar mass functions spanning the stellar mass range
$\sim 0.2 - 0.8$ M$_{\odot}$.  All of the
clusters are massive and of very old age, with total masses and ages
ranging from $\sim$ 10$^4$ - 10$^6$ M$_{\odot}$ and $\sim$ 10-12 Gyrs,
respectively 
\citep{harris96, deangeli05}.  This has allowed significant time for
their stellar mass functions to have been modified from their
primordial forms due to both stellar evolution and stellar dynamics.
However, most of the processes responsible for this evolution are now
largely understood.  Therefore, in principle, it is possible to use
current observations of old star clusters together with theoretical models
for their evolution to indirectly probe their IMFs.  

For most of the life of a star cluster, two-body relaxation is the
dominant physical mechanism driving its evolution
\citep[e.g.][]{heggie03, gieles11}.  
The term describes the cumulative effects of long-range gravitational
interactions that occur between pairs of
stars, which act to alter their orbits within the cluster.  Two-body
relaxation also acts to slowly modify the distribution of stellar masses
within clusters.  Among other things, it results in a 
phenomenon known as mass segregation.  This is the tendency for
heavier stars to accumulate in the central cluster regions and
low-mass stars to be dispersed to wider orbits.  This same process
also causes the continual escape of preferentially low-mass stars from
the cluster, which has been confirmed to occur observationally in real
star clusters \citep[e.g.][]{vonhippel98, demarchi10}.  

The time-scale for two-body relaxation to operate can range anywhere from
several million years to the age of the Universe or longer.  The rate
at which it occurs can be roughly approximated using the half-mass
relaxation time, which provides a rough average for the entire
cluster.  This is given by \citep{spitzer87}:  
\begin{equation}
\label{eqn:t-rh6}
t_{rh} = 1.7 \times 10^5[r_h(pc)]^{3/2}N^{1/2}[m/M_{\odot}]^{-1/2}
years,
\end{equation}
where $r_h$ is the half-mass radius (i.e. the radius enclosing half
the mass of the cluster), $N$ is the total number of stars
within $r_h$ and $m$ is the average stellar mass.  The half-mass radii
of MW GCs are remarkably similar independent of 
mass, and simulations have shown that $r_h$ changes by a factor of at
most a few
over the course of a cluster's lifetime \citep{henon73, murray09}.
The GCs that comprise the ACS sample show a range of masses spanning
roughly 3 orders of magnitude (10$^4$-10$^6$ M$_{\odot}$), and have
comparably old ages ($\sim$ 10-12 Gyrs) \citep{deangeli05}.  
Therefore, Equation~\ref{eqn:t-rh6} suggests that the total cluster
mass provides a rough proxy for the degree of
dynamical evolution (due to two-body relaxation).  In other words, the
effects of two-body relaxation on the evolution of 
the stellar mass function should be the most pronounced in the least
massive clusters in the ACS sample.  Said another way, dynamical
age increases with decreasing cluster mass.

In this chapter, we present a new technique to quantify
cluster-to-cluster variations 
in the observed stellar mass functions of a large sample of clusters
spanning a diverse range of properties.  
Our sample consists of 33 Milky Way globular clusters taken from the
ACS Survey for Globular Clusters \citep{sarajedini07}.  
With it, we constrain the universality of the IMF in a large
sample of old, metal-poor star 
clusters spanning a wider range of masses than ever before considered.  
In Section~\ref{method6}, we present our sample of stellar mass
functions and describe our technique.  The results of our analysis of
the ACS observations are presented in Section~\ref{results6}.  
Finally, we discuss the implications of our results for the universality of the
stellar IMF in Section~\ref{discussion6}, and describe how our results can be
compared to theoretical models for GC evolution. 

\section{Method} \label{method6}

In this section, we describe how we acquire our sample of mass
functions from the ACS data.

\subsection{The Data} \label{data6}

The data used in this study consists of a sample of 33 MW GCs taken
from the ACS Survey for Globular Clusters
\citep{sarajedini07}.\footnote[1]{The
data can be found at http://www.astro.ufl.edu/~ata/public\_hstgc/.}
The ACS Survey provides unprecedented deep photometry in the F606W ($\sim$
V) and F814W ($\sim$ I) filters
that is nearly complete down to $\sim 0.2$ M$_{\odot}$.  In other
words, the colour-magnitude diagrams (CMDs) extend reliably from the
HB all the way down to about 7 magnitudes below the main-sequence
turn-off (MSTO).  

Each cluster was centred in the ACS field, which
extends out to several core radii from the cluster
centre in most clusters.  Coordinates for the cluster centres were
taken from 
\citet{goldsbury10}.  These authors found their centres by fitting
a series of ellipses to the density distributions within the inner 2'
of the cluster centre, and computing an average value.  The core
radii were taken from \citet{harris96}.

\subsection{Mass Bin Selection Criteria} \label{criteria6}

In order to select the number of stars belonging to each mass bin, 
we fit theoretical isochrones taken from \citet{dotter07}
to the CMDs of every cluster in our sample.  Each isochrone
was generated using the metallicity and age of the cluster, and fit to
its CMD using the corresponding distance modulus and extinction
provided in \citet{dotter10}.
The MSTO was then defined using our isochrone fits by selecting the
bluest point along the MS.

We considered five mass bins along the main-sequence.  These ranged
from 0.25 - 0.75 M$_{\odot}$ in increments of 0.1 M$_{\odot}$.  This
range was chosen to ensure complete sampling in all bins since the
lowest MSTO mass in our sample corresponds to $\sim$ 0.75 M$_{\odot}$,
and the photometric errors remain small ($\lesssim$ 0.05 mag) within
the corresponding 
magnitude range for each cluster.  We have obtained number counts for
all mass bins within the core, as well as for within two and three core
radii.  We do not consider circles outside this since the spatial
coverage becomes incomplete for several clusters.  This greatly
reduces our sample size and causes the statistical significance of our
analysis to suffer.

We have obtained completeness fractions for each mass bin in all
three annuli for every cluster in our sample.  This was
done using the results of artificial star tests taken from
\citet{anderson08}.\footnote[2]{Artificial star tests were obtained
  directly from Ata
Sarajedini via private communication.}  Number counts for each mass
bin were then multiplied by their corresponding completeness
corrections.  The field of view of the ACS images is about
200'' on a side, which gives physical scales ranging between 1.5 and
16 pc (for the closest and furthest clusters in our sample).  Based on
this, we expect foreground contamination by field stars to be
negligible for most of the clusters in our sample given their current
locations in the Galaxy.  For example, \citet{dacosta82} considered
star count data 
in a similar area and over a comparable range of stellar masses for
three nearby globular clusters.  The author found that the 
corrections resulting from field contamination were always less than
10\% over nearly the entire range of stellar masses we are
considering.   



\subsection{Weighted Lines of Best-Fit} \label{lines}

In order to quantify cluster-to-cluster differences in the current
stellar mass functions of the clusters in our sample, we have obtained
lines of best-fit for 
(the logarithm of) the number of stars belonging to each mass bin
versus (the logarithm of) the total number of stars spanning all five
mass bins (which provides a rough proxy for the total cluster mass).
These lines have been weighted by adopting uncertainties for the
number of stars in each mass bin using Poisson statistics.  
We have performed this comparison for all three circles (i.e. within
one, two and three core radii).  Our motivation for adopting this
technique is as follows.  If the fraction of stars belonging to each
mass bin 
(relative to the total number of stars in all five mass bins), which
we denote by f$_{m_1-m_2}$, is
constant for all cluster masses, then we would expect the number of
stars in 
each mass bin N$_{m_1-m_2}$ to scale linearly with the total number of stars
spanning all five mass bins N$_{tot}$ (since f$_{m_1-m_2} =$
N$_{m_1-m_2}$/N$_{tot}$).  However, if there is any systematic
dependence of f$_{m_1-m_2}$ on the
total cluster mass, then we should find that N$_{m_1-m_2}$ does
\textit{not} scale linearly with N$_{tot}$.  In particular, the
power-law index for a given mass bin should be sub-linear if the
fraction of stars belonging to that mass bin systematically decreases
with increasing cluster mass.  Conversely, the
power-law index for a given mass bin should be super-linear if the
fraction of stars belonging to that mass bin systematically increases
with increasing cluster mass.  The slopes and y-intercepts
corresponding to the weighted lines of best-fit for each mass bin
provide a means of directly quantifying the number of stars belonging
to each mass bin as a function of the total cluster mass.  In other
words, our method provides a means of quantifying cluster-to-cluster
differences in the stellar mass function as a function of the total
cluster mass.

We have chosen to count the number of stars belonging to each mass bin
directly from the observations in order to quantify the dependence of
the fraction of stars belonging to each mass bin on the total cluster
mass (or, more specifically, the total number of stars spanning all
mass bins).  We note that an alternative, albeit considerably less
precise, 
way to go about this would be to characterize the stellar mass
functions using a log-normal function of the form given by
Equation~\ref{eqn:miller}.  Let this function be represented by
$\Psi$(m) = dN/dlnm.  We can normalize the distribution of stellar
masses over the entire mass range of interest (0.25-0.75 M$_{\odot}$)
by setting: 
\begin{equation}
\label{eqn:psi_norm}
\int_{m_{min}}^{m_{max}} \Psi(m) dlnm = 1,
\end{equation}
where, in our case, m$_{min} =$ 0.25 M$_{\odot}$ and m$_{max} =$ 0.75
M$_{\odot}$.  The fraction of stars belonging to a given mass bin
would then be given by:
\begin{equation}
\label{eqn:psi}
f_{m_1-m_2} = \int_{m_1}^{m_2} \Psi(m) dlnm,
\end{equation}
where m$_1$ and m$_2$ are the lower and upper mass limits,
respectively, of the mass bin under consideration.  


The reason why our technique provides a means of quantifying the
universality of the IMF, in addition to the effects had by 
two-body relaxation in modifying it, can be
understood as follows.  If we assume that all things are equal, such
as tidal effects from the Galaxy and the degree of primordial mass
segregation, two-body 
relaxation operates the fastest in low-mass clusters.  All of the
clusters in our sample are of comparably old age, which implies that the
dynamical age increases with decreasing cluster mass.  With increasing
dynamical age, the stellar mass function should be more severely
depleted of preferentially low-mass stars due to stellar evaporation
induced by two-body relaxation.  Based on this,
if all of the clusters in our sample were born with similar IMFs, we
would expect the slopes of their lines of best-fit to systematically
increase with decreasing 
mass bin.  This is because low-mass clusters should be more depleted
of their low-mass stars.  
Therefore, the weighted lines of best-fit and, in particular their
corresponding uncertainties, provide a means of quantifying
the universality of the stellar IMF in the range of cluster masses and
chemical compositions characteristic of Milky Way globular clusters.
Large uncertainties caused by a large degree of scatter could be
interpreted as evidence against a universal IMF.  This is because we
would only expect a systematic dependence of the number of stars
belonging to each mass bin on the total cluster mass, and
therefore small uncertainties for the slopes and y-intercepts
corresponding to their lines of best-fit, if all clusters
began with very similar IMFs in the first place.  We note that
although our method constrains the \textit{universality} of the
stellar IMF, it does not constrain its precise \textit{functional
  form}.  The reason for this is that we do not know how much dynamical
evolution has actually occurred in the clusters in our sample.
Therefore, we do not know how much the stellar mass function has been
modified by two-body relaxation.  However, as we will describe in
Section~\ref{discussion6}, our observational analysis is ideally suited
for comparison to theoretical models for globular cluster evolution,
and this will allow us to constrain the exact shape of the IMF, in
addition to the degree of primordial mass segregation.

Finally, we note that mass segregation
should also contribute to the systematic dependence on cluster mass we
have described in the previous paragraph for the slopes and
y-intercepts corresponding to the 
lines of best-fit for each mass bin.  This effect should be the most
severe for small circles (centred on the cluster centre), and should
become less important as increasingly larger circles are considered.
This is because as we consider progressively larger fractional areas
of the cluster, the number counts for each mass bin become less
sensitive to the stars' spatial distributions throughout the cluster,
and therefore less sensitive to the effects of mass segregation.

\section{Results} \label{results6}

In Figure~\ref{fig:Ncore_vs_Nms_rc}, Figure~\ref{fig:Ncore_vs_Nms_2rc}
and Figure~\ref{fig:Ncore_vs_Nms_3rc} we plot the number of
stars in each mass bin versus the total number of stars spanning all five
mass bins.  Slopes (a) and y-intercepts (b) for the weighted lines of best-fit
performed for each of these relations are shown in Table~\ref{table:bestfit6},
along with their corresponding uncertainties.  These were 
found using a bootstrap methodology in which we generated 1,000 fake
data sets by randomly sampling (with replacement) number counts from
the observations.  We obtained lines of best fit for each fake data
set, fit a Gaussian to the subsequent distribution and extracted its
standard deviation.  We have found that our results are insensitive to
both changes in bin width and bin centering.  Specifically, our slopes
and y-intercepts remain consistent to within one standard deviation
upon adjusting either the bin width or centering by up to roughly half
a bin width.

\begin{figure} [!h]
  \begin{center}
 \includegraphics[scale=0.5]{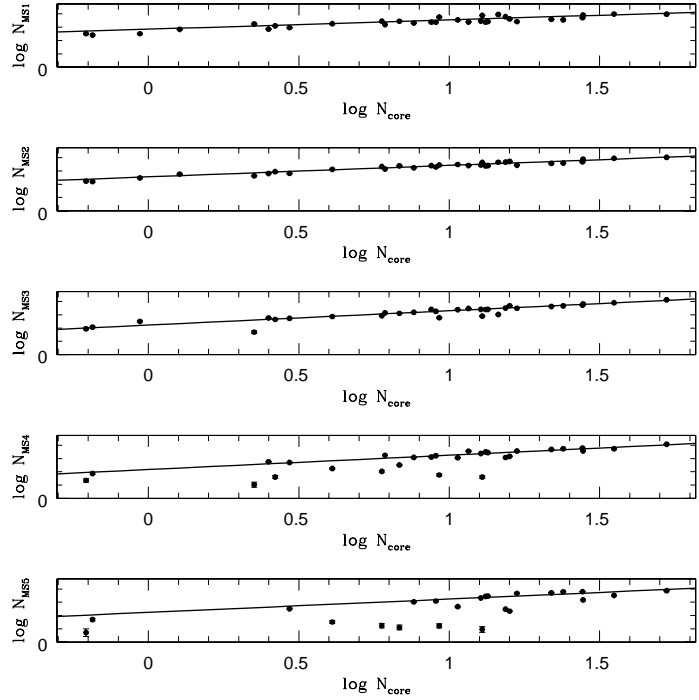}
\caption[Logarithm of the number of
stars belonging to each mass bin as a function of the
logarithm of the total number of stars spanning all five mass
bins in the core]{The logarithm of the number of
stars belonging to each mass bin as a function of the
logarithm of the total number of stars spanning all five mass
bins in the core.  MS1 corresponds to the mass range 0.65 - 0.75 M$_{\odot}$, MS2
to 0.55 - 0.65 M$_{\odot}$, MS3 to 0.45 - 0.55 M$_{\odot}$, MS4 to
0.35 - 0.45 M$_{\odot}$, and MS5 to 0.25 - 0.35 M$_{\odot}$.  Lines of
best fit are shown for each mass bin. 
\label{fig:Ncore_vs_Nms_rc}}
\end{center}
\end{figure}

\begin{figure} [!h]
  \begin{center}
 \includegraphics[scale=0.5]{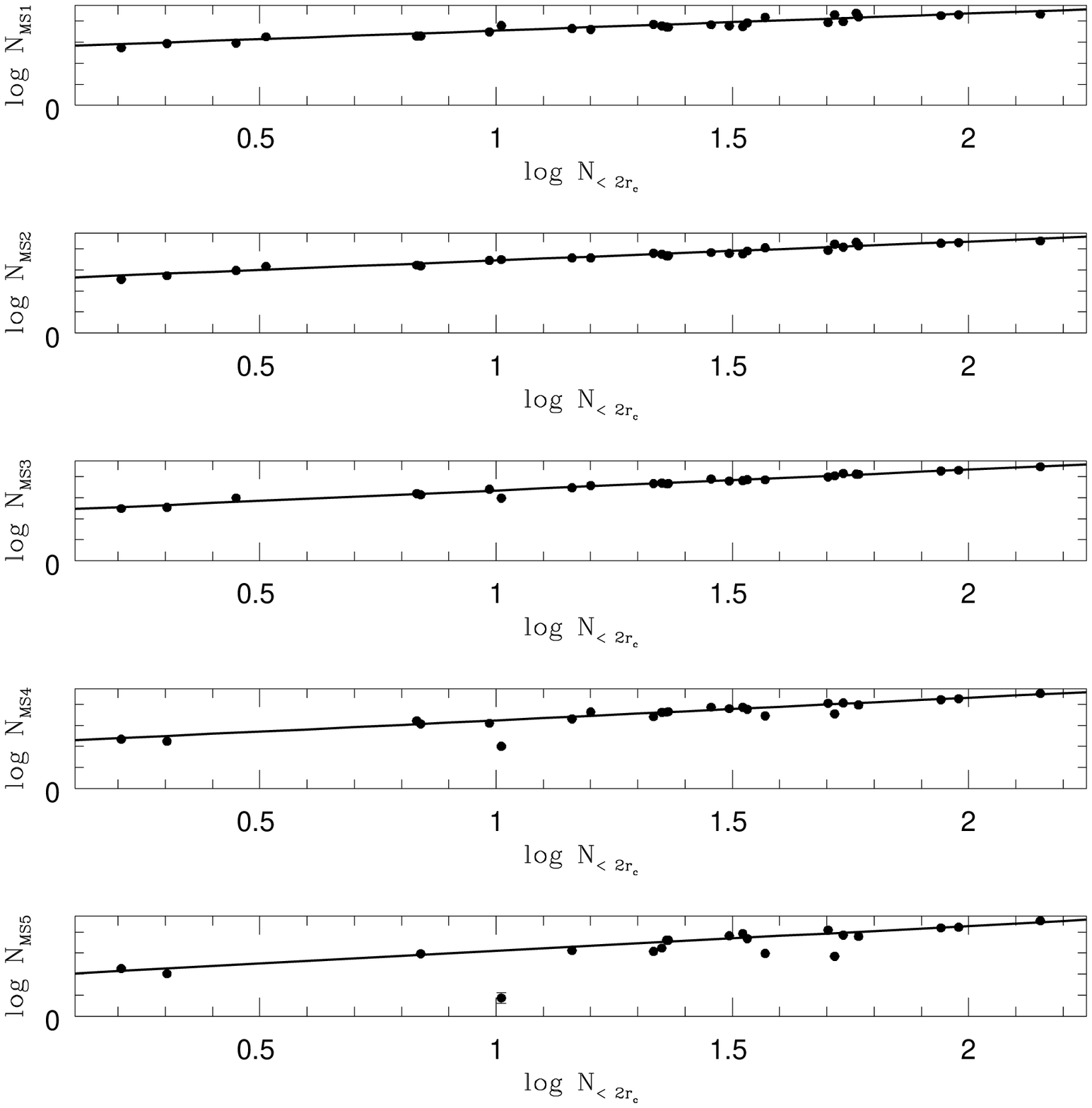}
\caption[Logarithm of the number of
stars belonging to each mass bin as a function of the
logarithm of the total number of stars spanning all five mass
bins within two core radii from the cluster centre]{The logarithm of the number of
stars belonging to each mass bin as a function of the
logarithm of the total number of stars spanning all five mass
bins within two core radii from the cluster centre.  The mass bins are
the same as in Figure~\ref{fig:Ncore_vs_Nms_rc}.
\label{fig:Ncore_vs_Nms_2rc}}
\end{center}
\end{figure}

\begin{figure} [!h]
  \begin{center}
 \includegraphics[scale=0.5]{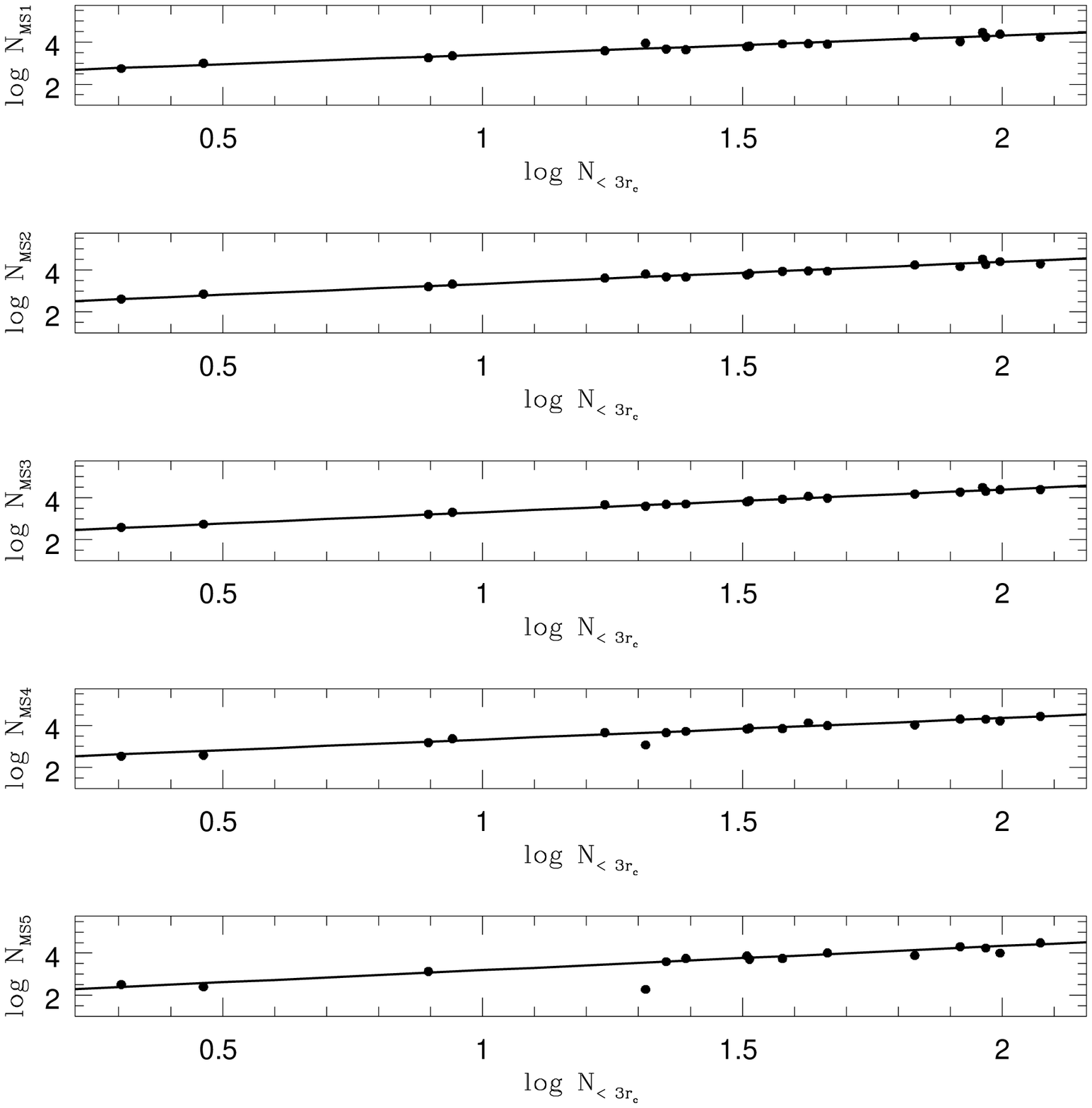}
 \caption[Logarithm of the number of
 stars belonging to each mass bin as a function of the
 logarithm of the total number of stars spanning all five mass
 bins within three core radii from the cluster centre]{The logarithm of
  the number of
stars belonging to each mass bin as a function of the
logarithm of the total number of stars spanning all five mass
bins within three core radii from the cluster centre.  The mass bins are
the same as in Figure~\ref{fig:Ncore_vs_Nms_rc}.
\label{fig:Ncore_vs_Nms_3rc}}
\end{center}
\end{figure}

\begin{sidewaystable}
\tiny
\centering
\caption{Lines of Best Fit for log N$_{MS}$ = (a $\pm$ $\Delta$a)log (N$_{tot}$/10$^3$) + (b $\pm$ $\Delta$b)
  \label{table:bestfit6}}
\begin{tabular}{|l|cccc|cccc|cccc|cccc|cccc|}
\hline
      &\multicolumn{4}{c|}{   MS1 (0.65-0.75 M$_{\odot}$)  }&\multicolumn{4}{c|}{   MS2 (0.55-0.65 M$_{\odot}$)  }&\multicolumn{4}{c|}{   MS3 (0.45-0.55 M$_{\odot}$)  }&\multicolumn{4}{c|}{   MS4 (0.35-0.45 M$_{\odot}$)  }&\multicolumn{4}{c|}{   MS5 (0.25-0.35 M$_{\odot}$)  }\\
\hline
Circle & a & $\Delta$a & b & $\Delta$b & a &$\Delta$a & b & $\Delta$b
& a & $\Delta$a & b & $\Delta$b & a & $\Delta$a & b & $\Delta$b & a &
$\Delta$a & b & $\Delta$b \\
\hline
$<$ r$_c$    & 0.70 & 0.07 & 2.85 & 0.09 & 0.86 & 0.05 & 2.56 & 0.05 & 1.08 & 0.05 & 2.22 & 0.07 & 1.08 & 0.09 & 2.17 & 0.12 & 1.00 & 0.20 & 2.24 & 0.25 \\
$<$ 2r$_c$   & 0.81 & 0.08 & 2.74 & 0.12 & 0.90 & 0.06 & 2.55 & 0.09 & 0.99 & 0.03 & 2.35 & 0.05 & 1.07 & 0.06 & 2.17 & 0.10 & 1.19 & 0.10 & 1.91 & 0.18 \\
$<$ 3r$_c$   & 0.91 & 0.12 & 2.50 & 0.19 & 1.04 & 0.11 & 2.30 & 0.17 & 1.08 & 0.08 & 2.23 & 0.12 & 1.02 & 0.05 & 2.31 & 0.10 & 1.15 & 0.10 & 2.04 & 0.16 \\
%
\hline
\end{tabular}
\end{sidewaystable}

As shown in Table~\ref{table:bestfit6}, the slopes tend to
systematically increase 
with decreasing mass bin.  For the comparison within two core radii
(shown in Figure~\ref{fig:Ncore_vs_Nms_2rc}), 
this trend applies to all mass bins.  For the comparisons within the core
and within three core radii (shown in Figure~\ref{fig:Ncore_vs_Nms_rc}
and Figure~\ref{fig:Ncore_vs_Nms_3rc}, respectively), however, only
the highest three mass bins (MS1, MS2, and MS3) 
follow this trend of increasing slope with decreasing mass bin.  After
the third mass bin, the slopes for the lowest two mass bins remain
about the same.  Note, however, that the uncertainties on the slopes
and y-intercepts are also higher for the lowest mass bins (MS4 and MS5).  
%
This is
likely to be due to the fact that the photometric errors are the
highest at these dim magnitudes, however they remain at most $\sim$
10\% of the width in magnitude of their corresponding mass bins.  The
completeness corrections are also the largest for MS4 and MS5,
and this also introduces additional uncertainty.  Although we have
taken the appropriate measures, it
is important that these effects be properly quantified in order to reliably
use our technique to constrain the degree of universality of the IMF.
We will return to this in Section~\ref{discussion6}.  

On the other hand, when we consider only the largest three mass bins
(MS1, MS2, and MS3) for which 
the photometric errors and level of incompleteness are the lowest, the
difference in slopes and y-intercepts between adjacent mass bins can
differ at better than the $2-\sigma$ confidence level.  Moreover, if we
compare non-adjacent mass bins, then this trend is typically
significant at the $3-\sigma$ confidence level.  In order to improve
upon these statistics, we have also calculated reduced chi-squared
values with added intrinsic dispersion for the relations for each mass
bin.  That is, for each mass bin we added a constant term to the uncertainty for each
data point, found the value that yielded a reduced chi-squared of one,
and looked at the subsequent effects on the uncertainties for the
line of best-fit.  Based on this, we appear to be slightly over-estimating the
uncertainties for the MS1, MS2 and MS3 mass bins using our bootstrap
approach, and slightly under-estimating them for the MS4 and MS5
bins.  This suggests that the slopes and y-intercepts differ at nearly
the $2-\sigma$ confidence level for the MS1, MS2 and MS3 mass bins,
but are consistent to within one standard deviation for the MS4 and
MS5 bins.  We conclude that our results are the most reliable (at nearly the
$2-\sigma$ confidence level) in the mass range 0.45-0.75 M$_{\odot}$. 

The change in the distribution of stellar masses as a function of the
total cluster mass can be illustrated using pie charts, as shown in
Figure~\ref{fig:pie_charts}.  Using the slopes and y-intercepts
provided in Table~\ref{table:bestfit6} for the comparison within two core
radii, we have generated pie charts for three total numbers of stars
(spanning all five mass bins), namely N$_{tot}$ = 10$^3$, 10$^4$,
10$^5$.  As is clear, low-mass stars are preferentially depleted, and
this effect becomes increasingly severe with decreasing total cluster
mass (or, equivalently, increasing dynamical age).  From right to
left, what the pie charts are showing are the mass
functions of progressively dynamically older clusters.  
If this is indeed the cause of the observed depletion of low-mass
stars in low-mass clusters, we are effectively looking at the
evolution of the stellar mass function in time.   

\begin{figure} [!h]
  \begin{center}
 \includegraphics[scale=0.5]{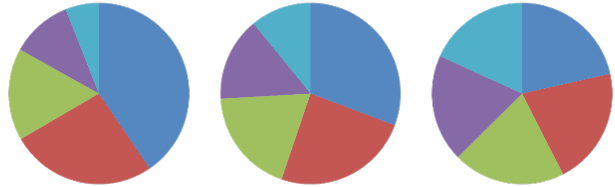}
    \caption[Mass Functions in Pie Chart Form]{
      Stellar mass functions depicted in pie chart form.  The total
      area of each circle corresponds to the total number of stars
      spanning all five mass bins, and each pie slice shows the
      fraction of this total corresponding to each mass bin.  Each of
      these fractions was calculated using the weighted lines of
      best-fit provided in Table~\ref{table:bestfit6}.  From left to
      right, the total number of stars used to generate each pie chart
      was 10$^3$, 10$^4$, and 10$^5$.  Dark blue corresponds to MS1,
      red to MS2, green to MS3, purple to MS4, and light blue to MS5.
      From right to left, the pie charts effectively show the mass functions of
      progressively dynamically older clusters.
      \label{fig:pie_charts}}
    \end{center}
\end{figure}

\section{Summary \& Discussion} \label{discussion6}

In this chapter, we have obtained completeness-corrected stellar mass
functions in the range 0.25-0.75 M$_{\odot}$ for a sample of 33
globular clusters using data taken from the ACS Survey for Globular
Clusters.  We have presented a new technique to quantify
cluster-to-cluster variations in the observed stellar mass functions.  
We have shown how our method can be used to quantify the universality
of the IMF for a large sample of clusters and, when used in
conjunction with theoretical models for globular cluster evolution,
can be used to constrain the degree to which two-body relaxation has
modified the currently observed stellar mass functions from their 
primordial form.  To this end, we have
obtained weighted lines of best-fit by comparing the number of stars in five
different mass bins to the total number of stars spanning the
entire mass range within several circles centred on the cluster
centre.  

As shown in Table~\ref{table:bestfit6}, our results suggest that the slopes
for the lines of best-fit tend to systematically increase
with decreasing mass bin.  Assuming all of the clusters in our sample
were born with similar IMFs, this is precisely what we expect from
two-body relaxation.  That is, we expect low-mass stars to become
preferentially depleted via stellar evaporation, and we expect this
effect to be the most severe in low-mass clusters.  This trend is
clearly seen in the observations, as demonstrated in 
Figure~\ref{fig:pie_charts}.  Interestingly, the fraction of stars
belonging to each mass bin is remarkably constant for all mass bins at
the high-mass end of our sample.  The total masses of these clusters
are sufficiently high that their half-mass relaxation times are
comparable to their ages \citep[e.g.][]{harris96}.  Therefore, we
expect that the highest mass clusters in our sample should have
been relatively unaffected by stellar evaporation due to two-body
relaxation.  It follows that their current mass functions should be
relatively unchanged from their primordial values (ignoring the
effects of stellar evolution).  If true, our
results could suggest that the slope IMF in Equation~\ref{eqn:kroupa}
was consistent with zero in the
mass range 0.25-0.75 M$_{\odot}$ for all of the clusters in our
sample.  This would be inconsistent with a ``standard'' Kroupa IMF.  This
hypothesis can be tested by comparing the results of the 
observational analysis presented in this chapter to simulations of
globular cluster evolution.  These can be used to directly quantify
the effects of two-body relaxation on the expected slopes and
y-intercepts for each mass bin given any set of initial conditions and
IMFs.  By comparing the results of these simulations to the results of
our observational analysis, the set of initial conditions that
best reproduces the observations can be found.  This will tell us
about the precise functional form of the IMF for the clusters in our
sample, in addition to the degree of primordial mass segregation.

The uncertainties for the slopes and 
y-intercepts are sufficiently large that they are often consistent with
those of their adjacent mass bins to within one standard deviation.
However, when we consider only the largest three mass bins for which
the photometric errors and level of incompleteness are the lowest, the
difference in slopes and y-intercepts between adjacent mass bins can
differ at better than the $2-\sigma$ confidence level.  Moreover, the
trend of increasing slope with decreasing mass is typically
significant at the $3-\sigma$ confidence level if we compare
non-adjacent mass bins.  Based on this, we conclude that our results
are consistent with a universal IMF for the clusters in our sample,
and that the currently observed mass functions have primarily been 
modified by internal two-body relaxation.  Having said that, it is
important to note that a higher degree of primordial mass segregation
effectively acts
to increase the rate of dynamical evolution and therefore stellar
evaporation \citep[e.g.][]{heggie03}.  Consequently, our results could
also be consistent with a non-universal IMF that depends on the total
cluster mass, provided the degree of primordial mass segregation also
depends systematically on the total cluster mass.  We also wish to
point out here 
that the large uncertainties found for the slopes and y-intercepts,
particularly for the two lowest mass bins, are primarily due to the
relatively large photometric errors at these dim magnitudes and
incompleteness resulting from crowding.  Despite the high quality of
the data used in this study, these issues are currently unavoidable
given the nature of the observations.  This will be a key challenge
for future studies to resolve, however the method we have presented
in this chapter offers a robust means of performing future analyses.

In terms of addressing the \textit{degree} of universality of the IMF,
our technique offers a considerable advantage over the standard
power-law and log-normal forms used to characterize the stellar mass
function shown in Equation~\ref{eqn:kroupa} and
Equation~\ref{eqn:miller}.  This is the case when comparing the mass
functions of a large sample of clusters.  The reason for this is that
our method
segments the mass function into mass bins, which characterizes
cluster-to-cluster differences in the stellar mass function for each
mass bin individually (i.e. as a function of stellar mass).  This
is done using the slopes and y-intercepts of the
relations found by correlating the number of stars in each mass bin
with the total number of stars spanning all mass bins.  Conversely,
standard Kroupa or Chabrier mass functions quantify the mass function
much more globally via functional fits to the data over considerably
larger mass ranges.  Of course, the standard forms
remain as useful as ever for characterizing the mass functions of
individual clusters, whereas our method is not appropriate for
this purpose.  That being
said, it is worth stating here that our results 
are consistent with those of \citet{demarchi10}, who fit a tapered
power-law distribution function with an exponential truncation to the
stellar mass functions of a sample of 30 clusters containing both
young and old members. 

There are several possible sources of additional uncertainty that
could have affected our analysis.  For 
instance, we have not yet considered the role played by binaries.
These tend to be unresolved in GCs, usually appearing as single
objects located above the MS in the 
cluster CMD.  Therefore, some of the objects in each mass bin are in
fact binaries masquerading as single stars.  Typical binary fractions
in Milky Way GCs are thought to be on the order of a few to a few tens
of a percent \citep[e.g.][]{rubenstein97, cool02, sollima08,
  davis08}.  This suggests that a non-negligible number of binaries
could often be included in our number counts for each mass bin.
However, there is no reason not to expect that binaries will
contribute to each mass bin in 
roughly equal proportions.  If the fraction of binaries is the same in
all mass bins, then the presence of binaries should not have
significantly affected the slopes.  Although they will have
affected the y-intercepts, this effect should not cause them to deviate
from within the uncertainties and 
should not have affected the interpretation of our results.  On the 
other hand, some observational evidence suggests that
the binary fraction could be inversely proportional to the total
cluster mass \citep[e.g.][]{sollima08, milone08, knigge09}.  It is not
clear how this might have affected our results since we do not know
how each mass bin should be affected.  It is our intent to address
this issue in a future study once binary fractions become available for
the majority of the clusters in our sample (Ata Sarajedini; private
communication).

Throughout our analysis, we have 
consistently compared two projected quantities.  Therefore, effects
related to projection should not have significantly affected
our results.  Moreover, these effects should become less severe upon
considering progressively larger circles, and we have performed 
our analysis for circles with three different sizes.  We note that
this could perhaps help to explain why 
the scatter in the lowest mass bins appears to become reduced for the
comparisons within two and three core radii (compared to the
comparison within one core radius), although it is not clear
why the agreement appears slightly better for the former comparison.  

Tidal effects from the Galaxy effectively act to reduce the time-scale
for two-body relaxation \citep[e.g.][]{heggie03}.  The same effect can also be had by
increasing the degree of primordial mass segregation
\citep[e.g.][]{spitzer87, portegieszwart01}.  Therefore, on average,
we would expect clusters with smaller Galactocentric radii and higher
initial concentrations to have experienced a higher degree of stellar
evaporation.  In an effort to quantify tidal effects from the Galaxy,
we performed several cuts in perigalacticon distance \citep{dinescu99,
  dinescu07} and re-performed our weighted lines of best-fit.  Despite
removing clusters from our sample with small perigalacticon distances
for which it is typically argued that tidal effects should be the most
severe \citep[e.g.][]{heggie03}, our slopes and 
y-intercepts, in addition to their uncertainties, remain more or less
unchanged.  This can be interpreted as rough evidence that tidal effects
from the Galaxy have not 
significantly affected our results.  It is much less clear which
clusters in our sample, if any, were born with high central 
concentrations.  Based on current observations, there is no
known reason not to expect a universal degree of primordial mass
segregation \citep[e.g.][]{portegieszwart10}.  If this was the case,
then our results could be interpreted as indicative of a
universal IMF for Milky Way globular clusters.  Unfortunately,
observational constraints for primordial 
binary populations are also lacking \citep[e.g.][]{mckee07}.  Binaries
play an important 
role in star cluster evolution by, for instance, providing an energy
source that serves to resist a cluster's tendency toward increasing 
its central density \citep{hut83}.  They are therefore an important consideration in
deciding the dynamical evolution of the stellar mass function, and
their role will need to be addressed in future studies.  

It is interesting and even surprising that, despite all of the
aforementioned factors 
expected to affect the dynamical evolution of globular clusters, we
observe relatively little scatter in the relations within two and
three core radii (Figure~\ref{fig:Ncore_vs_Nms_2rc} and
Figure~\ref{fig:Ncore_vs_Nms_3rc}).  Moreover, the slopes
systematically increase with decreasing mass bin.  Based on this, our
results are consistent with a single universal IMF for all of the
clusters in our sample that has been modified via internal two-body
relaxation by an amount determined by the total cluster mass.  
%
This is a new result given 
the old ages and therefore low metallicities ([Fe/H] $\sim$ -2.28 -
(-0.37)) of the clusters that comprise our sample.  However, the  
exact form of the IMF required to reproduce the current observations is
still not clear, nor is the role of primordial mass segregation.  To
what degree have low-mass clusters been depleted of 
their low-mass stars?  What combination of IMFs, Galactocentric
radii, initial concentrations and primordial binary fractions 
should evolve dynamically to best reproduce the currently observed
mass functions?  These questions can only be answered using 
simulations of globular cluster evolution that explore a range of
initial conditions and IMFs.  Our observational analysis of
the ACS data is ideally suited for comparison to these types of
models.  This would allow us to constrain the precise shape of the IMF
in the mass and metallicity range of interest, as well as to learn about
primordial mass segregation.  This will be the focus of a forthcoming
study. 

%

\section*{Acknowledgments}

We would like to thank Ata Sarajedini, Aaron Dotter and Roger Cohen
for providing the data on which this study is based and for their
extensive support in its analysis.  We would also like to thank Evert
Glebbeek for useful discussions.  This research has been
supported by NSERC and OGS.





\pagestyle{fancy}
\headheight 20pt
\lhead{Ph.D. Thesis --- N. Leigh }
\rhead{McMaster - Physics \& Astronomy}
\chead{}
\lfoot{}
\cfoot{\thepage}
\rfoot{}
\renewcommand{\headrulewidth}{0.1pt}
\renewcommand{\footrulewidth}{0.1pt}

\chapter{Summary \& Future Work} \label{chapter7}
\thispagestyle{fancy}

%
%
Within the last few decades, theoretical work has painted a
comprehensive picture for the 
various forms of gravitational interactions operating within star
clusters.  
And yet, direct observational confirmation that many of these processes
are actually occurring is still lacking.  
The results presented in 
this thesis have connected several of these processes to real
observations of star clusters, in many cases for the first time.  This
has allowed us to directly link the observed
properties of several peculiar stellar populations to the physical
processes responsible for their origins.  Given the importance of star
clusters for star formation, this also represents a key step toward
re-constructing the history of our Galaxy.  

In Chapter~\ref{chapter2}, we have presented a new adaptation of the
mean free path approximation.  Our technique provides analytic
time-scales for the rate of close gravitational encounters between
single, binary and triple stars in dense star systems.  With it, we
showed that encounters involving triple stars occur more commonly than
any other type of dynamical interaction in at least some star
clusters, and that these could be the dominant dynamical production
mechanism for stellar mergers.  This is a new result with important
implications for both star cluster evolution and the formation of
several types of stellar exotica.  Our method
can be generalized for application to systems composed of several
different types of particles that evolve under the influence of any
force-mediated interactions.  For example, our technique is
well suited for application to collisions between atoms and dust
grains in the interstellar medium (ISM) \citep[e.g.][]{spitzer41a,
  spitzer41b, spitzer42}.  In this case, the relevant forces governing 
the dynamics are electromagnetic instead of gravitational, however the
general form adopted for the mean free path approximation remains the same.
Specifically, we must only replace the gravitationally-focused
collisional cross-sections with their charge-focused analogues.
Recent observations have provided 
estimates for the concentrations of the different constituents of the
ISM, in addition to their temperatures \citep[e.g.][]{delburgo03,
  kiss06}.  These data provide the required ingredients 
to obtain analytic estimates for the rates of collisions between
atoms, molecules, as well as both small and large dust grains in the
ISM, and would allow us to make
predictions for the evolution of their relative concentrations.  This
is an important step in the development of our understanding of dust
grain coagulation and, by extension, the initial phases of star
formation in dense molecular clouds \citep[e.g.][]{mckee07}, the
interaction of winds from
evolved stars with the surrounding ISM \citep[e.g.][]{glassgold96},
and planet formation in protoplanetary disks \citep[e.g.][]{absil10}.

In Chapter~\ref{chapter3}, we introduced a statistical technique to
compare the relative sizes of different populations of stars in a
large sample of star clusters.  
We refined this technique in 
Chapter~\ref{chapter4}, and applied it to a large sample of clusters
collected using state-of-the-art observations.  Our results suggest
that dynamical effects do not significantly affect the relative sizes
of the different stellar populations.  This is the case for all
stellar populations above the main-sequence turn-off in the cluster
CMD.  Blue stragglers alone present a possible exception to this,
since we have identified a general trend in which more massive
clusters are home to proportionately smaller blue straggler
populations.  This provides compelling evidence in favour of a binary origin
for blue stragglers in even the dense 
cores of massive star clusters where direct collisions between stars are
expected to occur frequently \citep{knigge09}.  Although we have applied these
techniques to observations of dense star clusters throughout 
this thesis, they can be generalized for application to a number of
other studies related to population statistics.  For example, massive
elliptical galaxies have been identified in Galaxy Clusters and
Groups, and these could have formed from the mergers of smaller
galaxies.  Observational studies have now been 
performed that are allowing for the compilation of catalogues
providing population statistics for several different galaxy types in
a large number of Galaxy Clusters and Groups \citep[e.g.][]{abell58,
  abell89}.  Our technique is
well suited for application to these data, and would allow for a
systematic comparison between the relative population sizes of the
different galaxy types.  Among other things, this
would help to constrain the origins of massive elliptical galaxies in
Galaxy Clusters in
much the same way we have constrained the origins of BSs in globular
clusters.
In Chapter~\ref{chapter5}, we present an analytic model for blue
straggler formation in globular clusters.  Our model considers the
production of blue stragglers throughout the entire cluster, and
tracks their accumulation in the core due to dynamical friction (or,
equivalently, two-body relaxation).  Our results support the
conclusion that blue stragglers are descended from binary stars, as
first reported in \citet{knigge09} using the technique presented in
Chapter~\ref{chapter3} and Chapter~\ref{chapter4}.  Our model is applicable to
studying the radial distributions of other stellar and binary
populations in globular clusters.  This is a sorely needed theoretical
tool that can be used to address a number of recent observational 
studies that reported peculiarities among the radial distributions of
several different stellar and binary populations
\citep[e.g.][]{rood73, fusipecci93, ferraro04, lanzoni07}.

Finally, in Chapter~\ref{chapter6}, we extend the statistical
technique introduced in 
Chapter~\ref{chapter3} and Chapter~\ref{chapter4} to include the
entire main-sequence.  By applying our method to the ACS data, we have
obtained the first quantitative constraints for the degree of
universality of the stellar initial mass function for a large sample
of star clusters spanning a wider range of masses and chemical
compositions than ever before considered.  Given the very old ages of
the clusters in our sample, our results have important
implications for our understanding of star formation in the early
Universe.  Our results are 
consistent with a remarkably universal IMF in old
massive star clusters, and are well suited for
comparison to theoretical simulations of globular cluster evolution.
This would provide a simple means of directly quantifying
the extent to which the stellar mass function has been modified from
its primordial form by two-body relaxation.  In future studies, this
will allow us to obtain the very first constraints for the precise
functional form of the IMF and the degree of
primordial mass segregation in a large sample of old star clusters.

The results presented in this thesis have significantly advanced our
understanding of stellar dynamics, 
stellar evolution and, in particular, the interplay that occurs
between the two in dense star clusters.
But we
are not yet done.  Our results have raised several important questions
related to these topics and, by extension, the history of our Galaxy.
Once again, we are left asking:  Where do we go from here?   For
example, how else might we use observations of blue straggler
populations to learn about the dynamical histories of their host
clusters?  How can we use this information to constrain the
interactions that occur in clusters between stellar dynamics and
binary star evolution?  How do these interactions affect the
observed properties of their binary populations and the various types
of stellar exotica they are home to?  There is also the issue of the
universality of the stellar 
initial mass function.  What combination of initial concentrations and
mass functions should have evolved dynamically over the lifetimes of
star clusters to reproduce the currently observed mass functions?
What can this tell us about primordial mass segregation in massive
star clusters, and the exact form of the stellar IMF?  All of these
issues relate to the over-arching questions:  How do stars
form?; and:  How did the history of our Galaxy unfold?  
The next few years promise to hold exciting advances in the search for
answers to these questions.

\label{lastpage}

\end{doublespace}

\appendix
\pagestyle{fancy}
\headheight 20pt
\lhead{PhD Thesis - N. Leigh }
\rhead{McMaster Physics and Astronomy}
\chead{}
\lfoot{}
\cfoot{\thepage}
\rfoot{}
\renewcommand{\headrulewidth}{0.1pt}
\renewcommand{\footrulewidth}{0.1pt}

\chapter{Appendix A} \label{Appendix-A} 
\thispagestyle{fancy} 

In this appendix, we present the collisional cross sections and
time-scales used in Chapter~\ref{chapter2}.  The
gravitationally-focused cross sections for 1+1,
1+2, 2+2, 1+3, 2+3 and 3+3 collisions can be found using
Equation 6 from \citet{leonard89}.  Neglecting the first term and
assuming that binary and triple stars are on average twice and three
times as massive as single stars, respectively, this gives for the
various collisional cross sections:
\begin{equation}
\label{eqn:cs-1+1}
\sigma_{1+1} \sim \frac{8{\pi}GmR}{v_{rel}^2},
\end{equation}
\begin{equation}
\label{eqn:cs-1+2}
\sigma_{1+2} \sim \frac{3{\pi}Gma_b}{v_{rel}^2},
\end{equation}
\begin{equation}
\label{eqn:cs-2+2}
\sigma_{2+2} \sim \frac{8{\pi}Gma_b}{v_{rel}^2},
\end{equation}
\begin{equation}
\label{eqn:cs-1+3}
\sigma_{1+3} \sim \frac{4{\pi}Gma_t}{v_{rel}^2},
\end{equation}
\begin{equation}
\label{eqn:cs-2+3}
\sigma_{2+3} \sim \frac{5{\pi}Gm(a_b + a_t)}{v_{rel}^2},
\end{equation}
\begin{equation}
\label{eqn:cs-3+3}
\sigma_{3+3} \sim \frac{12{\pi}Gma_t}{v_{rel}^2}.
\end{equation}
Values for the
pericenters assumed for the various types of encounters are shown in
Table~\ref{table:peri}, where $R$ is the average stellar
radius, $a_b$ is the average binary semi-major axis and $a_t$ is the
average semi-major axis of the outer orbits of triples.

\begin{table}
\centering
\caption{Pericenters Assumed for Each Encounter Type
  \label{table:peri}}
\begin{tabular}{lc}
\hline
Encounter Type & Pericenter \\
\hline
1+1 & $2R$ \\
1+2 & $a_b/2$ \\
1+3 & $a_t/2$ \\
2+2 & $a_b$ \\
2+3 & $(a_b + a_t)/2$ \\
3+3 & $a_t$ \\
\hline
\end{tabular}
\end{table}

In general, the time between each of the different encounter
types can be found using Equation~\ref{eqn:coll-rate} and
the gravitationally-focused cross sections for collision given
above.  Following the derivation of \citet{leonard89}, we can write
the encounter rate in the general form:
\begin{equation}
\label{eqn:coll-rate-gen}
\Gamma_{x+y} = N_xn_y{\sigma}_{x+y}v_{x+y},
\end{equation}
where $N_x$ and $n_y$ are the number and number density, respectively,
of single, binary or triple stars and $v_{x+y}$ is the relative
velocity at infinity between objects $x$ and $y$.  For instance, the
time between binary-binary encounters in the core of a cluster is
given by:
\begin{equation}
\begin{gathered}
\label{eqn:coll2+2}
\tau_{2+2} = 1.3 \times 10^7f_b^{-2} \Big(\frac{1
  pc}{r_c}\Big)^3\Big(\frac{10^3
  pc^{-3}}{n_0}\Big)^2 \\
\Big(\frac{v_{rms}}{5 km/s}\Big)\Big(\frac{0.5
  M_{\odot}}{<m>}\Big)\Big(\frac{1
  AU}{a_{b}} \Big) \mbox{ years}.
\end{gathered}
\end{equation}
Similarly, the times between 1+1, 1+2, 1+3, 2+3 and 3+3 encounters are
given by:
\begin{equation}
\begin{gathered}
\label{eqn:coll1+1}
\tau_{1+1} = 1.1 \times 10^{10}(1-f_b-f_t)^{-2}\Big(\frac{1 pc}{r_c}
\Big)^3 \Big(\frac{10^3 pc^{-3}}{n_0} \Big)^2 \\
\Big(\frac{v_{rms}}{5 km/s} \Big) \Big(\frac{0.5 M_{\odot}}{<m>} \Big)
\Big(\frac{0.5 R_{\odot}}{<R>} \Big)\mbox{ years},
\end{gathered}
\end{equation}
\begin{equation}
\begin{gathered}
\label{eqn:coll1+2}
\tau_{1+2} = 3.4 \times 10^7(1-f_b-f_t)^{-1}f_b^{-1} \Big(\frac{1
  pc}{r_c}
\Big)^3 \Big(\frac{10^3 pc^{-3}}{n_0} \Big)^2 \\
\Big(\frac{v_{rms}}{5
  km/s} \Big) \Big(\frac{0.5 M_{\odot}}{<m>} \Big) \Big(\frac{1
  AU}{a_{b}} \Big)\mbox{ years},
\end{gathered}
\end{equation}
\begin{equation}
\begin{gathered}
\label{eqn:coll1+3}
\tau_{1+3} = 2.6 \times 10^7(1-f_b-f_t)^{-1}f_t^{-1} \Big(\frac{1
  pc}{r_c}
\Big)^3 \Big(\frac{10^3 pc^{-3}}{n_0} \Big)^2 \\
\Big(\frac{v_{rms}}{5
  km/s} \Big) \Big(\frac{0.5 M_{\odot}}{<m>} \Big) \Big(\frac{1
  AU}{a_{t}} \Big)\mbox{ years},
\end{gathered}
\end{equation}
\begin{equation}
\begin{gathered}
\label{eqn:coll2+3}
\tau_{2+3} = 2.0 \times 10^7f_b^{-1}f_t^{-1} \Big(\frac{1 pc}{r_c}
\Big)^3 \Big(\frac{10^3 pc^{-3}}{n_0} \Big)^2 \\
\Big(\frac{v_{rms}}{5
  km/s} \Big) \Big(\frac{0.5 M_{\odot}}{<m>} \Big) \Big(\frac{1
  AU}{a_{b}+a_{t}} \Big)\mbox{ years},
\end{gathered}
\end{equation}
and
\begin{equation}
\begin{gathered}
\label{eqn:coll3+3}
\tau_{3+3} = 8.3 \times 10^6f_t^{-2} \Big(\frac{1 pc}{r_c}
\Big)^3 \Big(\frac{10^3 pc^{-3}}{n_0} \Big)^2 \\
\Big(\frac{v_{rms}}{5
  km/s} \Big) \Big(\frac{0.5 M_{\odot}}{<m>} \Big) \Big(\frac{1
  AU}{a_{t}} \Big)\mbox{ years}.
\end{gathered}
\end{equation}

\pagestyle{fancy}
\headheight 20pt
\lhead{PhD Thesis - N. Leigh }
\rhead{McMaster Physics and Astronomy}
\chead{}
\lfoot{}
\cfoot{\thepage}
\rfoot{}
\renewcommand{\headrulewidth}{0.1pt}
\renewcommand{\footrulewidth}{0.1pt}

\chapter{Appendix B} \label{Appendix-B}
\thispagestyle{fancy}

In this appendix, we present our selection criteria for BS, RGB, HB
and MSTO stars used in Chapter~\ref{chapter4}.  Our method is similar
to that described in \citet{leigh07},
and we have used this as a basis for the selection criteria presented
in this chapter.  First, we define a location for the MSTO in the
(F606W-F814W)-F814W plane using our isochrone fits.  The MSTO is
chosen to be the bluest point along the MS of each isochrone, which we
denote by ((V-I)$_{MSTO}$,I$_{MSTO}$).  In order to distinguish BSs
from MSTO stars, we impose the conditions:

\begin{equation}
\label{eqn:bs_msto1}
F814W \le m_1(F606W-F814W) + b_{11},
\end{equation}
where the slope of this line is $m_1 = -9$ and its y-intercept is
given by:
\begin{equation}
\label{eqn:b_11}
b_{11} = (I_{MSTO}-0.10)-m_1((V-I)_{MSTO}-0.10)
\end{equation}

Similarly, we distinguish BSs from HB stars by defining the following
additional boundaries:
\begin{eqnarray}
\label{bs_hb1}
F814W &\ge& m_1(F606W-F814W) + b_{12} \\
F814W &\ge& m_2(F606W-F814W) + b_{21} \\
F814W &\le& m_2(F606W-F814W) + b_{22} \\
F814W &\ge& m_{HB}(F606W-F814W) + b_{HB} \\
(F606W-F814W) &\ge& (V-I)_{HB} \\
F814W &\le& I_{MSTO},
\end{eqnarray}
where $m_2 = 6$, $m_{HB} = -1.5$ and $(V-I)_{HB} = (V-I)_{MSTO} -
0.4$.  We also define:
\begin{eqnarray}
\label{b_12}
b_{12} &=& (I_{MSTO}-0.55)-m_1((V-I)_{MSTO}-0.55) \\
b_{21} &=& (I_{MSTO}-0.80)-m_2((V-I)_{MSTO}+0.10) \\
b_{22} &=& (I_{MSTO}+0.30)-m_2((V-I)_{MSTO}-0.20),
\end{eqnarray}
and $b_{HB} = I_{HB} + 1.2$, where $I_{HB}$ roughly corresponds to the
mid-point of points that populate the HB and is chosen
by eye for each cluster so that our selection criteria best fits the
HB in all of the CMDs in our sample.

We apply a similar set of conditions to the RGB in order to select
stars belonging to this stellar population.  These boundary conditions
are:
\begin{eqnarray}
\label{rgb_1}
F814W &\ge& m_{HB}(F606W-F814W) + b_{HB} \\
F814W &\ge& m_{RGB}(F606W-F814W) + b_{31} \\
F814W &\le& m_{RGB}(F606W-F814W) + b_{32} \\
F814W &\le& I_{RGB},
\end{eqnarray}
where $m_{RGB} = -23$, $I_{RGB}$ is defined as the F814W magnitude
corresponding to a core helium mass of 0.08 M$_{\odot}$ and:
\begin{eqnarray}
\label{b_31_and_b_32}
b_{31} &=& (I_{MSTO}-0.60)-m_{RGB}((V-I)_{MSTO}+0.05) \\
b_{32} &=& (I_{MSTO}-0.60)-m_{RGB}((V-I)_{MSTO}+0.25)
\end{eqnarray}

Core helium-burning stars, which we refer to as HB stars, are selected
if they satisfy one of the following sets of criteria:
\begin{eqnarray}
\label{hb1}
F814W &\ge& m_{HB}(F606W-F814W) + (b_{HB}-1.0) \\
F814W &\le& m_{HB}(F606W-F814W) + b_{HB} \\
(F606W-F814W) &\le& (V-I)_{MSTO} + (V-I)_{HB},
\end{eqnarray}
\begin{eqnarray}
\label{hb2}
F814W &>& m_{HB}(F606W-F814W) + b_{HB} \\
F814W &\le& I_{MSTO} + 2.5 \\
(F606W-F814W) &<& (V-I)_{MSTO} - 0.4,
\end{eqnarray}
or
\begin{eqnarray}
\label{hb3}
F814W &<& m_1(F606W-F814W) + b_{12} \\
F814W &>& m_{HB}(F606W-F814W) + b_{HB} \\
(F606W-F814W) &\ge& (V-I)_{MSTO} - 0.4
\end{eqnarray}
We define $(V-I)_{HB}$ on a cluster-by-cluster basis in order to
ensure that we do not over- or under-count the number of HB stars.
This is because the precise value of (F606W-F814W) at which the HB
becomes the RGB varies from cluster-to-cluster.  In addition, the
precise location of the transition in the cluster CMD between HB
and EHB stars remains poorly understood.  To avoid this ambiguity, we
consider HB and EHB stars
together throughout our analysis, and collectively refer to all core
helium-burning stars as HB stars throughout this chapter.

Finally, MSTO stars are selected according to the following criteria:
\begin{eqnarray}
\label{msto_1}
F814W &>& I_{RGB} \\
F814W &>& m_1(F606W-F814W) + b_{11} \\
F814W &\le& (V-I)_{MSTO}
\end{eqnarray}


\begin{singlespace}
  
  \pagestyle{fancy}
  \headheight 20pt
  \lhead{Ph.D. Thesis - N. Leigh}
  \rhead{McMaster - Physics and Astronomy}
  \chead{}
  \lfoot{}
  \cfoot{\thepage}
  \rfoot{}
  \renewcommand{\headrulewidth}{0.1pt}
  \renewcommand{\footrulewidth}{0.1pt}
  \thispagestyle{fancy}
  
  \bibliography{myrefs}
  \bibliographystyle{apj}
  \nocite{*}
  
  \addcontentsline{toc}{chapter}{Bibliography}
  
  \thispagestyle{fancy}
  
\end{singlespace}

\end{document}